\PassOptionsToPackage{unicode}{hyperref}
\PassOptionsToPackage{hyphens}{url}
\PassOptionsToPackage{dvipsnames,svgnames,x11names}{xcolor}
\documentclass[
  11pt]{article}
\usepackage{xcolor}
\usepackage[margin=1in]{geometry}
\usepackage{amsmath,amssymb}
\usepackage{iftex}
\ifPDFTeX
  \usepackage[T1]{fontenc}
  \usepackage[utf8]{inputenc}
  \usepackage{textcomp} % provide euro and other symbols
\else % if luatex or xetex
  \usepackage{unicode-math} % this also loads fontspec
  \defaultfontfeatures{Scale=MatchLowercase}
  \defaultfontfeatures[\rmfamily]{Ligatures=TeX,Scale=1}
\fi
\usepackage{lmodern}
\ifPDFTeX\else
\fi
\IfFileExists{upquote.sty}{\usepackage{upquote}}{}
\IfFileExists{microtype.sty}{% use microtype if available
  \usepackage[]{microtype}
  \UseMicrotypeSet[protrusion]{basicmath} % disable protrusion for tt fonts
}{}
\usepackage{setspace}
\makeatletter
\@ifundefined{KOMAClassName}{% if non-KOMA class
  \IfFileExists{parskip.sty}{%
    \usepackage{parskip}
  }{% else
    \setlength{\parindent}{0pt}
    \setlength{\parskip}{6pt plus 2pt minus 1pt}}
}{% if KOMA class
  \KOMAoptions{parskip=half}}
\makeatother
\usepackage{color}
\usepackage{fancyvrb}

\DefineVerbatimEnvironment{Highlighting}{Verbatim}{commandchars=\\\{\}}
\newenvironment{Shaded}{}{}

\newcommand{\CommentTok}[1]{\textcolor[rgb]{0.38,0.63,0.69}{\textit{#1}}}

\newcommand{\ExtensionTok}[1]{#1}

\newcommand{\FunctionTok}[1]{\textcolor[rgb]{0.02,0.16,0.49}{#1}}

\newcommand{\NormalTok}[1]{#1}
\newcommand{\OperatorTok}[1]{\textcolor[rgb]{0.40,0.40,0.40}{#1}}

\usepackage{longtable,booktabs,array}
\usepackage{calc} % for calculating minipage widths
\usepackage{etoolbox}
\makeatletter
\patchcmd\longtable{\par}{\if@noskipsec\mbox{}\fi\par}{}{}
\makeatother
\IfFileExists{footnotehyper.sty}{\usepackage{footnotehyper}}{\usepackage{footnote}}
\makesavenoteenv{longtable}
\usepackage{graphicx}
\makeatletter
\newsavebox\pandoc@box
\newcommand*\pandocbounded[1]{% scales image to fit in text height/width
  \sbox\pandoc@box{#1}%
  \Gscale@div\@tempa{\textheight}{\dimexpr\ht\pandoc@box+\dp\pandoc@box\relax}%
  \Gscale@div\@tempb{\linewidth}{\wd\pandoc@box}%
  \ifdim\@tempb\p@<\@tempa\p@\let\@tempa\@tempb\fi% select the smaller of both
  \ifdim\@tempa\p@<\p@\scalebox{\@tempa}{\usebox\pandoc@box}%
  \else\usebox{\pandoc@box}%
  \fi%
}
\def\fps@figure{htbp}
\makeatother
\providecommand{\tightlist}{%
  \setlength{\itemsep}{0pt}\setlength{\parskip}{0pt}}
\usepackage{newunicodechar}
\usepackage{etoolbox}
\usepackage{seqsplit}
\usepackage{microtype}
\usepackage{titlesec}
\usepackage{pifont}
\usepackage{abstract}

\titleformat{\section}{\normalfont\Large\bfseries}{\thesection}{0.6em}{}
\titleformat{\subsection}{\normalfont\large\bfseries}{\thesubsection}{0.5em}{}
\titleformat{\subsubsection}{\normalfont\normalsize\bfseries}{\thesubsubsection}{0.4em}{}
\titlespacing*{\section}{0pt}{1.2em}{0.5em}
\titlespacing*{\subsection}{0pt}{1em}{0.4em}
\titlespacing*{\subsubsection}{0pt}{0.8em}{0.3em}
\newcommand{\rkbest}{\ding{115}}
\newcommand{\rkworst}{\ding{116}}
\newcommand{\rkmid}{\ding{117}}
\newcommand{\bttt}[1]{\texttt{\seqsplit{#1}}}
\AtBeginEnvironment{longtable}{\small}
\newunicodechar{≤}{\ensuremath{\leq}}
\newunicodechar{≥}{\ensuremath{\geq}}
\newunicodechar{≈}{\ensuremath{\approx}}
\newunicodechar{≠}{\ensuremath{\neq}}
\newunicodechar{∞}{\ensuremath{\infty}}
\newunicodechar{→}{\ensuremath{\rightarrow}}
\newunicodechar{←}{\ensuremath{\leftarrow}}
\newunicodechar{↔}{\ensuremath{\leftrightarrow}}
\newunicodechar{×}{\ensuremath{\times}}
\newunicodechar{·}{\textperiodcentered}
\newunicodechar{▲}{\ding{115}}
\newunicodechar{◆}{\ding{117}}
\newunicodechar{▼}{\ding{116}}
\newunicodechar{°}{\ensuremath{^{\circ}}}
\newunicodechar{¹}{\ensuremath{^{1}}}
\newunicodechar{²}{\ensuremath{^{2}}}
\newunicodechar{³}{\ensuremath{^{3}}}
\newunicodechar{⁴}{\ensuremath{^{4}}}
\newunicodechar{⁵}{\ensuremath{^{5}}}
\newunicodechar{…}{\ldots}
\newunicodechar{–}{--}
\newunicodechar{—}{---}
\newunicodechar{§}{\S{}}
\newunicodechar{¶}{\P{}}
\newunicodechar{€}{\texteuro}
\newunicodechar{£}{\pounds}
\newunicodechar{±}{\ensuremath{\pm}}
\newunicodechar{α}{\ensuremath{\alpha}}
\newunicodechar{β}{\ensuremath{\beta}}
\newunicodechar{π}{\ensuremath{\pi}}
\newunicodechar{λ}{\ensuremath{\lambda}}
\newunicodechar{∈}{\ensuremath{\in}}
\newunicodechar{∩}{\ensuremath{\cap}}
\newunicodechar{∪}{\ensuremath{\cup}}
\newunicodechar{⊆}{\ensuremath{\subseteq}}
\newunicodechar{⊂}{\ensuremath{\subset}}
\newunicodechar{∅}{\ensuremath{\emptyset}}
\newunicodechar{∀}{\ensuremath{\forall}}
\newunicodechar{∃}{\ensuremath{\exists}}
\newunicodechar{Δ}{\ensuremath{\Delta}}
\newunicodechar{Σ}{\ensuremath{\Sigma}}
\newunicodechar{μ}{\ensuremath{\mu}}
\newunicodechar{σ}{\ensuremath{\sigma}}
\newunicodechar{∂}{\ensuremath{\partial}}
\newunicodechar{∇}{\ensuremath{\nabla}}
\usepackage{bookmark}
\IfFileExists{xurl.sty}{\usepackage{xurl}}{} % add URL line breaks if available
\hypersetup{
  pdftitle={AI Code Sandboxes: A Comparative Security Study Part 1 of 2 --- Engine-Level Properties (Attack Surface{,} Leakage{,} Stackability{,} CVE History{,} Patch Cadence{,} Fuzzing)},
  colorlinks=true,
  linkcolor={black},
  filecolor={Maroon},
  citecolor={black},
  urlcolor={black},
  pdfcreator={LaTeX via pandoc}}

\title{AI Code Sandboxes: A Comparative Security Study
\\[0.35em] \large Part 1 of 2 --- Engine-Level
Properties \\[0.2em] \large (Attack Surface, Leakage,
Stackability, CVE History, Patch Cadence, Fuzzing)}
\author{\normalsize
\begin{tabular}{c@{\hskip 4em}c}
  George Andronchik & Pavel Lokhmakov, PhD \\[0.2em]
  \small\itshape fellows.tech, orbitalab.dev & \small\itshape fellows.tech \\
  \small\texttt{george@orbitalab.dev} & \small\texttt{l.pavel.m@gmail.com} \\
\end{tabular}}
\date{2026-05-24}

\begin{document}
\maketitle
\begin{abstract}
This paper reads six engine-level measurements together --- 1.1 host
attack surface, 1.2 information leakage, 1.3 defense-in-depth
stackability, 1.4 public CVE history, 1.5 patch cadence, and 1.6
upstream fuzzing posture --- to describe how five AI-sandbox products
isolate guest code from the host kernel. No single axis is a sufficient
basis for a comparative judgement; the cross-axis reading is the
load-bearing analysis. Three high-level findings: (1) engine classes
(microVM, userspace kernel, OCI container) separate cleanly on every
architectural axis, but products within a class do not; (2) product pin
policy is the dominant operator-facing variable --- engine-side patch
latency aggregates to $\approx$0 days for coordinated disclosures, while
downstream lag spans 0 days to 471+ days to ``opaque'' to \(\infty\);
(3) fuzzing investment splits into three tiers, and the strongest
combination --- microVM \(\times\) continuous public fuzzer --- is
unoccupied in this set, leaving the ``0 published CVEs \(\times\) no
upstream fuzzer \(\times\) no academic study'' intersection structurally
unmeasured. We report per-axis orderings, per-product portraits, and a
threat-model qualification matrix; no overall ranking is proposed.
Companion repository (code, Apache-2.0):
\url{https://github.com/orbitalab/RnD-ai-sandboxes-sec-study-part-1}.
License: CC BY 4.0.
\end{abstract}

\setstretch{1.05}
\subsection{Background}\label{background}

The framing for this paper sits at the intersection of three lines of
work: (a) the agentic-AI security literature that establishes
containment as an open challenge; (b) the dangerous-capability research
that places the threat on a current rather than future timeline; and (c)
the execution-environment taxonomy that defines the engine classes under
test.

\textbf{Agentic-AI security frame.} A 2026 umbrella survey of agentic AI
security (Chhabra et al., IEEE Access --- arXiv:2510.23883) carves out
\emph{Sandboxing \& Capability Confinement} as one of four defense
families and names sandbox/emulated-environment fidelity measurement as
an open challenge (§VI.C(4)) --- the direct hand-off for this work. The
sibling SoK from Dehghantanha \& Homayoun (arXiv:2603.22928) treats
sandboxing as a single bullet in a 7-goal × 10-surface × 5-path
taxonomy; both surveys flag the absence of engine-level measurement at
this depth.

\textbf{Dangerous-capability urgency.} Three papers anchor the urgency
framing in a current-decade timeline rather than a future-decade one.
Phuong et al.~(arXiv:2403.13793, Google DeepMind, 27 authors) report no
strong dangerous capabilities in Gemini 1.0 Ultra/Pro/Nano but flag
early warning signs, with expert-forecast threshold-crossing windows of
2025--2029. Wei et al.~(arXiv:2505.18384v5) demonstrate that an
adversary can improve an offensive cybersecurity agent's capability on
InterCode CTF by more than 40\% relative to baseline within 8 H100
GPU-hours and ≤\$36 of compute, without external assistance. Folkerts et
al.~(arXiv:2603.11214, UK AISI) measure log-linear scaling of multi-step
cyber-attack capability with inference-time compute through Opus 4.6
with no observed plateau; their best run completes 22 of 32
corporate-network attack steps at a 100M-token budget. Containment
design at the engine layer is therefore a present-tense concern.

\textbf{Engine taxonomy lineage.} The four-class execution-environment
taxonomy (VMs · containers · gVisor · microVMs) we anchor against in §1
was first widely surveyed by Marin et al.~(arXiv:2107.03832, 2021/2022)
Table 2. Cloud Hypervisor and libkrun are absent from their taxonomy
(consistent with the 2021 publication date); we extend the microVM class
population to match the AI-agent-sandbox product set under test, but the
class structure is inherited.

\textbf{``Sandbox'' disambiguation.} The term \emph{sandbox} in the
AI-agent literature can mean either engine-level isolation (the
substrate this paper measures) or LM-emulated execution environments.
The earlier and widely-cited precedent for the second meaning is ToolEmu
(Ruan et al., ICLR 2024 / arXiv:2309.15817), where a GPT-4-based
emulator fabricates tool outputs from tool specifications without any
engine layer --- orthogonal to engine-level isolation but sharing the
term. This paper measures engine-level isolation; the LM-emulated line
is named for lineage disambiguation only.

\subsection{Methodology in brief}\label{methodology-in-brief}

The full methodology lives in
\href{https://github.com/orbitalab/RnD-ai-sandboxes-sec-study-part-1/blob/main/.docs/sandbox-isolation-methodology-v2.md}{\texttt{.docs/sandbox-isolation-methodology-v2.md}}
($\approx$1,200 lines, per-axis rationale and procedure); this section restates
the parts that govern how the synthesis should be read. The six source
papers (1.1--1.6, linked in the front-matter) ship as companion
artefacts in the public repo rather than in-document appendices;
§2.4--§2.6 intentionally inline their load-bearing material --- rollup
matrices, per-engine rankings, verdict tables, structural caveats --- so
the synthesis reads standalone.

\textbf{Six Tier-1 axes, three engine classes, five products.} The
Tier-1 set measures properties determined by the underlying engine: 1.1
host attack surface, 1.2 information leakage, 1.3 defense-in-depth
stackability, 1.4 public CVE history, 1.5 patch cadence, 1.6 upstream
fuzzing posture. Products are grouped by engine class (microVM /
userspace kernel / OCI container) so that within-class divergence on an
axis is itself a finding --- it means the property is product-driven,
not engine-driven.

\textbf{Five test genres.} Different axes produce different \emph{kinds}
of evidence, with different epistemic strength:

{\def\LTcaptype{none} % do not increment counter
\begin{longtable}[]{@{}
  >{\raggedright\arraybackslash}p{(\linewidth - 6\tabcolsep) * \real{0.2500}}
  >{\raggedright\arraybackslash}p{(\linewidth - 6\tabcolsep) * \real{0.2500}}
  >{\raggedright\arraybackslash}p{(\linewidth - 6\tabcolsep) * \real{0.2500}}
  >{\raggedright\arraybackslash}p{(\linewidth - 6\tabcolsep) * \real{0.2500}}@{}}
\toprule\noalign{}
\begin{minipage}[b]{\linewidth}\raggedright
Genre
\end{minipage} & \begin{minipage}[b]{\linewidth}\raggedright
What it does
\end{minipage} & \begin{minipage}[b]{\linewidth}\raggedright
Output
\end{minipage} & \begin{minipage}[b]{\linewidth}\raggedright
Axes
\end{minipage} \\
\midrule\noalign{}
\endhead
\bottomrule\noalign{}
\endlastfoot
Measurement & Quantifies an architectural property & Numbers,
distributions & 1.1, 1.5 (downstream) \\
Probe & Tests for a known-class hole in default config & Pass/fail per
probe & 1.2 \\
Composability test & Applies a hardening layer, checks if the engine
still works & Per-layer verdict & 1.3 \\
Desk research & Reads repos, advisories, dashboards & Tables, narrative
& 1.4, 1.6 \\
\end{longtable}
}

CVE replay is \emph{not} in the methodology --- single-shot replay tests
a specific PoC, not a bug class, and most cells in a 5-CVE × 3-class
replay matrix are trivial passes by architecture. The methodology
replaces replay with the meta-analysis trio (1.4 + 1.5 + 1.6).

\textbf{Pinning.} A single Linux host (Hetzner bare-metal, fixed kernel
+ distro), version-pinned SDKs, and \texttt{create()}-with-no-arguments
defaults. ``Default'' is what the product gives you out of the box;
deviations are recorded as product-level findings.

\textbf{Verdict semantics.} Each axis produces one of five verdicts:
\texttt{pass}, \texttt{fail}, \texttt{partial}, \texttt{inconclusive},
\texttt{skipped}. Hard rule: \texttt{inconclusive} ≠ \texttt{pass} and ≠
\texttt{fail}. The synthesis preserves this distinction (see the
\texttt{—} cell for libkrun on 1.4 in §2.4, and the §3.5 reading of
``unmeasured ≠ safe'').

\textbf{Threat model.} AISI \texttt{T0.H2.N2} --- a single-tenant
operator runs untrusted code (e.g., AI-agent-generated Python) on their
own infrastructure. The operator does not fully trust the code but does
trust the infrastructure. Explicitly out of scope: multi-tenant SaaS
(tenant-A vs tenant-B isolation), microarchitectural side channels,
orchestrator/control-plane attacks, live exploitation of unpublished
bugs. An operator reading these portraits with a different threat model
needs a different evidence base.

\textbf{No composite score.} The methodology forbids combining per-axis
verdicts into a single ranking. Different operators weight axes
differently; the per-axis numbers are presented in their own column, the
threat-model qualification matrix (§4) splits the single threat into
four operator-facing sub-questions, and no cross-axis sum is computed.

\textbf{Methodology revisions in effect.} v2.1 added the §1.1.4
primitive reachability matrix (14 entries × 5 products --- the
load-bearing artefact of 1.1's ``primitives reachable / 14'' ordering)
and §2.7 compositional kill-chains. v2.2 expanded 1.3 from 5 layers to 7
(added \texttt{no\_new\_privs} and \texttt{pids.max}). v2.3 split each
1.3 cell into \texttt{applied} × \texttt{stack-effort} (with a five-rung
ladder
\texttt{trivial\ /\ policy\ /\ redeploy\ /\ not-exposed\ /\ engine-blocks})
--- the cross-class meaningful 1.3 ordering in §2.3 uses
\texttt{not-exposed} count for this reason.

\subsection{1. The five products under
test}\label{the-five-products-under-test}

The methodology evaluates five products that map cleanly to three engine
classes:

\begin{figure}
\centering
\includegraphics[width=1\linewidth,height=\textheight,keepaspectratio,alt={Three engine classes; the five products tested across them}]{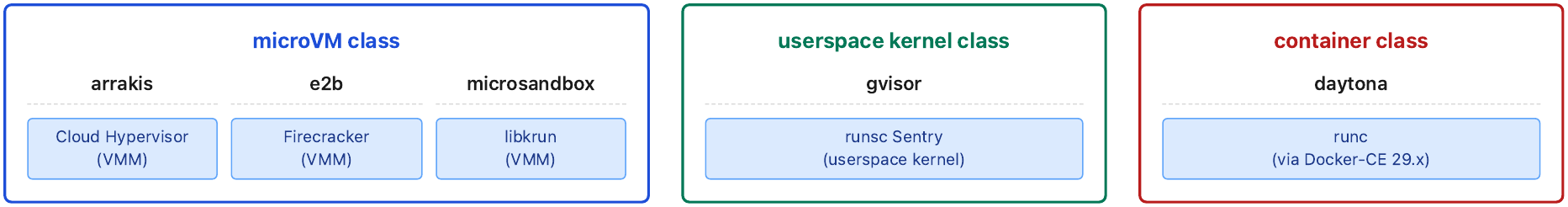}
\caption{Three engine classes; the five products tested across them}
\end{figure}

{\def\LTcaptype{none} % do not increment counter
\begin{longtable}[]{@{}
  >{\raggedright\arraybackslash}p{(\linewidth - 8\tabcolsep) * \real{0.2000}}
  >{\raggedright\arraybackslash}p{(\linewidth - 8\tabcolsep) * \real{0.2000}}
  >{\raggedright\arraybackslash}p{(\linewidth - 8\tabcolsep) * \real{0.2000}}
  >{\raggedright\arraybackslash}p{(\linewidth - 8\tabcolsep) * \real{0.2000}}
  >{\raggedright\arraybackslash}p{(\linewidth - 8\tabcolsep) * \real{0.2000}}@{}}
\toprule\noalign{}
\begin{minipage}[b]{\linewidth}\raggedright
Product
\end{minipage} & \begin{minipage}[b]{\linewidth}\raggedright
Engine
\end{minipage} & \begin{minipage}[b]{\linewidth}\raggedright
Engine class
\end{minipage} & \begin{minipage}[b]{\linewidth}\raggedright
Engine language
\end{minipage} & \begin{minipage}[b]{\linewidth}\raggedright
Pin model
\end{minipage} \\
\midrule\noalign{}
\endhead
\bottomrule\noalign{}
\endlastfoot
\textbf{arrakis} & Cloud Hypervisor & microVM & Rust & Per-release pin
in product's \texttt{Cargo.lock}-equivalent (currently \texttt{v44.0},
last bumped 2025-02-04 per 1.5) \\
\textbf{e2b} & Firecracker & microVM & Rust & Self-hosted: orchestrator
default pin in \texttt{e2b-dev/infra} (currently
\texttt{v1.14.1\_458ca91}, 399 days since bump per 1.5); the separate
\bttt{e2b-dev/fc-versions} release-artefact registry has shipped newer
manual builds in May 2026 that the orchestrator default does not
consume. Cloud-hosted: opaque \\
\textbf{microsandbox} & libkrun & microVM (lightweight) & Rust &
Per-release pin, 1--3 month cadence per 1.5 \\
\textbf{gvisor} & runsc & userspace kernel & Go & Rolling main; the
product \emph{is} upstream \\
\textbf{daytona} & runc & OCI container & Go & Indirect: pulls
\texttt{containerd.io} from Docker-CE 29.x on the default deploy path;
runc 1.3.5 bundled \\
\end{longtable}
}

The three-class structure (microVM / userspace kernel / OCI container)
is consistent with Marin et al.'s 2021/2022 execution-environment
taxonomy (arXiv:2107.03832 Table 2 --- VMs · containers · gVisor ·
microVMs); we adopt that class structure and update the microVM
population with Cloud Hypervisor and libkrun, which post-date their
survey.

Two operational notes:

\begin{enumerate}
\def\labelenumi{\arabic{enumi}.}
\tightlist
\item
  \textbf{Product is not engine.} The synthesis distinguishes the engine
  row (the upstream code that does isolation work) from the product row
  (the deployable artefact an operator runs). Most axes have engine-side
  measurements (1.1 attack surface, 1.4 CVE history, 1.6 fuzzing
  posture) and product-side measurements (1.3 stackability for the
  deploy shape, 1.5 downstream lag for the pin policy). Per-axis
  orderings in §2 are reported at whichever row the axis measures.
\item
  \textbf{daytona's default install path is Docker Compose.} Per 1.5,
  the default daytona deploy is
  \texttt{docker\ compose\ -f\ docker/docker-compose.yaml\ up\ -d},
  which pulls Docker-CE 29.x with runc 1.3.5 bundled via
  \texttt{containerd.io}. This includes all in-window runc CVE fixes.
  The alternative path (operator runs \texttt{apt\ install\ runc} from
  Ubuntu's archive) lands on the ``Ignored --- backport too intrusive''
  Ubuntu posture for the 2025-11 runc trio, but this is not the
  documented default and the \texttt{won\textquotesingle{}t-fix} finding
  survives only for operators who deviate from daytona's docs.
\end{enumerate}

\subsection{2. Per-axis orderings}\label{per-axis-orderings}

This section reports the five products' relative positions on each of
the six axes. Orderings use \rkbest{} (best on this axis) / \rkmid{} (middle) / \rkworst{}
(worst on this axis). Ties are marked at the same level. \textbf{These
are per-axis orderings, not weights for an overall ranking} --- the
methodology forbids combining them into a composite score.

The orderings at a glance:

\begin{figure}
\centering
\includegraphics[width=1\linewidth,height=\textheight,keepaspectratio,alt={Per-axis orderings across the five products (\rkbest{} best · \rkmid{} middle · \rkworst{} worst · --- unmeasured)}]{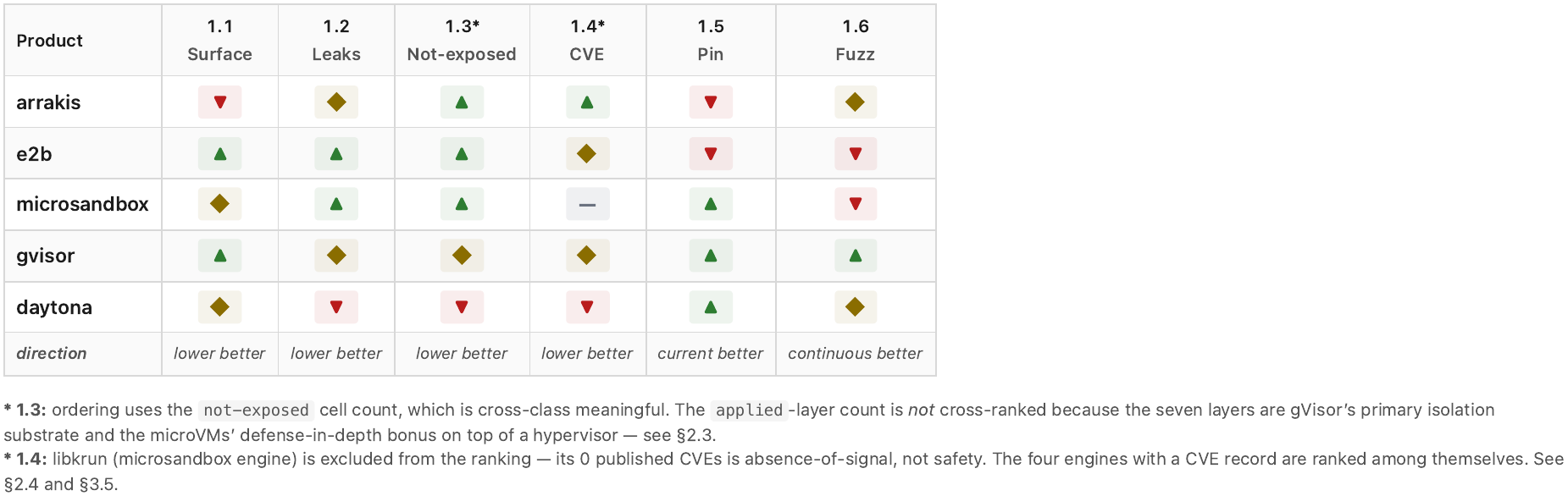}
\caption{Per-axis orderings across the five products (\rkbest{} best · \rkmid{} middle
· \rkworst{} worst · --- unmeasured)}
\end{figure}

\subsubsection{2.1 Axis 1.1 --- Host attack
surface}\label{axis-1.1-host-attack-surface}

The four-layer model that anchors the axis: an in-sandbox attacker sits
at L4 (guest userspace), and any escape to the host kernel at L1 must
travel through L2 (the runtime / mediator). 1.1's probes measure L2's
footprint on L1 --- the only surface a successful escape can cross.

\begin{figure}
\centering
\includegraphics[width=1\linewidth,height=\textheight,keepaspectratio,alt={Four-layer model and where each 1.1 probe attaches}]{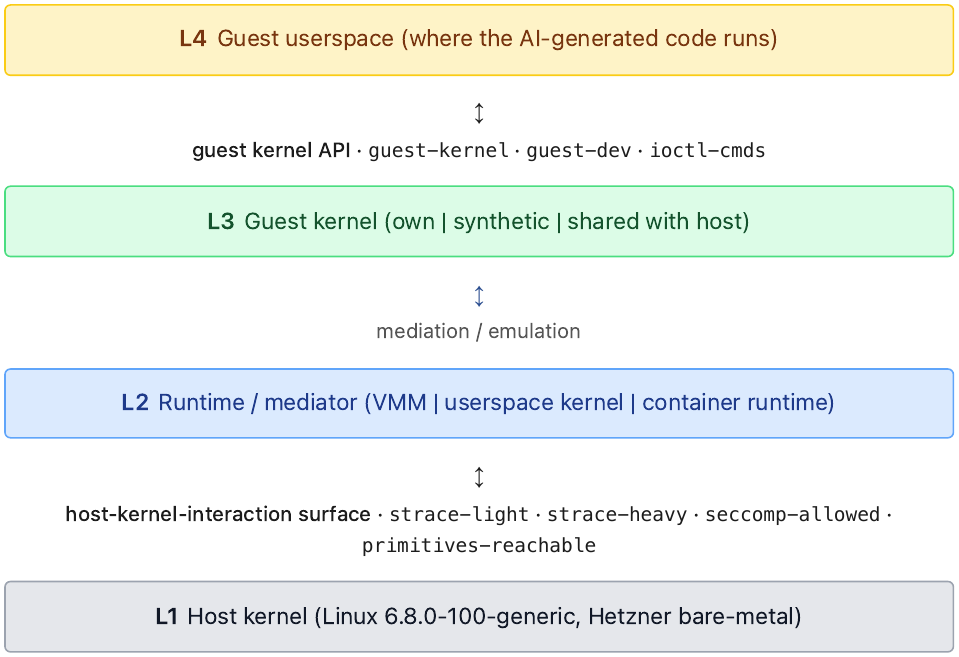}
\caption{Four-layer model and where each 1.1 probe attaches}
\end{figure}

\paragraph{2.1.1 Observed VMM/runtime strace
surface}\label{observed-vmmruntime-strace-surface}

The cross-engine baseline. Same fixed workloads
(\texttt{echo\ benchmark} × 100 for the light tier;
\texttt{python3\ -c\ "import\ pandas\ as\ pd,\ numpy\ as\ np;\ …"} × 100
for the heavy tier) for every product; distinct-syscall counts are
directly comparable across classes, total-call counts are not (microVMs'
guest syscalls happen inside the guest kernel at L3 and never surface at
the host strace level; gVisor and runc funnel guest syscalls through L2
and they all surface).

{\def\LTcaptype{none} % do not increment counter
\begin{longtable}[]{@{}
  >{\raggedright\arraybackslash}p{(\linewidth - 10\tabcolsep) * \real{0.1364}}
  >{\raggedleft\arraybackslash}p{(\linewidth - 10\tabcolsep) * \real{0.1818}}
  >{\raggedleft\arraybackslash}p{(\linewidth - 10\tabcolsep) * \real{0.1818}}
  >{\raggedleft\arraybackslash}p{(\linewidth - 10\tabcolsep) * \real{0.1818}}
  >{\raggedleft\arraybackslash}p{(\linewidth - 10\tabcolsep) * \real{0.1818}}
  >{\raggedright\arraybackslash}p{(\linewidth - 10\tabcolsep) * \real{0.1364}}@{}}
\toprule\noalign{}
\begin{minipage}[b]{\linewidth}\raggedright
Product
\end{minipage} & \begin{minipage}[b]{\linewidth}\raggedleft
Light --- distinct
\end{minipage} & \begin{minipage}[b]{\linewidth}\raggedleft
Light --- total
\end{minipage} & \begin{minipage}[b]{\linewidth}\raggedleft
Heavy --- distinct
\end{minipage} & \begin{minipage}[b]{\linewidth}\raggedleft
Heavy --- total
\end{minipage} & \begin{minipage}[b]{\linewidth}\raggedright
Trace target
\end{minipage} \\
\midrule\noalign{}
\endhead
\bottomrule\noalign{}
\endlastfoot
arrakis & 5 & 812 & 8 & 2,658 & \texttt{cloud-hypervisor} \\
e2b & 13 & 15,652 & 13 & 16,658 & \texttt{firecracker} \\
microsandbox & 22 & 7,158 & 25 & 15,834 & \texttt{msb} (libkrun host) \\
gvisor & 33 & 74,592 & 41 & 1,300,859 & \texttt{runsc-sandbox}
(Sentry) \\
daytona & 57 & 59,705 & 62 & 908,163 & \texttt{runc} container init \\
\end{longtable}
}

For microVMs the light-to-heavy delta in distinct syscalls is small
(Firecracker does not move; arrakis adds 3, microsandbox adds 3) --- the
VMM's hot path is workload-invariant under these probes. For Sentry and
runc the delta is larger and the total-call count grows by orders of
magnitude --- guest syscalls funnel through L2 and surface in the trace.
The 1.3 M heavy-tier total for gVisor vs 17 K for Firecracker is a
methodological artefact of where the trace attaches, not a security
finding on its own.

\paragraph{2.1.2 Seccomp filter ceiling}\label{seccomp-filter-ceiling}

The architectural upper bound on what syscalls L2 \emph{can ever} reach,
regardless of workload. Read from
\texttt{/proc/\textless{}pid\textgreater{}/task/*/status} (per-thread
mode matrix) and
\texttt{seccomp-tools\ dump\ \textless{}pid\textgreater{}}
(allowed-syscall list).

{\def\LTcaptype{none} % do not increment counter
\begin{longtable}[]{@{}
  >{\raggedright\arraybackslash}p{(\linewidth - 8\tabcolsep) * \real{0.1875}}
  >{\raggedright\arraybackslash}p{(\linewidth - 8\tabcolsep) * \real{0.1875}}
  >{\raggedleft\arraybackslash}p{(\linewidth - 8\tabcolsep) * \real{0.2500}}
  >{\raggedright\arraybackslash}p{(\linewidth - 8\tabcolsep) * \real{0.1875}}
  >{\raggedright\arraybackslash}p{(\linewidth - 8\tabcolsep) * \real{0.1875}}@{}}
\toprule\noalign{}
\begin{minipage}[b]{\linewidth}\raggedright
Product
\end{minipage} & \begin{minipage}[b]{\linewidth}\raggedright
Leader mode
\end{minipage} & \begin{minipage}[b]{\linewidth}\raggedleft
Allowed syscalls
\end{minipage} & \begin{minipage}[b]{\linewidth}\raggedright
Per-thread (mode × N / total)
\end{minipage} & \begin{minipage}[b]{\linewidth}\raggedright
Notes
\end{minipage} \\
\midrule\noalign{}
\endhead
\bottomrule\noalign{}
\endlastfoot
e2b & filter (mode 2) & 55 & mode2 × 5 / 5 & Firecracker's upstream
default filter on every thread \\
gvisor & filter (mode 2) & 84 & mode2 × 30 / 30 & runsc Sentry filter on
every thread \\
arrakis & disabled (mode 0) & --- (leader) & mode0 × 1, mode2 × 32 / 33
& Cloud Hypervisor installs seccomp on worker threads (vmm / vcpu* /
virtio / iou-wrk) \emph{after} leader arg parsing \\
microsandbox & disabled (mode 0) & --- & mode0 × 16 / 16 & No BPF filter
on any \texttt{msb} thread \\
daytona & disabled (mode 0) & --- & mode0 × 15 / 15 & No BPF filter on
any runc-init thread \\
\end{longtable}
}

The per-thread matrix is load-bearing for arrakis: the leader thread
reads mode 0, but the 32 security-critical worker threads carry a BPF
filter. A leader-only read would have classified arrakis identically to
microsandbox/daytona; the per-thread read places arrakis with e2b/gvisor
on enforced filter posture. The exact allowed-syscall count for
arrakis's per-thread filter is not yet surfaced ---
\texttt{seccomp-tools\ dump} cannot read a filter program off a thread
that isn't the install point.

\paragraph{2.1.3 Guest-visible surface --- devices and
kernel}\label{guest-visible-surface-devices-and-kernel}

\texttt{/dev} entry count from \texttt{ls\ /dev} inside the sandbox;
kernel string from \texttt{uname\ -a}.

{\def\LTcaptype{none} % do not increment counter
\begin{longtable}[]{@{}
  >{\raggedright\arraybackslash}p{(\linewidth - 6\tabcolsep) * \real{0.2500}}
  >{\raggedleft\arraybackslash}p{(\linewidth - 6\tabcolsep) * \real{0.2500}}
  >{\raggedright\arraybackslash}p{(\linewidth - 6\tabcolsep) * \real{0.2500}}
  >{\raggedright\arraybackslash}p{(\linewidth - 6\tabcolsep) * \real{0.2500}}@{}}
\toprule\noalign{}
Product & \texttt{/dev} entries & Guest kernel string & Class \\
\midrule\noalign{}
\endhead
\bottomrule\noalign{}
\endlastfoot
gvisor & 15 & \texttt{4.19.0-gvisor} & synthetic \\
microsandbox & 116 & \texttt{6.12.68} & own \\
e2b & 128 & \texttt{6.1.158} & own \\
arrakis & 170 & \texttt{6.12.8+} & own \\
daytona & 268 & \texttt{6.8.0-100-generic} & \textbf{host kernel
verbatim} \\
\end{longtable}
}

Two readings. (a) gVisor exposes the smallest \texttt{/dev} (15 entries)
--- the Sentry virtualises devices aggressively and the guest cannot see
what the Sentry decided not to expose. (b) daytona's guest kernel string
is \emph{bit-for-bit identical to the host kernel} --- by design for
runc-class containers (no separate guest kernel). Every published kernel
CVE that requires only userspace triggering applies inside this sandbox
at the same patch level as the host.

\paragraph{2.1.4 ioctl-cmds catalog
probe}\label{ioctl-cmds-catalog-probe}

The probe opens each character device with
\texttt{O\_RDONLY\textbar{}O\_NONBLOCK} and issues a fixed 10-cmd
catalog (\texttt{FIONREAD}, \texttt{FIONBIO}, \texttt{FIOCLEX},
\texttt{FIONCLEX}, \texttt{FIOQSIZE}, \texttt{TIOCGWINSZ},
\texttt{TCGETS}, \texttt{BLKGETSIZE}, \texttt{BLKSSZGET},
\texttt{\_IOC\_NULL}). A cmd that returns \texttt{0} / \texttt{EINVAL} /
\texttt{EPERM} / \texttt{EBUSY} / etc. is \emph{acknowledged} (the
driver's ioctl switch table knows the cmd); \texttt{ENOTTY} is \emph{not
supported}.

{\def\LTcaptype{none} % do not increment counter
\begin{longtable}[]{@{}
  >{\raggedright\arraybackslash}p{(\linewidth - 8\tabcolsep) * \real{0.1579}}
  >{\raggedleft\arraybackslash}p{(\linewidth - 8\tabcolsep) * \real{0.2105}}
  >{\raggedleft\arraybackslash}p{(\linewidth - 8\tabcolsep) * \real{0.2105}}
  >{\raggedleft\arraybackslash}p{(\linewidth - 8\tabcolsep) * \real{0.2105}}
  >{\raggedleft\arraybackslash}p{(\linewidth - 8\tabcolsep) * \real{0.2105}}@{}}
\toprule\noalign{}
\begin{minipage}[b]{\linewidth}\raggedright
Product
\end{minipage} & \begin{minipage}[b]{\linewidth}\raggedleft
Acknowledged / 10
\end{minipage} & \begin{minipage}[b]{\linewidth}\raggedleft
Chardevs probed / sampled
\end{minipage} & \begin{minipage}[b]{\linewidth}\raggedleft
Open errors
\end{minipage} & \begin{minipage}[b]{\linewidth}\raggedleft
Total chardevs
\end{minipage} \\
\midrule\noalign{}
\endhead
\bottomrule\noalign{}
\endlastfoot
arrakis & 9/10 & 24/25 & 1 & 113 \\
microsandbox & 9/10 & 17/25 & 8 & 100 \\
e2b & 9/10 & 7/25 & 18 & 106 \\
gvisor & 3/10 & 6/7 & 1 & 7 \\
daytona & 3/10 & 4/25 & 21 & 167 \\
\end{longtable}
}

The three microVMs all acknowledge the same 9 cmds --- chardev
passthrough is real: a hostile guest can reach the host driver's ioctl
handler for whichever chardevs the image makes openable. They differ in
how many chardevs are openable (arrakis 24, microsandbox 17, e2b 7 ---
e2b's image restricts non-root access most aggressively). gVisor and
daytona both acknowledge only 3/10 but for opposite reasons: gVisor's
Sentry implements an ioctl table that honours only the generic file-fd
ioctls and returns \texttt{ENOTTY} for everything else; daytona's runc
shows the same 3-cmd union because cgroup \texttt{devices.deny} blocks
21 of 25 sampled chardevs from being openable, and the 4 chardevs that
\emph{do} open only expose generic file-fd ioctls.

\paragraph{2.1.5 Primitive-reachability ---
ranking}\label{primitive-reachability-ranking}

Per-product ordering on the primitive-reachability count (14 kernel-LPE
/ container-escape primitives anchored to one CVE family each; per the
methodology, lower is better):

{\def\LTcaptype{none} % do not increment counter
\begin{longtable}[]{@{}
  >{\raggedright\arraybackslash}p{(\linewidth - 6\tabcolsep) * \real{0.2500}}
  >{\raggedright\arraybackslash}p{(\linewidth - 6\tabcolsep) * \real{0.2500}}
  >{\raggedright\arraybackslash}p{(\linewidth - 6\tabcolsep) * \real{0.2500}}
  >{\raggedright\arraybackslash}p{(\linewidth - 6\tabcolsep) * \real{0.2500}}@{}}
\toprule\noalign{}
\begin{minipage}[b]{\linewidth}\raggedright
Rank
\end{minipage} & \begin{minipage}[b]{\linewidth}\raggedright
Product
\end{minipage} & \begin{minipage}[b]{\linewidth}\raggedright
Reachable / 14
\end{minipage} & \begin{minipage}[b]{\linewidth}\raggedright
Defensible basis
\end{minipage} \\
\midrule\noalign{}
\endhead
\bottomrule\noalign{}
\endlastfoot
\rkbest{} & \textbf{gvisor} & 5 / 14 & 84-syscall mode-2 filter + Sentry returns
\texttt{ENOSYS} on \texttt{io\_uring\_setup}, \texttt{userfaultfd}, and
\texttt{quotactl} regardless of the seccomp posture \\
\rkbest{} & \textbf{e2b} & 7 / 14 & Tightest seccomp ceiling in the set (55
syscalls allowlisted) under Firecracker's mode-2 filter applied to all
threads \\
\rkmid{} & \textbf{microsandbox} & 11 / 14 & libkrun runs mode-0 across all 16
threads; no BPF filter on the VMM; primitives gated only by what the VMM
happens not to call \\
\rkmid{} & \textbf{daytona} & 11 / 14 & runc shares the host kernel verbatim
with mode-0 across all 15 threads; no engine-side seccomp \\
\rkworst{} & \textbf{arrakis} & 12 / 14 & Cloud Hypervisor leader-thread mode-0
plus the live \texttt{/dev/kvm} ioctl surface reachable to guest
userland (\bttt{KVM\_GET\_API\_VERSION=12};
\bttt{KVM\_CHECK\_EXTENSION(KVM\_CAP\_NR\_VCPUS)=6}) \\
\end{longtable}
}

Two cells need a footnote. arrakis ranks \rkworst{} on the primitive-reachability
count (12/14), but per §2.1.2 the per-thread seccomp matrix shows 32 of
arrakis's 33 worker threads carry a mode-2 BPF filter --- only the
leader thread is mode-0. The matrix lifts arrakis's \emph{stackability}
cell in 1.3 from \texttt{stack-redeploy} to \texttt{pass} for seccomp;
on 1.1 the per-thread matrix narrows the surface significantly versus a
leader-only read. §2.6 notes that ``the strongest combination ---
microVM × continuous fuzzer --- is unoccupied,'' and Cloud Hypervisor's
in-tree \texttt{cargo-fuzz} harness positions arrakis as the candidate
to fill that empty cell --- but on 1.1 specifically, the nested-KVM lead
anchors arrakis at \rkworst{} (12/14) rather than at \rkmid{} alongside microsandbox and
daytona (11/14), and removing the \texttt{/dev/kvm} exposure would close
arrakis's one-primitive gap to the \rkmid{} cluster but still leave a
four-primitive gap to e2b's \rkbest{} at 7/14.

\paragraph{2.1.6 Primitive-reachability
catalog}\label{primitive-reachability-catalog}

The 14 primitives, indexed by exploit class. Reachability is \emph{not}
the same as exploitability --- a reachable primitive means the kernel
routed to the relevant handler; whether the exploit succeeds depends on
the in-guest kernel version (§2.1.3) and the engine's filter / namespace
posture beyond the primitive itself. The catalog's value over a gross
syscall count is that the operator can read the matrix as ``this engine
gates the BPF family but not the io\_uring family'' rather than guessing
from totals.

{\def\LTcaptype{none} % do not increment counter
\begin{longtable}[]{@{}
  >{\raggedright\arraybackslash}p{(\linewidth - 4\tabcolsep) * \real{0.3333}}
  >{\raggedright\arraybackslash}p{(\linewidth - 4\tabcolsep) * \real{0.3333}}
  >{\raggedright\arraybackslash}p{(\linewidth - 4\tabcolsep) * \real{0.3333}}@{}}
\toprule\noalign{}
\begin{minipage}[b]{\linewidth}\raggedright
Primitive
\end{minipage} & \begin{minipage}[b]{\linewidth}\raggedright
Category
\end{minipage} & \begin{minipage}[b]{\linewidth}\raggedright
Exploit class anchor
\end{minipage} \\
\midrule\noalign{}
\endhead
\bottomrule\noalign{}
\endlastfoot
\texttt{bpf} & syscall & BPF JIT/verifier LPE family (CVE-2020-8835,
CVE-2021-3490, CVE-2022-23222) \\
\texttt{io\_uring\_setup} & syscall & io\_uring use-after-free family
(CVE-2023-21400, CVE-2023-2598, CVE-2024-0582) \\
\texttt{userfaultfd} & syscall & Heap-spray race-window gadget
(CVE-2016-2384 class) \\
\bttt{unprivileged\_userns\_clone} & syscall & User-namespace LPE
family (CVE-2022-0185, CVE-2023-32233, CVE-2024-1086) \\
\texttt{keyctl\_add\_key} & syscall & Keyring LPE (CVE-2022-1664;
trigger surface for CVE-2022-0185) \\
\texttt{perf\_event\_open} & syscall & perf subsystem LPE family
(CVE-2013-2094, CVE-2023-6111 class) \\
\texttt{ptrace\_traceme} & syscall & ptrace-based credential injection
(CVE-2019-13272) \\
\texttt{mount\_tmpfs} & syscall & OCI mount-target traversal
(CVE-2019-5736, CVE-2024-21626) \\
\texttt{clone3} & syscall & clone3 seccomp-filter-bypass vector (older
filters not updated for clone3) \\
\texttt{dev\_kvm} & device-open & Nested-virt primitives, full KVM ioctl
surface (CVE-2023-3640 class) \\
\texttt{dev\_fuse} & device-open & FUSE-mediated path-confusion
(CVE-2018-10906 class) \\
\texttt{dev\_net\_tun} & device-open & TUN/TAP network manipulation,
ARP-poisoning vectors \\
\bttt{cgroup\_v1\_release\_agent\_w} & fs-access & Classic
container-escape primitive (CVE-2022-0492) \\
\bttt{proc\_sys\_kernel\_printk\_w} & fs-access & Sysctl-tampering
escape gadgets (CVE-2016-1583 class) \\
\end{longtable}
}

\paragraph{\texorpdfstring{2.1.7 arrakis \texttt{/dev/kvm} follow-up ---
confirmed nested-KVM
ABI}{2.1.7 arrakis /dev/kvm follow-up --- confirmed nested-KVM ABI}}\label{arrakis-devkvm-follow-up-confirmed-nested-kvm-abi}

The \texttt{primitive-reachability} probe opens \texttt{/dev/kvm} from
guest userland on arrakis without privilege escalation. A guest open of
a chardev could in principle be a passthrough that fails on first ioctl,
so a follow-up one-off probe issued three read-only KVM ioctls:

\begin{itemize}
\tightlist
\item
  \texttt{KVM\_GET\_API\_VERSION} returns \texttt{12} (current upstream
  KVM API)
\item
  \texttt{KVM\_GET\_VCPU\_MMAP\_SIZE} returns \texttt{12288}
\item
  \texttt{KVM\_CHECK\_EXTENSION(KVM\_CAP\_NR\_VCPUS)} returns \texttt{6}
\end{itemize}

All three return real, non-stub values --- the device behind the fd is a
live KVM instance with the modern ABI surface. \texttt{KVM\_CREATE\_VM}
was deliberately not exercised; the implication is that a guest userland
process could call it without privilege escalation, materially expanding
the kernel-LPE surface of the product relative to the rest of the
shortlist. The shipped guest image makes \texttt{/dev/kvm} reachable
with the device-node owner on the host-side \texttt{kvm} group;
nested-virt is enabled in the default image. Disclosure status: see §8.

What 1.1 does not measure: information leakage (deferred to 1.2);
defense-in-depth composition (1.3); what bugs exist \emph{inside} the
surface (1.4). A small surface bounds \emph{reachable code paths}; it
does not bound \emph{the bug rate in those paths}.

\textbf{Methodology lineage.} Two prior papers anchor the
syscall-limitation methodology this axis extends. Zhan et al.~(TSC 2023
/ arXiv:2510.03720) hybrid-static-and-dynamic syscall limitation over
100 Docker-Hub utilities establishes the runtime-instrumented refinement
layer that 1.1's static reachability count does not exercise (a future
axis 2.x candidate). Wan et al.~(arXiv:1712.05493, 2017) ``Mining
Sandboxes for Linux Containers'' anchors the historical
Docker-default-seccomp baseline; their verbatim observation that ``by
default, Docker disallows 44 system calls out of 300+'' ---
cross-checked 2026-05-30 against moby/moby \texttt{docker-v29.5.2} and
confirmed to still hold (361 syscalls unconditionally allowed, 426
unique names across all ALLOW blocks; see §References) --- frames the
§5.5 reading of daytona's \texttt{Privileged:\ true} choice as
foreclosing the layer the wider literature treats as the entry-level
container hardening lever. A 2025 systematic mapping (Sroor et al.,
arXiv:2512.11940) covering 129 container-security studies (2000--2024)
finds the image / host / runtime phases receiving comparatively less
attention than build-time phases --- an empirical-attention-gap framing
for why an engine-level measurement at this depth is timely.

\subsubsection{2.2 Axis 1.2 --- Information
leakage}\label{axis-1.2-information-leakage}

The same four-layer model, re-read for leakage rather than escape: a
\texttt{cat\ /proc/cpuinfo} read at L4 returns whatever L3 (own kernel
\textbar{} Sentry-acting-as-L3 \textbar{} shared host \texttt{/proc})
decides to put there. A leak is a property of how completely L2/L3
substitutes its own content for the host's.

\begin{figure}
\centering
\includegraphics[width=1\linewidth,height=\textheight,keepaspectratio,alt={Four-layer model and where 1.2 leak probes read (from axis 1.2 §2.1)}]{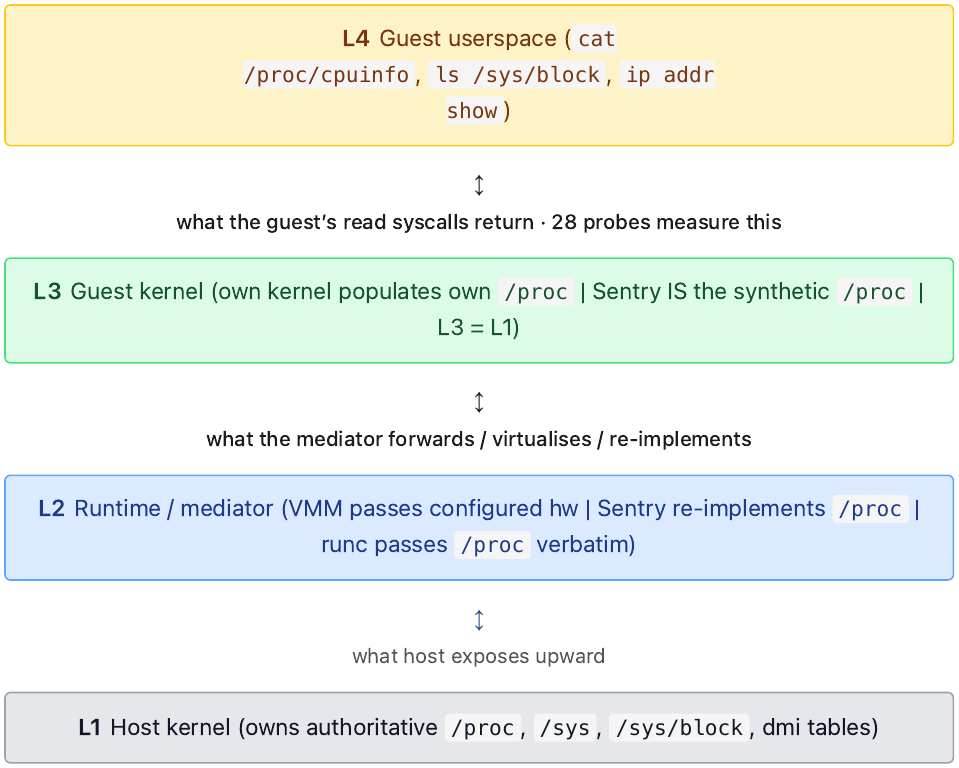}
\caption{Four-layer model and where 1.2 leak probes read (from axis 1.2
§2.1)}
\end{figure}

\paragraph{2.2.1 Verdict matrix}\label{verdict-matrix}

All 140 sub-checks (5 products × 28 probes) executed on a single Hetzner
bare-metal host within a 401 ms window --- the narrow window guarantees
the host-side baseline snapshots are interchangeable across the matrix.
Two products' rollups are \texttt{pass} (leaks = 0); three are
\texttt{partial} (leaks ≥ 1).

{\def\LTcaptype{none} % do not increment counter
\begin{longtable}[]{@{}
  >{\raggedright\arraybackslash}p{(\linewidth - 12\tabcolsep) * \real{0.1200}}
  >{\raggedright\arraybackslash}p{(\linewidth - 12\tabcolsep) * \real{0.1200}}
  >{\raggedright\arraybackslash}p{(\linewidth - 12\tabcolsep) * \real{0.1200}}
  >{\raggedleft\arraybackslash}p{(\linewidth - 12\tabcolsep) * \real{0.1600}}
  >{\raggedleft\arraybackslash}p{(\linewidth - 12\tabcolsep) * \real{0.1600}}
  >{\raggedleft\arraybackslash}p{(\linewidth - 12\tabcolsep) * \real{0.1600}}
  >{\raggedleft\arraybackslash}p{(\linewidth - 12\tabcolsep) * \real{0.1600}}@{}}
\toprule\noalign{}
\begin{minipage}[b]{\linewidth}\raggedright
Product
\end{minipage} & \begin{minipage}[b]{\linewidth}\raggedright
Engine class
\end{minipage} & \begin{minipage}[b]{\linewidth}\raggedright
Rollup
\end{minipage} & \begin{minipage}[b]{\linewidth}\raggedleft
Leaks
\end{minipage} & \begin{minipage}[b]{\linewidth}\raggedleft
Virtualized
\end{minipage} & \begin{minipage}[b]{\linewidth}\raggedleft
Inconclusive
\end{minipage} & \begin{minipage}[b]{\linewidth}\raggedleft
Unreadable
\end{minipage} \\
\midrule\noalign{}
\endhead
\bottomrule\noalign{}
\endlastfoot
e2b & microvm & \textbf{pass} & 0 & 25 & 1 & 2 \\
microsandbox & microvm & \textbf{pass} & 0 & 24 & 2 & 2 \\
arrakis & microvm & partial & \textbf{1} & 23 & 2 & 2 \\
gvisor & userspace-kernel & partial & \textbf{2} & 21 & 1 & 4 \\
daytona & container & partial & \textbf{10} & 17 & 1 & 0 \\
\end{longtable}
}

The class spread (0 / 0 / 1 / 2 / 10) separates the three engine classes
cleanly. Within the microVM class, two-of-three are zero-leak and one is
one-leak (arrakis's Cloud Hypervisor \texttt{cpu\ =\ host} passthrough).
Across classes, the userspace kernel sits between microVMs and
containers --- Sentry virtualises most files but has known
implementation gaps. The \texttt{inconclusive} cells are root-on-root
coincidences (host harness runs root with full cap mask, microVM
defaults also run root with full mask --- match is structural, not a
leak); the \texttt{unreadable} cells are correct outcomes for microVMs'
empty \texttt{/sys/block/*/device/} sub-trees and gvisor's sparse
\texttt{/sys/class/net}.

\paragraph{2.2.2 Per-product leak
ranking}\label{per-product-leak-ranking}

Per-product ordering on the host-identifying leak count (28 sub-probes;
per the methodology, lower is better):

{\def\LTcaptype{none} % do not increment counter
\begin{longtable}[]{@{}
  >{\raggedright\arraybackslash}p{(\linewidth - 6\tabcolsep) * \real{0.2500}}
  >{\raggedright\arraybackslash}p{(\linewidth - 6\tabcolsep) * \real{0.2500}}
  >{\raggedright\arraybackslash}p{(\linewidth - 6\tabcolsep) * \real{0.2500}}
  >{\raggedright\arraybackslash}p{(\linewidth - 6\tabcolsep) * \real{0.2500}}@{}}
\toprule\noalign{}
\begin{minipage}[b]{\linewidth}\raggedright
Rank
\end{minipage} & \begin{minipage}[b]{\linewidth}\raggedright
Product
\end{minipage} & \begin{minipage}[b]{\linewidth}\raggedright
Leaks
\end{minipage} & \begin{minipage}[b]{\linewidth}\raggedright
Defensible basis
\end{minipage} \\
\midrule\noalign{}
\endhead
\bottomrule\noalign{}
\endlastfoot
\rkbest{} & \textbf{e2b} & 0 & Firecracker's minimal device set + \texttt{cpuid}
virtualisation defaults leak nothing on the 28-probe matrix \\
\rkbest{} & \textbf{microsandbox} & 0 & libkrun's lightweight VMM also leaks
nothing under the same matrix \\
\rkmid{} & \textbf{arrakis} & 1 & Cloud Hypervisor default
\texttt{cpu\ =\ host} passthrough leaks the host CPU brand string via
\texttt{/proc/cpuinfo}; configurable but not by default \\
\rkmid{} & \textbf{gvisor} & 2 & Sentry implementation gaps in
\texttt{/proc/meminfo} (\texttt{MemTotal} = host RAM) and
\texttt{/sys/class/dmi/*} (host BIOS product name) \\
\rkworst{} & \textbf{daytona} & 10 & Full shared-kernel signature (CPU brand,
RAM, kernel version, IRQ map, BIOS product, microcode revision) plus
four disk-hardware identifiers (partitions, diskstats, disk-model,
disk-serial) \\
\end{longtable}
}

An asymmetry caveat applies before reading the count as a deeper
architectural claim. A microVM leak is a VMM-configuration property (the
arrakis \texttt{cpuinfo-model} leak is Cloud Hypervisor's
\texttt{cpu\ =\ host} default --- configurable); a gVisor leak is a
Sentry implementation gap (\texttt{MemTotal} and
\texttt{dmi-product-name} are files the Sentry has not virtualised); a
runc leak is the shared-kernel architecture asserting itself (10 leaks
are not gaps but the structural signature of running a workload inside a
container that shares the host kernel). A 2-leak gvisor and 1-leak
arrakis reading do not mean arrakis is more isolated than gvisor in a
deeper sense --- they mean a particular VMM config knob is set one way
and the Sentry has two specific unimplemented files.

\paragraph{2.2.3 Per-probe leak
distribution}\label{per-probe-leak-distribution}

Ten of the 28 probes leaked on at least one product. The other 18
virtualised correctly across all five products on this host.

{\def\LTcaptype{none} % do not increment counter
\begin{longtable}[]{@{}
  >{\raggedright\arraybackslash}p{(\linewidth - 4\tabcolsep) * \real{0.3333}}
  >{\raggedright\arraybackslash}p{(\linewidth - 4\tabcolsep) * \real{0.3333}}
  >{\raggedright\arraybackslash}p{(\linewidth - 4\tabcolsep) * \real{0.3333}}@{}}
\toprule\noalign{}
\begin{minipage}[b]{\linewidth}\raggedright
Probe
\end{minipage} & \begin{minipage}[b]{\linewidth}\raggedright
Engines that leaked
\end{minipage} & \begin{minipage}[b]{\linewidth}\raggedright
What the leak discloses
\end{minipage} \\
\midrule\noalign{}
\endhead
\bottomrule\noalign{}
\endlastfoot
\texttt{cpuinfo-model} & arrakis, daytona & Host CPU brand string
verbatim (\bttt{AMD\ Ryzen\ 5\ 3600\ 6-Core\ Processor}) \\
\texttt{meminfo-total} & gvisor, daytona & Host RAM total in kB (the
full 64 GiB, not the guest's configured allocation) \\
\texttt{dmi-product-name} & gvisor, daytona &
\bttt{/sys/class/dmi/id/product\_name} reads the host BIOS product
string \\
\texttt{kernel-version} & daytona & \texttt{/proc/version} is the host's
verbatim --- runc shares the host kernel \\
\texttt{cpu-microcode} & daytona & Host CPU microcode revision passed
through \\
\texttt{interrupts} & daytona & \texttt{/proc/interrupts} shows the
host's IRQ device names + per-CPU counts \\
\texttt{partitions} & daytona & \texttt{/proc/partitions} matched the
host this run (NBD-attachment state aligned --- see §2.2.5) \\
\texttt{diskstats} & daytona & Host's full block-device list visible
(loop / md / nbd / nvme*) --- shared-kernel signature \\
\texttt{disk-model} & daytona & \bttt{/sys/block/*/device/model}
returns \bttt{SAMSUNG\ MZVL2512HCJQ-00B00} verbatim \\
\texttt{disk-serial} & daytona & \bttt{/sys/block/*/device/serial}
returns host NVMe serials (\texttt{S675NU0TB52819},
\texttt{S675NU0TB52833}) --- persistent host fingerprint \\
\end{longtable}
}

The probes that virtualised across all five products: \texttt{loadavg},
\texttt{uptime}, \texttt{iomem}, \texttt{modules},
\texttt{self-mountinfo}, \texttt{kernel-random-uuid},
\texttt{kernel-hostname-sysctl}, \texttt{dmi-board-serial},
\texttt{machine-id}, \texttt{etc-hostname}, \texttt{hostname-cmd},
\texttt{ip-addr} (sysfs fallback, \texttt{unreadable} on gvisor),
\texttt{ip-route}, \texttt{arp-table}, \texttt{date-tz},
\texttt{disk-size} (subject to the §2.2.5 NBD caveat). The two
\texttt{self-status} sub-probes are \texttt{inconclusive} for
root-on-root reasons rather than leak findings.

\paragraph{2.2.4 Per-probe-family
readings}\label{per-probe-family-readings}

\textbf{CPU identity (\texttt{/proc/cpuinfo}, \texttt{cpu-microcode}).}
Two of five leak the CPU brand string verbatim (arrakis, daytona). e2b
and microsandbox genericise it to \texttt{AMD\ EPYC}-class strings the
VMM injects independently of the underlying CPU. gVisor returns
\texttt{"unknown"} for the CPU model --- the Sentry treats the field as
unsupported rather than passing through. Daytona also leaks the CPU
microcode revision (which, combined with the kernel version,
fingerprints the patch state of the host's Spectre / Meltdown
mitigations).

\textbf{Memory and load (\texttt{/proc/meminfo}, \texttt{/proc/loadavg},
\texttt{/proc/uptime}).} gvisor and daytona leak \texttt{meminfo-total}
(host RAM). microVMs report their configured guest RAM (e2b 493 572 kB,
microsandbox 528 972 kB, arrakis 19 264 892 kB), which is what the guest
kernel allocated, not the host's 64 GiB. \texttt{loadavg} and
\texttt{uptime} virtualised across all five --- none of the products
return the host's loadavg or uptime values.

\textbf{Kernel build (\texttt{/proc/version}, \texttt{/proc/interrupts},
\texttt{/proc/modules}, \texttt{/proc/iomem}).} daytona leaks
\texttt{kernel-version} and \texttt{interrupts} (shared-kernel
signature). The other four products report their own kernel build
(microVMs: 6.1.158 / 6.12.8+ / 6.12.68; gvisor: 4.19.0-gvisor synthetic)
and their own IRQ device lists. \texttt{modules} and \texttt{iomem}
virtualise across all five --- every product either has its own
loaded-module list or returns empty.

\textbf{Self-introspection (\texttt{/proc/self/\{status,mountinfo\}}).}
All five are \texttt{inconclusive} on \texttt{self-status-seccomp}
because the host harness itself runs at mode 0; the harness gates the
probe to emit \texttt{inconclusive} when the host is unrestricted rather
than counting a structural match as a leak (the deeper seccomp question
is what 1.1 measures via per-thread strace). microsandbox and arrakis
are additionally \texttt{inconclusive} on \texttt{self-status-caps}
because both default sandboxes run the workload as root with
\texttt{CapEff\ =\ 000001ffffffffff}, matching the host-side harness
which also runs as root with the unrestricted mask --- a root-on-root
coincidence, not a leak.

\textbf{Hardware identity (\texttt{/sys/class/dmi/*},
\texttt{machine-id}, \texttt{kernel-random-uuid}).} gvisor and daytona
leak \texttt{dmi-product-name} (the host's BIOS product string). The
other three identity probes (\texttt{dmi-board-serial},
\texttt{machine-id}, \texttt{kernel-random-uuid}) virtualised across all
five --- every product either has its own machine-id or returns empty
for these. The dmi leak is the strongest cloud-provider fingerprint a
hostile workload can read passively (AWS stamps \texttt{Amazon\ EC2};
GCP stamps \texttt{Google\ Compute\ Engine}); on this bare-metal Hetzner
box the leaked product name is the chassis model.

\textbf{Hostname triplet (\texttt{hostname}, \texttt{/etc/hostname},
sysctl).} All five virtualised consistently across the three sources,
with no triplet-internal disagreements. A disagreement between the three
sources within a single product would be a methodology-relevant
inconsistency (partial UTS namespacing where \texttt{/etc/hostname} was
not bind-mounted to a synthetic file) --- none surfaced this run.

\textbf{Network and locale (\texttt{ip-addr}, \texttt{ip-route},
\texttt{arp-table}, \texttt{date-tz}).} All five virtualised. gvisor's
\texttt{ip-addr} is \texttt{unreadable} because the Sentry's
\texttt{/sys/class/net} emulation is sparse enough that the harness's
sysfs fallback returns nothing --- methodology-correct unreadable, not a
leak. The host timezone (CEST) differed from every guest's UTC, which is
the expected virtualised outcome.

\textbf{Disk (\texttt{/proc/partitions}, \texttt{/proc/diskstats},
\texttt{/sys/block/*/\{model,serial,size\}}).} daytona leaks 4 of 5 disk
probes (\texttt{partitions}, \texttt{diskstats}, \texttt{disk-model},
\texttt{disk-serial}); \texttt{disk-size} virtualised only because the
host's snapshot captured an \texttt{nbd13} attachment with non-zero size
that the container's mount-namespace did not re-expose. microVMs
virtualise \texttt{partitions} + \texttt{diskstats} + \texttt{disk-size}
(own virtio/ram block devices populate these) and report
\texttt{disk-model} / \texttt{disk-serial} as \texttt{unreadable} (the
microVM kernels do not expose \texttt{/sys/block/*/device/} sub-trees,
so the glob matches nothing). gvisor reports \texttt{partitions} +
\texttt{diskstats} as virtualised (Sentry's sparse listing differs from
the host) and all three glob disk probes as \texttt{unreadable}
(Sentry's \texttt{/sys/block} has no device sub-trees).

\paragraph{2.2.5 daytona disk NBD-attachment race
window}\label{daytona-disk-nbd-attachment-race-window}

The disk-side \texttt{partitions} and \texttt{disk-size} verdicts for
daytona depend on whether the host's \texttt{/proc/partitions} and
\texttt{/sys/block/*/size} views at snapshot time include leftover NBD
(Network Block Device) attachments from prior sandbox activity that the
freshly-created container's mount namespace does not re-expose. Three
runs on the same host produced three different cells of the partitions /
disk-size 2×2 matrix:

{\def\LTcaptype{none} % do not increment counter
\begin{longtable}[]{@{}
  >{\raggedright\arraybackslash}p{(\linewidth - 6\tabcolsep) * \real{0.2500}}
  >{\raggedright\arraybackslash}p{(\linewidth - 6\tabcolsep) * \real{0.2500}}
  >{\raggedright\arraybackslash}p{(\linewidth - 6\tabcolsep) * \real{0.2500}}
  >{\raggedright\arraybackslash}p{(\linewidth - 6\tabcolsep) * \real{0.2500}}@{}}
\toprule\noalign{}
\begin{minipage}[b]{\linewidth}\raggedright
Run
\end{minipage} & \begin{minipage}[b]{\linewidth}\raggedright
partitions
\end{minipage} & \begin{minipage}[b]{\linewidth}\raggedright
disk-size
\end{minipage} & \begin{minipage}[b]{\linewidth}\raggedright
Reason
\end{minipage} \\
\midrule\noalign{}
\endhead
\bottomrule\noalign{}
\endlastfoot
run 3 (prior) & virtualized & virtualized & host had \texttt{nbd13}
attached with size 2 793 392 from prior activity; container saw nbd13
absent (partitions) and size 0 (size) \\
run 4 (current) & \textbf{leak} & virtualized & host had no leftover NBD
activity in \texttt{/proc/partitions}; both sides saw same device names;
\texttt{disk-size} still divergent at index 13 \\
hypothetical & leak & leak & both kernel-state and NBD device sizes
aligned across snapshot window \\
\end{longtable}
}

Daytona's \texttt{partitions} / \texttt{disk-size} results should be
read as a \emph{range of outcomes} rather than a fixed cell. The
shared-kernel signature surfaces concretely through \texttt{diskstats} +
\texttt{disk-model} + \texttt{disk-serial} regardless of NBD state
(those three are leaks in every run). A future axis 2.x covering the
attachment-layer surface --- can an in-sandbox process attach an NBD
that becomes visible in a sibling sandbox? --- is the right place for a
stable measurement; 1.2 measures namespace virtualisation, not
block-attachment isolation.

What 1.2 does not measure: post-escape capability to read host memory
(\texttt{/dev/mem} requires \texttt{CAP\_SYS\_RAWIO}); raw
disk-attachment isolation (deferred to a future axis 2.x --- namespace
virtualisation is not the same as block-attachment isolation); active
probing or microarchitectural side channels (per the methodology's
threat-model scope).

\subsubsection{2.3 Axis 1.3 --- Defense-in-depth
stackability}\label{axis-1.3-defense-in-depth-stackability}

\textbf{This axis does not cross-rank products by \texttt{applied}
count.} The seven layers (seccomp / AppArmor / SELinux / user-ns /
cap-drop / \texttt{no\_new\_privs} / \texttt{pids.max}) are the
\emph{substrate} of the userspace-kernel class --- gVisor's Sentry is
constructed from a seccomp-mode-2 ceiling, a reduced cap mask, user-ns,
and \texttt{no\_new\_privs}; the layers \emph{are} the engine's
isolation. For the microVM class, the same layers are \emph{defense in
depth on top of} an already-isolating hypervisor; for the container
class, they are the engine's whole isolation again but applied through a
fundamentally different (shared-kernel) substrate. Comparing the count
of applied layers across classes therefore measures something different
in each row --- and ``gvisor applies 4, microsandbox applies 0'' is not
a like-for-like comparison.

\paragraph{2.3.1 Verdict matrix}\label{verdict-matrix-1}

All 45 sub-checks (5 products × 9 sub-probes --- 7 per-layer + 2
pre-checks) executed against runtime PIDs on the same Hetzner host. Two
pre-checks (\texttt{host-capabilities}, \texttt{workload-baseline}) pass
for every product. Each per-layer cell carries two fields under the v2.3
verdict scheme: \texttt{applied} (engine-defaults posture, observed via
\texttt{/proc}) and \texttt{stack-effort} (operator's lift to attach the
layer, inferred from product source). The rendered cell label combines
both: \texttt{pass} (applied at defaults), \texttt{stack-redeploy}
(operator path via systemd-wrap / daemon-config edit without source
patch), \texttt{not-exposed} (would require patching product source),
\texttt{skipped} (SELinux on an AppArmor host).

{\def\LTcaptype{none} % do not increment counter
\begin{longtable}[]{@{}
  >{\raggedright\arraybackslash}p{(\linewidth - 14\tabcolsep) * \real{0.1250}}
  >{\raggedright\arraybackslash}p{(\linewidth - 14\tabcolsep) * \real{0.1250}}
  >{\raggedright\arraybackslash}p{(\linewidth - 14\tabcolsep) * \real{0.1250}}
  >{\raggedright\arraybackslash}p{(\linewidth - 14\tabcolsep) * \real{0.1250}}
  >{\raggedright\arraybackslash}p{(\linewidth - 14\tabcolsep) * \real{0.1250}}
  >{\raggedright\arraybackslash}p{(\linewidth - 14\tabcolsep) * \real{0.1250}}
  >{\raggedright\arraybackslash}p{(\linewidth - 14\tabcolsep) * \real{0.1250}}
  >{\raggedright\arraybackslash}p{(\linewidth - 14\tabcolsep) * \real{0.1250}}@{}}
\toprule\noalign{}
\begin{minipage}[b]{\linewidth}\raggedright
Product
\end{minipage} & \begin{minipage}[b]{\linewidth}\raggedright
seccomp
\end{minipage} & \begin{minipage}[b]{\linewidth}\raggedright
apparmor
\end{minipage} & \begin{minipage}[b]{\linewidth}\raggedright
selinux
\end{minipage} & \begin{minipage}[b]{\linewidth}\raggedright
user-ns
\end{minipage} & \begin{minipage}[b]{\linewidth}\raggedright
cap-drop
\end{minipage} & \begin{minipage}[b]{\linewidth}\raggedright
no-new-privs
\end{minipage} & \begin{minipage}[b]{\linewidth}\raggedright
pids-max
\end{minipage} \\
\midrule\noalign{}
\endhead
\bottomrule\noalign{}
\endlastfoot
gvisor (runsc/Sentry) & \textbf{pass} & not-exposed & skipped &
stack-redeploy & \textbf{pass} & \textbf{pass} & \textbf{pass} \\
e2b (Firecracker) & \textbf{pass} & stack-redeploy & skipped &
stack-redeploy & stack-redeploy & \textbf{pass} & stack-redeploy \\
daytona (runc DinD) & not-exposed & not-exposed & skipped & not-exposed
& \textbf{pass} & not-exposed & not-exposed \\
microsandbox (libkrun) & stack-redeploy & stack-redeploy & skipped &
stack-redeploy & stack-redeploy & stack-redeploy & stack-redeploy \\
arrakis (cloud-hypervisor) & \textbf{pass} & stack-redeploy & skipped &
stack-redeploy & stack-redeploy & stack-redeploy & stack-redeploy \\
\end{longtable}
}

Aggregate counts across the 35 per-layer cells: \textbf{8
\texttt{pass}}, \textbf{16 \texttt{stack-redeploy}}, \textbf{6
\texttt{not-exposed}}, \textbf{5 \texttt{skipped}} (SELinux column),
\textbf{0 \texttt{inconclusive}}, \textbf{0 \texttt{engine-blocks}}
(Path A baseline; Path B confirmation deferred). The class spread
separates the three engine classes cleanly: the userspace-kernel class
(gvisor) sweeps four layers as defaults; the microVM class applies 0--2
layers per product (e2b 2; arrakis 1; microsandbox 0) and renders the
remainder as \texttt{stack-redeploy}; the container class (daytona)
applies one layer (cap-drop on init) and renders five as
\texttt{not-exposed}.

\paragraph{2.3.2 Layers applied at
defaults}\label{layers-applied-at-defaults}

Layers applied at engine defaults (out of 7), reported without ranking:

{\def\LTcaptype{none} % do not increment counter
\begin{longtable}[]{@{}
  >{\raggedright\arraybackslash}p{(\linewidth - 4\tabcolsep) * \real{0.3333}}
  >{\raggedright\arraybackslash}p{(\linewidth - 4\tabcolsep) * \real{0.3333}}
  >{\raggedright\arraybackslash}p{(\linewidth - 4\tabcolsep) * \real{0.3333}}@{}}
\toprule\noalign{}
\begin{minipage}[b]{\linewidth}\raggedright
Product
\end{minipage} & \begin{minipage}[b]{\linewidth}\raggedright
Layers applied / 7
\end{minipage} & \begin{minipage}[b]{\linewidth}\raggedright
What's applied at defaults
\end{minipage} \\
\midrule\noalign{}
\endhead
\bottomrule\noalign{}
\endlastfoot
\textbf{gvisor} & 4 & seccomp mode-2 (84-syscall ceiling on Sentry),
cap-drop (reduced \texttt{0008001f} --- Sentry's deliberate mask),
\texttt{no\_new\_privs} (Docker default propagates), \texttt{pids.max}
(Docker scope \texttt{76946}, generous but bounded). These four are how
the Sentry isolates --- they are not stacked on top of a separate
isolation primitive. \\
\textbf{e2b} & 2 & seccomp mode-2 (Firecracker upstream filter),
\texttt{no\_new\_privs} (Firecracker prctl). Applied on top of the
Firecracker hypervisor boundary. \\
\textbf{arrakis} & 1 & seccomp mode-2 on 32/33 worker threads
(\texttt{pass} under the per-thread matrix). Applied on top of the Cloud
Hypervisor hypervisor boundary. \\
\textbf{daytona} & 1 & runc empty-cap default on the container init PID
(zero capabilities); offset by the \texttt{Privileged:\ true} choice in
the runner that disables AppArmor and would re-add caps to the post-exec
workload. \\
\textbf{microsandbox} & 0 & libkrun ships no engine-side hardening; all
seven layers require operator-side stacking. The libkrun model treats
the hardening as the operator's job. \\
\end{longtable}
}

\paragraph{\texorpdfstring{2.3.3 Cross-class ranking ---
\texttt{not-exposed}
count}{2.3.3 Cross-class ranking --- not-exposed count}}\label{cross-class-ranking-not-exposed-count}

The cross-class meaningful ordering on this axis is the
\textbf{\texttt{not-exposed} count} --- layers an operator literally
cannot attach without patching product source. This count measures
product-side foreclosure of the operator's hardening lever, which is
architecture-independent. Fewer \texttt{not-exposed} cells is better:

{\def\LTcaptype{none} % do not increment counter
\begin{longtable}[]{@{}
  >{\raggedright\arraybackslash}p{(\linewidth - 6\tabcolsep) * \real{0.2500}}
  >{\raggedright\arraybackslash}p{(\linewidth - 6\tabcolsep) * \real{0.2500}}
  >{\raggedright\arraybackslash}p{(\linewidth - 6\tabcolsep) * \real{0.2500}}
  >{\raggedright\arraybackslash}p{(\linewidth - 6\tabcolsep) * \real{0.2500}}@{}}
\toprule\noalign{}
\begin{minipage}[b]{\linewidth}\raggedright
Rank
\end{minipage} & \begin{minipage}[b]{\linewidth}\raggedright
Product
\end{minipage} & \begin{minipage}[b]{\linewidth}\raggedright
\texttt{not-exposed} cells
\end{minipage} & \begin{minipage}[b]{\linewidth}\raggedright
Defensible basis
\end{minipage} \\
\midrule\noalign{}
\endhead
\bottomrule\noalign{}
\endlastfoot
\rkbest{} & \textbf{e2b} / \textbf{microsandbox} / \textbf{arrakis} & 0 each &
Every applicable layer reachable via operator-side stacking \\
\rkmid{} & \textbf{gvisor} & 1 & AppArmor cell (runsc strips the OCI spec
AppArmor field; operator cannot reach the Sentry PID without patching
\texttt{specutils.go}) \\
\rkworst{} & \textbf{daytona} & 5 & Concentrated downstream of the runner
hardcoding \texttt{Privileged:\ true}; four cells (seccomp, AppArmor,
user-ns, \texttt{no\_new\_privs}) blocked by that choice;
\texttt{pids.max} blocked independently by the runner not populating
\bttt{HostConfig.PidsLimit} \\
\end{longtable}
}

\paragraph{2.3.4 Per-layer applicability across
products}\label{per-layer-applicability-across-products}

{\def\LTcaptype{none} % do not increment counter
\begin{longtable}[]{@{}
  >{\raggedright\arraybackslash}p{(\linewidth - 8\tabcolsep) * \real{0.2000}}
  >{\raggedright\arraybackslash}p{(\linewidth - 8\tabcolsep) * \real{0.2000}}
  >{\raggedright\arraybackslash}p{(\linewidth - 8\tabcolsep) * \real{0.2000}}
  >{\raggedright\arraybackslash}p{(\linewidth - 8\tabcolsep) * \real{0.2000}}
  >{\raggedright\arraybackslash}p{(\linewidth - 8\tabcolsep) * \real{0.2000}}@{}}
\toprule\noalign{}
\begin{minipage}[b]{\linewidth}\raggedright
Layer
\end{minipage} & \begin{minipage}[b]{\linewidth}\raggedright
\texttt{pass} (applied at defaults)
\end{minipage} & \begin{minipage}[b]{\linewidth}\raggedright
\texttt{stack-redeploy} (operator path)
\end{minipage} & \begin{minipage}[b]{\linewidth}\raggedright
\texttt{not-exposed} (source patch)
\end{minipage} & \begin{minipage}[b]{\linewidth}\raggedright
Skipped
\end{minipage} \\
\midrule\noalign{}
\endhead
\bottomrule\noalign{}
\endlastfoot
\texttt{seccomp} & gvisor, e2b, arrakis & microsandbox & daytona &
--- \\
\texttt{apparmor} & --- & e2b, microsandbox, arrakis & gvisor, daytona &
--- \\
\texttt{selinux} & --- & --- & --- & all 5 (AppArmor host) \\
\texttt{user-ns} & --- & gvisor, e2b, microsandbox, arrakis & daytona &
--- \\
\texttt{cap-drop} & gvisor, daytona & e2b, microsandbox, arrakis & --- &
--- \\
\texttt{no-new-privs} & gvisor, e2b & microsandbox, arrakis & daytona &
--- \\
\texttt{pids-max} & gvisor & e2b, microsandbox, arrakis & daytona &
--- \\
\end{longtable}
}

\paragraph{2.3.5 Per-layer readings}\label{per-layer-readings}

\textbf{seccomp.} Three of five \texttt{pass}: gvisor (Sentry installs a
BPF filter at \texttt{runsc/cmd/sandboxsetup/caps.go:39-71} --- measured
TGID leader \texttt{mode=filter,\ 1\ filter}), e2b (Firecracker
hardcodes \texttt{prctl(PR\_SET\_NO\_NEW\_PRIVS)} and applies its BPF
filter at \texttt{firecracker/src/vmm/src/seccomp.rs:112-137}), arrakis
(Cloud Hypervisor's \texttt{-\/-seccomp\ true} upstream default at
\texttt{src/main.rs:432-436} confines the worker threads --- 32 of 33
worker threads in \texttt{mode=filter} this run, the TGID leader runs
\texttt{mode=0} because the filter is applied to per-thread-spawned
workers after the leader fork; per-thread reading via
\texttt{readSeccompPerThread} is what surfaces this --- a
TGID-leader-only read misclassified the cell as \texttt{stack-redeploy}
in earlier runs). microsandbox lands on \texttt{stack-redeploy} (msb
runs \texttt{mode=0} on leader and all 16 worker threads). daytona's
\texttt{not-exposed} is the load-bearing one: the runner hardcodes
\texttt{Privileged:\ true} which disables Docker's default seccomp
profile, and the runner doesn't populate \texttt{SecurityOpt} either.

\textbf{AppArmor.} Zero of five \texttt{pass}. Ubuntu loads AppArmor by
default but only enforces profiles where one is explicitly assigned ---
the test host has no profile loaded on any of the five engine processes.
The three \texttt{stack-redeploy} cells (e2b, microsandbox, arrakis) are
operator-deploy paths: wrap the orchestrator in a systemd unit with
\texttt{AppArmorProfile=docker-default} and the kernel applies it to the
engine PID on \texttt{execve(2)}. The two \texttt{not-exposed} cells:
gvisor (runsc silently drops the OCI spec's AppArmor field at
\texttt{runsc/specutils/specutils.go:151-154} before the Sentry sees it;
even if Docker passes
\texttt{-\/-security-opt\ apparmor=docker-default}, the profile is
discarded), daytona (\texttt{Privileged:\ true} disables AppArmor
regardless of profile inventory).

\textbf{SELinux.} All five \texttt{skipped} because the test host is
Ubuntu 24.04 with AppArmor (SELinux and AppArmor are mutually exclusive
at the host level). A re-test on a RHEL/Fedora/Rocky box would populate
this column; runsc would render \texttt{not-exposed} because of the
matching \texttt{specutils.go:147-149} strip, the other four would
render \texttt{stack-redeploy} (systemd-wrap path attaches a
\texttt{container\_t} context).

\textbf{User namespace.} Zero of five \texttt{pass}. All five runtime
PIDs map \texttt{uid\ 0\ →\ uid\ 0} (the engine's host-side process runs
as real host root). Four cells render \texttt{stack-redeploy} (gvisor's
\texttt{-\/-rootless} flag at \texttt{runsc/config/config.go:270-275};
e2b / microsandbox / arrakis via systemd-wrap with
\texttt{PrivateUsers=yes}); daytona's \texttt{not-exposed} is structural
--- \texttt{Privileged:\ true} is incompatible with user-ns. Path B
confirmation for the four microVM / userspace-kernel cells is the most
uncertain item in this matrix --- KVM access typically requires host
capabilities that user-ns strips, so several of these may upgrade to
\texttt{engine-blocks} under active stacking.

\textbf{cap-drop.} Two of five \texttt{pass}: gvisor (Sentry's reduced
mask \texttt{000000000008001f} is deliberate --- \texttt{chroot},
\texttt{ptrace}, \texttt{setuid}, \texttt{setgid},
\texttt{dac\_override}, \texttt{audit\_write} are what Sentry needs to
function; reducing further would break the engine), daytona
(\texttt{runc\ init} runs with \texttt{0000000000000000} because
Docker's default-cap-drop applies before the privileged workload exec
--- see the structural-narrowness caveat in §2.3.7). Three
\texttt{stack-redeploy}: e2b / microsandbox / arrakis all run with
\texttt{CapEff=000001ffffffffff} (the full 41-bit mask as of kernel 6.8
--- full root); operator wraps the orchestrator with
\texttt{CapabilityBoundingSet=} and the bounding set inherits to the
child VMM.

\textbf{\texttt{no\_new\_privs}.} Two of five \texttt{pass}: e2b
(Firecracker sets the prctl explicitly at \texttt{seccomp.rs:112-137}),
gvisor (Docker default propagates to the runsc Sentry). Two
\texttt{stack-redeploy}: microsandbox / arrakis ship
\texttt{NoNewPrivs=0}; operator opts in via systemd
\texttt{NoNewPrivileges=yes}. Daytona's \texttt{not-exposed}:
\texttt{Privileged:\ true} effectively pins
\texttt{no-new-privileges=false}; no operator override exists short of
patching the runner.

\textbf{\texttt{pids.max} cgroup.} One of five \texttt{pass}: gvisor's
Docker scope has \texttt{pids.max=76946} (the systemd \texttt{TasksMax}
default × Docker accounting --- generous but bounded; the methodology
treats any finite cap as \texttt{pass} because the compatibility
question is ``does operator-applied \texttt{-\/-pids-limit\ 1024}
work?'', and a pre-existing cap of 76946 means yes). Three
\texttt{stack-redeploy}: e2b (orchestrator cgroup
\texttt{/sys/fs/cgroup/e2b/sbx-\textless{}id\textgreater{}/} exists but
doesn't list \texttt{pids} in \texttt{cgroup.subtree\_control} --- the
v2 controller-not-delegated case, posture-equivalent to \texttt{max};
operator caps via systemd \texttt{TasksMax=} on the orchestrator unit),
microsandbox (\texttt{pids.max=max} on the user session scope), arrakis
(same). Daytona's \texttt{not-exposed}: the runner doesn't populate
\texttt{HostConfig.PidsLimit}; the nested cgroup
\texttt{/system.slice/docker-\textless{}outer\textgreater{}.scope/docker/\textless{}inner\textgreater{}}
confirms DinD topology and Docker's per-container \texttt{PidsLimit}
does \emph{not} inherit from outer to inner.

\paragraph{2.3.6 daytona --- wrapper-source foreclosure
pattern}\label{daytona-wrapper-source-foreclosure-pattern}

Five of daytona's seven cells render \texttt{not-exposed} --- the most
foreclosed row in the matrix. The root cause is wrapper-source, not the
runc engine itself (runc composes with all seven layers cleanly when its
OCI spec carries the right fields). Four cells trace to the runner
hardcoding \texttt{Privileged:\ true} at
\texttt{apps/runner/pkg/docker/container\_configs.go:134}, which
disables Docker's default seccomp profile, disables AppArmor, makes
user-ns incompatible, and pins \texttt{no-new-privileges=false}. The
fifth cell (\texttt{pids-max}) is independent of \texttt{Privileged} but
is also runner-blocked: \texttt{CreateSandboxDTO} at
\texttt{apps/runner/pkg/api/dto/sandbox.go:6-30} exposes CpuQuota /
GpuQuota / MemoryQuota / StorageQuota fields but no PidsLimit field, and
the hardcoded daemon.json at \texttt{apps/runner/Dockerfile:68} only
sets \texttt{\{"insecure-registries":{[}"registry:6000"{]}\}} --- the
inner dockerd never receives a \texttt{default-pids-limit}.

Two distinct runner-source patch paths would close the pids-max gap
independently of the \texttt{Privileged:\ true} blocker that locks the
other four: per-sandbox (add \texttt{PidsLimit\ *int64} to
\texttt{CreateSandboxDTO} and populate \texttt{HostConfig.PidsLimit}
alongside the existing \texttt{container.Resources} block --- one DTO
field + one HostConfig line, the smallest patch surface of daytona's
five \texttt{not-exposed} gaps), or daemon-wide (extend the hardcoded
daemon.json with
\texttt{"default-pids-limit":\ \textless{}N\textgreater{}} so the inner
dockerd caps every spawned container regardless of per-call
configuration). The four \texttt{Privileged:\ true}-blocked cells
require a larger refactor --- removing the hardcoded privilege cascades
through the runner's assumptions about what it can do inside the
container.

\paragraph{2.3.7 Structural caveats}\label{structural-caveats}

\textbf{\texttt{cap-drop\ pass} on daytona is structurally narrow.} The
\texttt{runc\ init} PID has the expected zero-cap mask
(\texttt{CapEff=0000000000000000}), but \texttt{Privileged:\ true}
re-adds the full cap set to the workload that init execs into. The
methodology probes the \emph{runtime PID} (init) --- that's the
layer-stacking surface --- so the \texttt{pass} is methodology-correct;
readers interpreting the matrix for \emph{workload-level} posture should
pair this cell with the cap-drop interpretation: gvisor's
\texttt{0008001f} is Sentry's deliberate engineering-bounded mask,
daytona's \texttt{0000000000000000} is Docker's default-cap-drop applied
to init before privileged exec, e2b / microsandbox / arrakis
\texttt{000001ffffffffff} is the full Linux capability set as of kernel
6.8.

\textbf{\texttt{user-ns\ applied=no} uniform but means different
things.} All five runtime PIDs map \texttt{uid\ 0\ →\ uid\ 0}. Reading
this as ``the sandbox has no user-ns isolation'' would be wrong for the
microVMs (the \emph{guest workload} runs in a separate guest kernel;
user-ns is irrelevant inside the VM) and for daytona (the container's
\emph{workload} may run as a non-root container user even though
\texttt{runc\ init} is host root). The probe targets the engine's
runtime process because that is the layer-stacking surface --- the §1.3
question is ``does the engine compose with rootless / user-ns stacking
on top?'', and the honest answer for ``host process runs as real root''
is ``no, not at default config.''

\textbf{Path A inference vs Path B confirmation.} All
\texttt{stack-redeploy} cells assume the operator-facing systemd-wrap
(or daemon-config-edit) path works at runtime. Some may upgrade to
\texttt{engine-blocks} under Path B --- most likely candidates: e2b /
microsandbox / arrakis user-ns (microVMs typically need
\texttt{/dev/kvm} privileges that user-ns strips). Path A inference is
grounded in product source (file:line citations in the fixture report),
but it does not prove the workload would still run under the layer ---
only that the configuration path exists at the cited tier.
\texttt{stack-redeploy} is a \emph{path exists} claim, not a \emph{path
works} claim. Path B is the only route to upgrade an inference to
measurement-grade evidence, and the only route to
\texttt{engine-blocks}.

What 1.3 does not measure: whether each layer \emph{actually blocks
attacks} (the methodology defers active negative-control measurement to
a deferred axis); performance regressions under stacking; whether the
operator's CI process bumps the stacked layer when the engine updates;
cross-class comparability on the \texttt{applied} count (see the
architecture caveat above).

\subsubsection{2.4 Axis 1.4 --- Public CVE
history}\label{axis-1.4-public-cve-history}

The historical signal that 1.1--1.3 architectural measurements cannot
express: how often the engine has failed on public record, which bug
classes recur, and how the count weighs against the engine's age and
eyeballs. The methodology classifies each CVE per the gVisor schema
(Escape / HostLeak / HostDoS / PeerDoS / InternalEsc) and orders the
four scope-eligible engines by the rollup-window count; libkrun is
excluded from the ranking because its 0 published CVEs is the absence of
a finding, not the presence of soundness (cross-axis read in §3.5).

\paragraph{2.4.1 Rollup matrix --- last 24
months}\label{rollup-matrix-last-24-months}

The 24-month rollup (2024-05-20 → 2026-05-20) over six channels (OSV.dev
primary; GHSA per-repo; NVD 2.0; CVE.org tie-breaker;
\texttt{lore.kernel.org/linux-cve-announce}; vendor trackers). 11 engine
CVEs land in the window; the Linux kernel as shared dependency
contributes \textasciitilde3,500 in the same window --- two and a half
orders of magnitude larger than all five engines combined.

{\def\LTcaptype{none} % do not increment counter
\begin{longtable}[]{@{}
  >{\raggedright\arraybackslash}p{(\linewidth - 12\tabcolsep) * \real{0.1154}}
  >{\raggedright\arraybackslash}p{(\linewidth - 12\tabcolsep) * \real{0.1154}}
  >{\raggedleft\arraybackslash}p{(\linewidth - 12\tabcolsep) * \real{0.1538}}
  >{\raggedleft\arraybackslash}p{(\linewidth - 12\tabcolsep) * \real{0.1538}}
  >{\raggedleft\arraybackslash}p{(\linewidth - 12\tabcolsep) * \real{0.1538}}
  >{\raggedleft\arraybackslash}p{(\linewidth - 12\tabcolsep) * \real{0.1538}}
  >{\raggedleft\arraybackslash}p{(\linewidth - 12\tabcolsep) * \real{0.1538}}@{}}
\toprule\noalign{}
\begin{minipage}[b]{\linewidth}\raggedright
Engine
\end{minipage} & \begin{minipage}[b]{\linewidth}\raggedright
Class
\end{minipage} & \begin{minipage}[b]{\linewidth}\raggedleft
CVEs (window)
\end{minipage} & \begin{minipage}[b]{\linewidth}\raggedleft
Escape
\end{minipage} & \begin{minipage}[b]{\linewidth}\raggedleft
HostLeak
\end{minipage} & \begin{minipage}[b]{\linewidth}\raggedleft
HostDoS
\end{minipage} & \begin{minipage}[b]{\linewidth}\raggedleft
InternalEsc
\end{minipage} \\
\midrule\noalign{}
\endhead
\bottomrule\noalign{}
\endlastfoot
Firecracker & microVM & 2 & 2† & 0 & 0 & 0 \\
Cloud Hypervisor & microVM & 2 & 1‡ & 1 & 0 & 0 \\
libkrun & microVM & 0 & 0 & 0 & 0 & 0 \\
gVisor (runsc) & userspace kernel & 3 & 0 & 2 & 0 & 1 \\
runc & container & 4 & 4 & 0 & 0 & 0 \\
Linux kernel & (shared dep) & \textasciitilde3,500 NVD total & --- & ---
& --- & --- \\
\end{longtable}
}

† Firecracker baseline shift --- both 2026 advisories carry escape-class
primitives; the methodology's prior baseline of ``no published
hypervisor-escape'' is contradicted (see §2.4.5).\\
‡ Cloud Hypervisor baseline shift --- CVE-2026-45782 (virtio-block
async-I/O UAF, CVSS v4 8.9 high) is the engine's first published
Escape-class advisory; combined with the Firecracker shift, two of three
microVM engines now carry escape-class CVEs on public record within a
\textasciitilde4-month period.

The class spread separates the three classes cleanly: runc's
Escape-heavy distribution (4/4) is endemic to the container class's
shared-kernel threat model; gVisor's information-disclosure-heavy
distribution (2 HostLeak + 1 InternalEsc, 0 Escape) is consistent with
the Sentry intercepting most syscalls before they reach the host kernel;
the microVM class sits in a 0--2 CVE per-engine range with the 2026
cluster shifting the historical baseline.

\textbf{Literature framing --- landscape only.} A 2025 systematic
mapping (Sroor et al., arXiv:2512.11940) consolidates 129
container-vulnerability studies (2000--2024) across a six-phase
life-cycle taxonomy with 66 distinct risks; the mapping enumerates no
CVE IDs and is cited as landscape, not as a CVE-comparator against the
per-engine counts above. Two other engine-overlapping empirical studies
measure a different channel and are listed for completeness: Weissman et
al.~(arXiv:2311.15999, Spectre/MDS PoCs against Firecracker v1.0.0 and
v1.4.0) and Dipta et al.~(arXiv:2404.10715, CNN-based cpufreq
fingerprinting against Docker / gVisor / Firecracker / Gramine /
AMD-SEV). Both are \textbf{out of scope per AISI T0.H2.N2}
(microarchitectural side channels --- single-tenant operator threat
model excludes cross-tenant μarch leakage); their findings do not enter
the rollup table.

\paragraph{2.4.2 Per-engine ranking on the rollup-window
count}\label{per-engine-ranking-on-the-rollup-window-count}

Per-engine ordering, lower is better. libkrun is reported but not ranked
(see §3.5):

{\def\LTcaptype{none} % do not increment counter
\begin{longtable}[]{@{}
  >{\raggedright\arraybackslash}p{(\linewidth - 8\tabcolsep) * \real{0.2000}}
  >{\raggedright\arraybackslash}p{(\linewidth - 8\tabcolsep) * \real{0.2000}}
  >{\raggedright\arraybackslash}p{(\linewidth - 8\tabcolsep) * \real{0.2000}}
  >{\raggedright\arraybackslash}p{(\linewidth - 8\tabcolsep) * \real{0.2000}}
  >{\raggedright\arraybackslash}p{(\linewidth - 8\tabcolsep) * \real{0.2000}}@{}}
\toprule\noalign{}
\begin{minipage}[b]{\linewidth}\raggedright
Rank
\end{minipage} & \begin{minipage}[b]{\linewidth}\raggedright
Engine
\end{minipage} & \begin{minipage}[b]{\linewidth}\raggedright
In-window count
\end{minipage} & \begin{minipage}[b]{\linewidth}\raggedright
Escape-class subset
\end{minipage} & \begin{minipage}[b]{\linewidth}\raggedright
Defensible basis
\end{minipage} \\
\midrule\noalign{}
\endhead
\bottomrule\noalign{}
\endlastfoot
\rkbest{} & \textbf{Cloud Hypervisor} & 2 & 1 Escape (potential): CVE-2026-45782
virtio-block UAF, 8.9 v4 & First Escape-class CVE landed 2026-05-14;
small engine, mature design, small attack surface \\
\rkmid{} & \textbf{gVisor} & 3 & 0 Escape, 2 HostLeak, 1 InternalEsc & No
Escape-class CVEs in the rollup; HostLeak and InternalEsc are
operator-relevant but not host-boundary \\
\rkmid{} & \textbf{Firecracker} & 2 & 2 escape-class (1 Escape (potential):
CVE-2026-5747 virtio-pci OOB, 8.7 v4; 1 Escape: CVE-2026-1386 jailer
symlink host-write, 6.0 v4) & First two escape-class CVEs landed in
2026, contradicting the methodology's prior baseline of ``no published
hypervisor-escape'' \\
\rkworst{} & \textbf{runc} & 4 & 4 Escape (incl.~1 `Escape (limited)') &
In-window cluster: CVE-2025-52565, CVE-2025-52881, CVE-2025-31133
(2025-11-05 procfs / mount-race trio), CVE-2024-45310 (`Escape
(limited)', filesystem trick); same bug class as historical
CVE-2019-19921, CVE-2023-27561, CVE-2023-28642 \\
--- & \textbf{libkrun} & 0 & 0 & Excluded from the ranking. Zero
published CVEs is the absence of a finding, not the presence of
soundness --- see §3.5 \\
\end{longtable}
}

\paragraph{2.4.3 Bug-pattern recurrence and class
fit}\label{bug-pattern-recurrence-and-class-fit}

\textbf{runc --- procfs / mount-race recurrence.} At least 3 historical
runc CVEs trace to \emph{procfs mount races} (CVE-2019-19921,
CVE-2023-27561, CVE-2023-28642). The 2025-11-05 cluster (CVE-2025-52565
\texttt{/dev/console} mount + races, CVE-2025-52881 arbitrary-write
gadgets + procfs write redirects, CVE-2025-31133 `masked path' abuse +
mount races) extends the same pattern --- three Escape CVEs published
simultaneously, all variants of procfs / mount-namespace handling under
runtime configuration. The pattern is endemic to the container class's
threat model: runc shares the host kernel, so any mount / procfs /
symlink race the runtime constructs is a potential boundary crossing. 13
of 16 historical runc CVEs classify as Escape; the class fit is the
load-bearing pattern, not the count.

\textbf{gVisor --- information-disclosure-heavy distribution.} The 3
in-window CVEs (CVE-2025-2713 InternalEsc; CVE-2024-10603 HostLeak;
CVE-2024-10026 HostLeak) carry no host-impact category beyond
information disclosure. This is methodology-expected for the
userspace-kernel class --- the Sentry intercepts most syscalls before
they reach the host kernel, so direct host-impact CVEs are structurally
harder to land. The sole historical gVisor Escape (CVE-2018-16359,
early-2018 seccomp gap allowing \texttt{renameat} on host files) sits
outside the 24-month window; modern gVisor's class fit is the
\emph{absence} of Escape entries in the rollup. CVE-2024-10026 (weak
hashing + small seed sizes leaking local IP + per-boot identifier)
credits academic researchers Inon Kaplan, Ron Even, Amit Klein (Hebrew U
/ Bar-Ilan) --- the only external-academic attribution across all five
engines in window.

\textbf{microVM class --- 0--2 CVE per-engine, 2026 cluster.}
Firecracker's 5 historical CVEs split between pre-2021 (HostDoS via
serial buffer growth; microVM-internal DoS; Escape (potential) vsock
buffer overflow) and the 2026 cluster (CVE-2026-5747 Escape (potential),
OOB write in virtio-pci; CVE-2026-1386 Escape, jailer symlink
host-write). Cloud Hypervisor's three historical CVEs cluster tightly:
CVE-2026-27211 (HostLeak, QCOW backing-file abuse) and CVE-2026-45782
(Escape (potential), virtio-block UAF) sit 84 days apart in 2026 ---
both reported by parties external to the core maintainers (DemiMarie /
Qubes OS for the UAF, with Meta engineers landing both fixes). May
signal the engine has entered an active disclosure window in 2026 rather
than reflecting a long-term shift in shallow-bug rate.

\paragraph{2.4.4 Per-class transitivity --- the Linux kernel as shared
dependency}\label{per-class-transitivity-the-linux-kernel-as-shared-dependency}

The Linux kernel contributes \textasciitilde3,500 in-window CVEs across
the same rollup, two and a half orders of magnitude larger than all five
engines combined. Every product carries the kernel surface that 1.1
measures (with very different surface sizes --- gVisor exposes 84
syscalls via the Sentry filter; daytona's runc exposes all 400+).

{\def\LTcaptype{none} % do not increment counter
\begin{longtable}[]{@{}
  >{\raggedright\arraybackslash}p{(\linewidth - 2\tabcolsep) * \real{0.5000}}
  >{\raggedright\arraybackslash}p{(\linewidth - 2\tabcolsep) * \real{0.5000}}@{}}
\toprule\noalign{}
\begin{minipage}[b]{\linewidth}\raggedright
Engine class
\end{minipage} & \begin{minipage}[b]{\linewidth}\raggedright
Kernel CVE impact
\end{minipage} \\
\midrule\noalign{}
\endhead
\bottomrule\noalign{}
\endlastfoot
Container (runc) & \textbf{Full surface.} Every host-kernel CVE
reachable from unprivileged userspace applies. \\
microVM (Firecracker, Cloud Hypervisor, libkrun) & \textbf{Partial via
guest kernel only.} Host kernel CVE → escape only if the VMM has its own
surface bug. Guest kernel pin shifts the risk profile (operators choose
the guest kernel independently of the host kernel). \\
Userspace kernel (gVisor) & \textbf{Minimal.} Sentry intercepts most
syscalls; host-kernel CVE generally not reachable via the syscall path.
Direct host-kernel reach is limited to the few syscalls the Sentry
forwards. \\
\end{longtable}
}

Three syzkaller-typical kernel UAFs in the rollup window illustrate the
asymmetry. CVE-2024-50264 (vsock / virtio dangling pointer → UAF, local
priv-esc primitive, CVSS 7.8) and CVE-2025-21756 (vsock binding lost
during transport-reassignment race → UAF, CVSS 7.8) apply to runc
directly, apply to microVMs only via the operator-chosen guest kernel,
and largely don't apply to gVisor (the Sentry doesn't forward the vsock
socket family). CVE-2026-31431 (crypto/algif\_aead in-place /
out-of-place bug; reachable via AF\_ALG, CVSS 7.8) follows the same
pattern.

Engine-level CVE counts do not include the shared-kernel exposure; per
the methodology, per-CVE attribution against the kernel's set is
infeasible at this scale and is reported as an aggregate baseline. The
aggregate also drifts upward each rerun as NVD continues to ingest
kernel-CNA filings post-publication --- the May 2025 sample alone moved
28 → 86 in six weeks across the prior and current compiles; the
magnitude conclusion (thousands of CVEs) is unchanged.

Ghimire et al.~(arXiv:2511.18155, eBPF-PATROL) state the structural
implication verbatim: \emph{``As containerized workloads share the same
kernel, any vulnerability that can be triggered via system calls
\ldots{} can potentially compromise the host or other co-resident
containers.''} This is the shared-kernel transitivity that the table
above splits by engine class.

\paragraph{2.4.5 Baseline-shifting findings this
rerun}\label{baseline-shifting-findings-this-rerun}

Three live findings from the 2026-05-20 rerun re-set the methodology's
prior baselines.

\textbf{Firecracker first published escape-class CVEs (CVE-2026-5747,
CVE-2026-1386).} The methodology's pre-existing baseline read
``Firecracker has no published hypervisor-escape CVE.'' Two 2026
advisories carry escape-class primitives: CVE-2026-5747 (OOB write into
VMM heap, CVSS v4 8.7 high --- path from there to host RCE still
requires gadget construction, ASLR/CFI bypass on a hardened VMM build)
and CVE-2026-1386 (arbitrary host file overwrite via symlink in jailer,
CVSS v4 6.0 medium --- easier to weaponize for specific files but
narrower than a code-execution primitive). Neither is a demonstrated
full RCE on host; both cross the host boundary. The
\texttt{Escape\ (potential)} label exists exactly for this case ---
primitive exists, no end-to-end exploit published. The methodology
baseline has been revised in
\texttt{sandbox-isolation-methodology-v2.md} §1.4 alongside the source
paper.

\textbf{Cloud Hypervisor first published Escape-class CVE
(CVE-2026-45782).} Virtio-block async-I/O use-after-free in the default
config (io\_uring / aio path), CVSS v4 8.9 high, CWE-416. The UAF write
primitive lands in the VMM heap; the GHSA wording explicitly states it
``can be escalated to arbitrary code execution and a full guest-to-host
(VM) escape.'' Reporter DemiMarie (Qubes OS security team); remediation
by Dylan Reid (Meta) --- both external to the Cloud Hypervisor core
maintainers. Combined with the Firecracker shift, two of three microVM
engines now carry escape-class CVEs on public record within a
\textasciitilde4-month period.

\textbf{CVE.org tie-breaker rule fires its first concrete case
(CVE-2026-27211).} First cross-source severity disagreement under the
new methodology. Same CVE, same CNA (\texttt{GitHub\_M}), divergent
vectors:

{\def\LTcaptype{none} % do not increment counter
\begin{longtable}[]{@{}
  >{\raggedright\arraybackslash}p{(\linewidth - 4\tabcolsep) * \real{0.3333}}
  >{\raggedright\arraybackslash}p{(\linewidth - 4\tabcolsep) * \real{0.3333}}
  >{\raggedright\arraybackslash}p{(\linewidth - 4\tabcolsep) * \real{0.3333}}@{}}
\toprule\noalign{}
\begin{minipage}[b]{\linewidth}\raggedright
Source
\end{minipage} & \begin{minipage}[b]{\linewidth}\raggedright
Score
\end{minipage} & \begin{minipage}[b]{\linewidth}\raggedright
Vector
\end{minipage} \\
\midrule\noalign{}
\endhead
\bottomrule\noalign{}
\endlastfoot
GHSA REST API (per-repo + global advisory DB) & 7.2 high &
\bttt{CVSS:4.0/AV:L/AC:L/AT:P/PR:N/UI:N/VC:H/VI:N/VA:N/SC:H/SI:H/SA:H} \\
CVE.org official record & 9.1 critical &
\bttt{CVSS:4.0/AV:N/AC:L/AT:P/PR:N/UI:N/VC:H/VI:N/VA:N/SC:H/SI:H/SA:H} \\
\end{longtable}
}

The Attack Vector field diverges (\texttt{AV:L} Local vs \texttt{AV:N}
Network); all other metrics agree. Methodology §1.4's tie-breaker rule
resolves it in CVE.org's favor → critical (9.1 v4) stands. Without the
rule the choice between ``high'' and ``critical'' would be arbitrary;
with it the resolution is deterministic and audit-recordable.

\paragraph{2.4.6 Structural caveats}\label{structural-caveats-1}

\textbf{``0 CVEs'' is not the strongest signal it appears, and the
methodology excludes libkrun from the ordering.} libkrun's 0 published
CVEs is \emph{the absence of a finding}, not the presence of soundness.
The same number could mean (a) no bugs exist (improbable for any
non-trivial codebase), (b) bugs exist but no one has searched, (c) bugs
were found and silently fixed without CVE assignment. Cross-axis with
1.6's ``no upstream fuzzer'' finding for libkrun, interpretation (b) is
the most likely --- see §3.5. Putting libkrun at \rkbest{} on this axis would
invert the signal; the engine is \texttt{—} (unmeasured) instead.

\textbf{Eyeballs-proxy normalization is qualitative.} SLOC / age / stars
/ known production users bound the surface and the discovery cadence but
do not compose into a single ``expected CVE rate.'' Reading the table
without context (runc 4-in-window against 310k Go LOC + 10.9 years +
13.2k stars; Firecracker 2-in-window against 85k Rust LOC + 8.6 years +
34.2k stars; libkrun 0 against 72k Rust LOC + 5.7 years + 2.1k stars)
misses the structural argument: lower code volume × lower eyeballs is
the worst combination for the ``0 CVEs means quality'' reading.

\textbf{Days-to-patch $\approx$ 0 by construction across all
coordinated-disclosure CVEs.} Every in-window runc / Firecracker / Cloud
Hypervisor advisory shipped the patched release on advisory day; P50 /
P95 evaluate to 0 across the entire rollup table. This is
methodology-correct but a \emph{dataset artefact}, not a
\emph{project-quality signal}. The metric becomes meaningful only when
uncoordinated disclosures enter the dataset. 1.5 promotes the
downstream-side measurement to first-class because of this artefact (see
§2.5).

\textbf{Source-priority is split by engine ecosystem after OSV.dev came
back empty for four of five engines.} Despite the methodology promoting
OSV.dev to primary aggregator on the basis of broad coverage, this
rerun's per-engine queries returned 0 results across crates.io and
bare-name forms for Firecracker, Cloud Hypervisor, and libkrun, and 0
for \texttt{gvisor.dev/gvisor} (Google CNA filings don't flow through
OSV's Go ecosystem feed). Only runc's Go ecosystem feed produces useful
signal --- and it does add real value there (surfacing
\texttt{GHSA-g54h-m393-cpwq} / \texttt{GO-2022-0396}, an advisory
without a CVE that the NVD-keyword pass missed). The methodology now
splits priority by ecosystem (Go → OSV primary; Rust-VMM → GHSA-per-repo
+ NVD primary; Google-CNA → NVD-keyword primary).

What 1.4 does not measure: whether the engine is actually safer (per
§1.4 of the axis paper, ``a higher CVE count is compatible with more
security work, not less''); forward-looking bug-rate prediction;
unpublished or embargoed bugs; whether the disclosed CVEs were actually
exploited in production (that's KEV-list / EPSS scoring, not in scope).

\subsubsection{2.5 Axis 1.5 --- Patch
cadence}\label{axis-1.5-patch-cadence}

The axis splits into upstream and downstream readings, governed by the
disclosure-model classification 1.4 introduces in §1.4 v2 and reads back
here as a measurement-design constraint. The methodology promotes the
downstream reading to first-class measurement; the operator-facing
signal is the downstream one.

\paragraph{2.5.1 Disclosure-model classification --- the schema that
controls the
reading}\label{disclosure-model-classification-the-schema-that-controls-the-reading}

Before any percentile aggregates, each engine carries a per-engine
disclosure-model classification that determines how the upstream
\texttt{days-to-patch} row should be \emph{read}:

{\def\LTcaptype{none} % do not increment counter
\begin{longtable}[]{@{}
  >{\raggedright\arraybackslash}p{(\linewidth - 4\tabcolsep) * \real{0.3333}}
  >{\raggedright\arraybackslash}p{(\linewidth - 4\tabcolsep) * \real{0.3333}}
  >{\raggedright\arraybackslash}p{(\linewidth - 4\tabcolsep) * \real{0.3333}}@{}}
\toprule\noalign{}
\begin{minipage}[b]{\linewidth}\raggedright
Model
\end{minipage} & \begin{minipage}[b]{\linewidth}\raggedright
Signature
\end{minipage} & \begin{minipage}[b]{\linewidth}\raggedright
Implication for upstream P-x
\end{minipage} \\
\midrule\noalign{}
\endhead
\bottomrule\noalign{}
\endlastfoot
\textbf{Coordinated} (GHSA + patched release published together) &
advisory date $\approx$ patched release date & P-x $\approx$ 0d by construction; reads
as process maturity, not patch speed \\
\textbf{Silent-fix-first} (fix commit lands in a release; CVE assigned
later, often months) & commit date $\ll$ disclosure date; first reachable
release tag $\ll$ disclosure date & disclosure → release \textbf{negative};
operators on the live channel had the fix before the CVE existed \\
\textbf{Uncoordinated} (researcher publishes PoC or write-up before
maintainer ships fix) & disclosure date \textless{} patched release date
& P-x meaningful as project-quality signal --- the case the metric was
originally designed for \\
\textbf{Mixed} & per-CVE varies & report per-CVE in the per-engine
table; do not aggregate percentiles across models in a single row \\
\end{longtable}
}

Engine classifications under this rerun: runc, Firecracker, Cloud
Hypervisor --- \textbf{coordinated} (every in-window CVE ships advisory
+ patched release same day; one runc exception, CVE-2023-27561, was a
known regression referencing public prior art). gVisor ---
\textbf{silent-fix-first} across all four modern (post-2023) CVEs (the 3
older 2018-era rows are right-censored against the earliest reachable
git tag \texttt{release-20190304.1}, March 2019, and are not
representative). libkrun --- n/a (0 CVEs). No engine in scope renders
\texttt{mixed} or \texttt{uncoordinated}; the schema fires on its first
concrete test (see §2.5.7).

The fix-first disclosure pattern observed for gVisor is documented in
the broader literature as the dominant ecosystem-level pattern. Heng et
al.~(arXiv:2411.07480) hand-code 312 Java-ecosystem CVEs post-2017 into
a six-event lifecycle (Reserved · Vendor-Fix · Vendor-Disclose ·
CVE-Published · Community-Fix · Community-Disclose) and report fix-first
as dominant at 85.89\%. Cited as methodology-precedent for the
lifecycle-event framing; \textbf{not} quoted as a numerical comparator
--- the authors' own External Validity disclaimer reads \emph{``the
findings are confined to the Java software ecosystem, suggesting that
the observed trends and reactions may not be applicable to
vulnerabilities in other platforms.''} Our container-engine corpus is
disjoint from theirs, and the 85.89\% figure should not be transported
across the ecosystem boundary.

\paragraph{2.5.2 Upstream cadence --- per-engine distribution and gVisor
silent-fix
crosswalk}\label{upstream-cadence-per-engine-distribution-and-gvisor-silent-fix-crosswalk}

{\def\LTcaptype{none} % do not increment counter
\begin{longtable}[]{@{}
  >{\raggedright\arraybackslash}p{(\linewidth - 14\tabcolsep) * \real{0.1034}}
  >{\raggedright\arraybackslash}p{(\linewidth - 14\tabcolsep) * \real{0.1034}}
  >{\raggedleft\arraybackslash}p{(\linewidth - 14\tabcolsep) * \real{0.1379}}
  >{\raggedleft\arraybackslash}p{(\linewidth - 14\tabcolsep) * \real{0.1379}}
  >{\raggedleft\arraybackslash}p{(\linewidth - 14\tabcolsep) * \real{0.1379}}
  >{\raggedleft\arraybackslash}p{(\linewidth - 14\tabcolsep) * \real{0.1379}}
  >{\raggedleft\arraybackslash}p{(\linewidth - 14\tabcolsep) * \real{0.1379}}
  >{\raggedright\arraybackslash}p{(\linewidth - 14\tabcolsep) * \real{0.1034}}@{}}
\toprule\noalign{}
\begin{minipage}[b]{\linewidth}\raggedright
Engine
\end{minipage} & \begin{minipage}[b]{\linewidth}\raggedright
Disclosure model
\end{minipage} & \begin{minipage}[b]{\linewidth}\raggedleft
N
\end{minipage} & \begin{minipage}[b]{\linewidth}\raggedleft
P50
\end{minipage} & \begin{minipage}[b]{\linewidth}\raggedleft
P95
\end{minipage} & \begin{minipage}[b]{\linewidth}\raggedleft
Max
\end{minipage} & \begin{minipage}[b]{\linewidth}\raggedleft
Min
\end{minipage} & \begin{minipage}[b]{\linewidth}\raggedright
Reading
\end{minipage} \\
\midrule\noalign{}
\endhead
\bottomrule\noalign{}
\endlastfoot
runc & coordinated & 16 & 0d & 0d & 26d & 0d & embargo pipeline working;
15/16 coordinated; outlier CVE-2023-27561 is a regression \\
Firecracker & coordinated & 5 & 0d & 0d & 0d & 0d & 100\% coordinated;
pipeline working \\
Cloud Hypervisor & coordinated & 3 & 0d & 0d & 0d & 0d & 100\%
coordinated; N small \\
libkrun & n/a & 0 & --- & --- & --- & --- & no published CVEs (1.4
negative finding) \\
gVisor (runsc) & silent-fix-first & 7 & −165d & −458d & −460d & −10d &
all-negative for modern (post-2023) CVEs; 2018-era values
right-censored \\
Linux kernel & excluded & \textasciitilde3,500 & --- & --- & --- & --- &
excluded --- kernel CNA assigns CVEs \emph{after} fix ships \\
\end{longtable}
}

A \texttt{0d} row under coordinated disclosure and a \texttt{0d} row
under uncoordinated disclosure carry opposite meanings; the
classification beside each row is required to read the percentile
correctly. Under v1 of the methodology, Cloud Hypervisor's \texttt{0d}
and gVisor's \texttt{−165d} both rendered as ``fast patches'' --- a
category error the v2 schema fixes.

The gVisor silent-fix-first crosswalk (fix commit → release tag → CVE
assignment for the four modern CVEs):

{\def\LTcaptype{none} % do not increment counter
\begin{longtable}[]{@{}
  >{\raggedright\arraybackslash}p{(\linewidth - 8\tabcolsep) * \real{0.1875}}
  >{\raggedright\arraybackslash}p{(\linewidth - 8\tabcolsep) * \real{0.1875}}
  >{\raggedright\arraybackslash}p{(\linewidth - 8\tabcolsep) * \real{0.1875}}
  >{\raggedright\arraybackslash}p{(\linewidth - 8\tabcolsep) * \real{0.1875}}
  >{\raggedleft\arraybackslash}p{(\linewidth - 8\tabcolsep) * \real{0.2500}}@{}}
\toprule\noalign{}
\begin{minipage}[b]{\linewidth}\raggedright
CVE
\end{minipage} & \begin{minipage}[b]{\linewidth}\raggedright
Fix commit
\end{minipage} & \begin{minipage}[b]{\linewidth}\raggedright
Commit date
\end{minipage} & \begin{minipage}[b]{\linewidth}\raggedright
First release
\end{minipage} & \begin{minipage}[b]{\linewidth}\raggedleft
Disclosure → release
\end{minipage} \\
\midrule\noalign{}
\endhead
\bottomrule\noalign{}
\endlastfoot
CVE-2025-2713 & \texttt{586c38d70081} (``Apply image's file
capabilities'') & 2024-03-15 & release-20240325.0 (2024-03-23) &
\textbf{−370d} \\
CVE-2024-10603 & \texttt{cbdb2c61b1f7} (``Randomize TCP source port'') +
follow-ups & 2023-11-10 & release-20231113.0 (2023-11-12) &
\textbf{−445d} \\
CVE-2024-10026 & \texttt{83f75082e5b0} (``netstack: cryptographically
secure RNG'') + chain & 2023-10-28 & release-20231030.0 (2023-10-30) &
\textbf{−458d} \\
CVE-2023-7258 & \texttt{6a112c60a257} (``Check mounts marked as
unmounted before propagating'') & 2023-11-30 & release-20231204.0
(2023-12-02) & \textbf{−165d} \\
\end{longtable}
}

Operators consuming gVisor's \texttt{release} bucket would have had each
of these fixes for the better part of a year before the CVE went public.
CVE-2024-10026 and CVE-2024-10603 share fix commits (the
\texttt{math/rand} weakness class); a follow-up cleanup commit
\texttt{305818798} (April 2026) extends the fix to five additional code
paths still using \texttt{math/rand}. The bug class is not fully closed
and operators should watch for follow-up CVEs.

\paragraph{2.5.3 Downstream verdict matrix --- the operator-facing
signal}\label{downstream-verdict-matrix-the-operator-facing-signal}

Downstream ordering (per the methodology, verdicts are \texttt{current}
/ \texttt{moderate} / \texttt{frozen} /
\texttt{won\textquotesingle{}t-fix} / \texttt{opaque}):

{\def\LTcaptype{none} % do not increment counter
\begin{longtable}[]{@{}
  >{\raggedright\arraybackslash}p{(\linewidth - 6\tabcolsep) * \real{0.2500}}
  >{\raggedright\arraybackslash}p{(\linewidth - 6\tabcolsep) * \real{0.2500}}
  >{\raggedright\arraybackslash}p{(\linewidth - 6\tabcolsep) * \real{0.2500}}
  >{\raggedright\arraybackslash}p{(\linewidth - 6\tabcolsep) * \real{0.2500}}@{}}
\toprule\noalign{}
\begin{minipage}[b]{\linewidth}\raggedright
Rank
\end{minipage} & \begin{minipage}[b]{\linewidth}\raggedright
Product
\end{minipage} & \begin{minipage}[b]{\linewidth}\raggedright
Verdict
\end{minipage} & \begin{minipage}[b]{\linewidth}\raggedright
Defensible basis
\end{minipage} \\
\midrule\noalign{}
\endhead
\bottomrule\noalign{}
\endlastfoot
\rkbest{} & \textbf{microsandbox} & \texttt{current} & 1--3 month bump cadence;
libkrun has 0 CVEs, so exposure is undefined \\
\rkbest{} & \textbf{gvisor product} & \texttt{current} & Rolling main / release
bucket; zero pin lag by construction \\
\rkbest{} & \textbf{daytona (default)} & \texttt{current} & Docker Compose path
pulls Docker-CE 29.x with runc 1.3.5 via \texttt{containerd.io}; all
in-window runc CVEs included \\
\rkworst{} & \textbf{arrakis} & \texttt{frozen} & Pinned at Cloud Hypervisor
\texttt{v44.0} for 471+ days across 12 upstream releases, including
CVE-2026-45782 (≥7d unpatched) and CVE-2026-27211 (≥90d unpatched) \\
\rkworst{} & \textbf{e2b self-hosted} & \texttt{frozen} & Pinned at Firecracker
\texttt{v1.14.1\_458ca91} for 399 days; CVE-2026-5747 (Escape-class 8.7
v4) ≥44d unpatched at the orchestrator-default flag \\
\rkworst{} & \textbf{e2b cloud-hosted} & \texttt{opaque} & Pin is not publicly
observable \\
\end{longtable}
}

The downstream lag spectrum at a glance:

\begin{figure}
\centering
\includegraphics[width=1\linewidth,height=\textheight,keepaspectratio,alt={Downstream pin lag spectrum across the five products (axis 1.5)}]{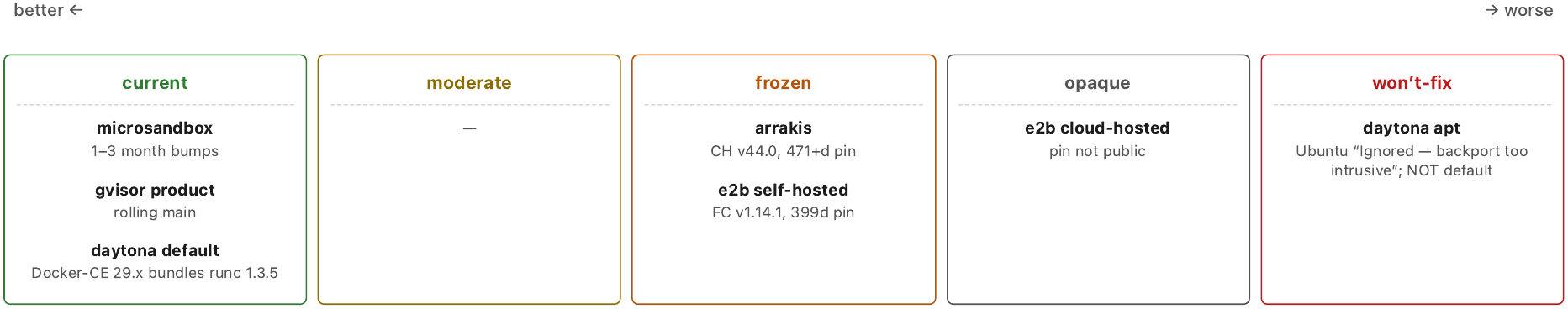}
\caption{Downstream pin lag spectrum across the five products (axis
1.5)}
\end{figure}

\paragraph{2.5.4 arrakis pin-freeze --- 471+ days on v44.0 across 12
upstream
releases}\label{arrakis-pin-freeze-471-days-on-v44.0-across-12-upstream-releases}

arrakis pinned Cloud Hypervisor at \texttt{v44.0} in commit
\texttt{024f74a} (2025-02-04) and has not bumped since. The pin is a
single hard-coded version in \texttt{setup/install-images.py:45} with no
automated bump policy. Upstream has cut 12 tagged releases in the
interval (eight majors v45.0 through v52.0, plus point releases v50.1 /
v50.2 / v51.1 / v51.2), two of which ship patches for in-window
critical/high CVEs:

{\def\LTcaptype{none} % do not increment counter
\begin{longtable}[]{@{}
  >{\raggedright\arraybackslash}p{(\linewidth - 6\tabcolsep) * \real{0.2308}}
  >{\raggedright\arraybackslash}p{(\linewidth - 6\tabcolsep) * \real{0.2308}}
  >{\raggedleft\arraybackslash}p{(\linewidth - 6\tabcolsep) * \real{0.3077}}
  >{\raggedright\arraybackslash}p{(\linewidth - 6\tabcolsep) * \real{0.2308}}@{}}
\toprule\noalign{}
\begin{minipage}[b]{\linewidth}\raggedright
Tag
\end{minipage} & \begin{minipage}[b]{\linewidth}\raggedright
Date
\end{minipage} & \begin{minipage}[b]{\linewidth}\raggedleft
Days since arrakis pin (2025-02-04)
\end{minipage} & \begin{minipage}[b]{\linewidth}\raggedright
Patches
\end{minipage} \\
\midrule\noalign{}
\endhead
\bottomrule\noalign{}
\endlastfoot
\textbf{v50.1 / v51.0} & 2026-02-20 & +381 & CVE-2026-27211 (HostLeak,
CVSS v4 9.1 critical) \\
\textbf{v52.0 / v51.2} & 2026-05-14 & +464 & CVE-2026-45782 (Escape
(potential), CVSS v4 8.9 high) \\
\end{longtable}
}

The pin policy is the structural failure mode here, not the engine.
Cloud Hypervisor's upstream cadence is fast (12 releases in 464 days,
two carrying same-day coordinated CVE fixes). arrakis's exposure window
is a product of the absence of a bump policy, not of upstream behaviour.
The v52.0 / v51.2 tags were confirmed locally this rerun via
\texttt{git\ fetch\ -\/-tags} against \texttt{.repos/cloud-hypervisor}
(both dated 2026-05-14 20:59 UTC / 21:49 UTC respectively).

\paragraph{2.5.5 e2b orchestrator default --- 399 days unchanged,
fc-versions iteration is
non-security}\label{e2b-orchestrator-default-399-days-unchanged-fc-versions-iteration-is-non-security}

e2b pins Firecracker in two places: the \texttt{e2b-dev/infra}
orchestrator default flag (\texttt{v1.14.1\_458ca91}, pinned
2025-04-17), and the \texttt{e2b-dev/fc-versions} release artefact
registry. The orchestrator default has not changed in 399 days,
confirmed this rerun against
\texttt{e2b-dev/infra/packages/orchestrator/README.md}.
\texttt{fc-versions} has shipped two newer manual builds in May 2026:

{\def\LTcaptype{none} % do not increment counter
\begin{longtable}[]{@{}
  >{\raggedright\arraybackslash}p{(\linewidth - 6\tabcolsep) * \real{0.2500}}
  >{\raggedright\arraybackslash}p{(\linewidth - 6\tabcolsep) * \real{0.2500}}
  >{\raggedright\arraybackslash}p{(\linewidth - 6\tabcolsep) * \real{0.2500}}
  >{\raggedright\arraybackslash}p{(\linewidth - 6\tabcolsep) * \real{0.2500}}@{}}
\toprule\noalign{}
\begin{minipage}[b]{\linewidth}\raggedright
Tag
\end{minipage} & \begin{minipage}[b]{\linewidth}\raggedright
Date
\end{minipage} & \begin{minipage}[b]{\linewidth}\raggedright
Commit subject
\end{minipage} & \begin{minipage}[b]{\linewidth}\raggedright
CVE backport?
\end{minipage} \\
\midrule\noalign{}
\endhead
\bottomrule\noalign{}
\endlastfoot
\texttt{v1.14.1\_639196c} & 2026-05-18 & ``fix: saving/restoring async
IO engine transport state'' (cherry-picked from upstream, Babis Chalios)
& no --- snapshot/restore bug fix \\
\texttt{v1.14.1\_f0a35a1} & 2026-05-13 & manual build (commit subject
not surfaced via release notes) & unknown \\
\texttt{v1.14.1\_458ca91} & 2025-04-17 & \textbf{orchestrator default} &
CVE-2026-1386 unverified \\
\end{longtable}
}

The 2026-05-18 build's commit subject is dispositive ---
snapshot/restore, not a CVE backport. CVE-2026-5747 (Escape (potential),
CVSS v4 8.7 high, patched upstream in Firecracker 1.14.4 / 1.15.1 on
2026-04-07) is not on any of the three e2b builds; self-hosted operators
using the default infra pin are ≥44 days unpatched. CVE-2026-1386
remains genuinely unverifiable from public artefacts ---
\texttt{458ca91} is an e2b-applied patch on top of a 1.14.1 base, and
\texttt{fc-versions} release notes don't enumerate which backports a
manual build carries. Verification would require reading the patch file
in e2b's Firecracker fork. \textbf{Cloud-hosted verdict is
\texttt{opaque}} --- cloud customers may run a different pin than the
public infra default; the self-hosted finding does not transfer.

\paragraph{2.5.6 daytona install-path correction --- Docker-bundled
runc, not
apt-runc}\label{daytona-install-path-correction-docker-bundled-runc-not-apt-runc}

A material correction to the prior 1.5 compile (2026-05-10). The prior
compile assumed \texttt{apt\ install\ runc} as the default daytona
install path and concluded \texttt{won\textquotesingle{}t-fix} against
Ubuntu Noble's ``Ignored --- backport too intrusive'' stance on the
2025-11 runc trio. Verification this rerun against
\texttt{.repos/daytona/docker/README.md} and
\texttt{daytona.io/docs/oss-deployment} confirms the default OSS deploy
is:

\begin{verbatim}
docker compose -f docker/docker-compose.yaml up -d
\end{verbatim}

with stated prerequisite ``Install Docker and Docker Compose'' (linking
to docker.com's install pages). The runc binary daytona actually invokes
is the one bundled with Docker-CE 29.x via \texttt{containerd.io}, not
the apt \texttt{runc} package.

{\def\LTcaptype{none} % do not increment counter
\begin{longtable}[]{@{}
  >{\raggedright\arraybackslash}p{(\linewidth - 4\tabcolsep) * \real{0.3333}}
  >{\raggedright\arraybackslash}p{(\linewidth - 4\tabcolsep) * \real{0.3333}}
  >{\raggedright\arraybackslash}p{(\linewidth - 4\tabcolsep) * \real{0.3333}}@{}}
\toprule\noalign{}
\begin{minipage}[b]{\linewidth}\raggedright
Docker Engine
\end{minipage} & \begin{minipage}[b]{\linewidth}\raggedright
Bundled runc
\end{minipage} & \begin{minipage}[b]{\linewidth}\raggedright
Includes
\end{minipage} \\
\midrule\noalign{}
\endhead
\bottomrule\noalign{}
\endlastfoot
29.0.0 & runc 1.3.3 & 2025-11 trio (CVE-2025-52565/52881/31133),
CVE-2024-45310, CVE-2024-21626 \\
29.4.0 & runc 1.3.5 & same + follow-up fixes \\
29.5.2 & runc 1.3.5 (2026-05-20 latest) & same \\
\end{longtable}
}

The default daytona-on-Docker path is therefore \textbf{current} across
all in-window runc CVEs. The \texttt{won\textquotesingle{}t-fix} finding
survives but only on a non-default path (operator deviates from
daytona's docs and installs \texttt{docker.io} from Ubuntu's archive,
which pulls apt \texttt{runc} as a dependency). The Ubuntu Noble per-CVE
status preserved for the alt-path matrix:

{\def\LTcaptype{none} % do not increment counter
\begin{longtable}[]{@{}
  >{\raggedright\arraybackslash}p{(\linewidth - 4\tabcolsep) * \real{0.3333}}
  >{\raggedright\arraybackslash}p{(\linewidth - 4\tabcolsep) * \real{0.3333}}
  >{\raggedright\arraybackslash}p{(\linewidth - 4\tabcolsep) * \real{0.3333}}@{}}
\toprule\noalign{}
\begin{minipage}[b]{\linewidth}\raggedright
CVE
\end{minipage} & \begin{minipage}[b]{\linewidth}\raggedright
Noble apt-runc status
\end{minipage} & \begin{minipage}[b]{\linewidth}\raggedright
Noble snap-runc status
\end{minipage} \\
\midrule\noalign{}
\endhead
\bottomrule\noalign{}
\endlastfoot
CVE-2024-45310 & \textbf{Vulnerable --- no USN issued} (\textgreater625
days open) & same \\
CVE-2025-52565 & ``Ignored --- backport too intrusive'' & USN-7851-1
(snap only), 2025-11-04 \\
CVE-2025-52881 & ``Ignored --- backport too intrusive'' & USN-7851-1
(snap only), 2025-11-04 \\
CVE-2025-31133 & ``Ignored --- backport too intrusive'' & USN-7851-1
(snap only), 2025-11-04 \\
\end{longtable}
}

The methodology addition: for host-distro-pattern products, the install
path must be verified against the product's own docs before distro-lag
is computed, otherwise an alternative-path failure mode can be misread
as the default. This is now \texttt{Update\ procedure} step 5 in the
methodology v2.

\paragraph{2.5.7 Structural --- downstream dominates upstream by orders
of
magnitude}\label{structural-downstream-dominates-upstream-by-orders-of-magnitude}

The cross-cutting finding from this rerun is that operator-visible
patch-propagation is dominated by \textbf{product-side pin policy}, not
by \textbf{engine-side patch speed}. The upstream cadence column (with
the disclosure-model classification) maxes out at \textasciitilde26 days
for runc's single regression case and 0 days for everything else under
coordinated disclosure. The downstream column spans \textbf{0 days
(rolling channel) to ∞ (won't-fix on a host-distro path) to 471+ days
(frozen pin)} --- four orders of magnitude wider than the upstream
column.

Two cases anchor the spread. \textbf{arrakis is frozen on a fast-moving
engine} --- Cloud Hypervisor's upstream cadence is among the fastest in
the set (12 releases in 464 days, two carrying same-day coordinated CVE
patches); arrakis's downstream lag (471+ days, ≥90d on CVE-2026-27211)
is entirely attributable to the absence of a pin-bump policy.
\textbf{gvisor-product is current on a silent-fix engine} --- gVisor's
upstream cadence is structurally inverse (negative disclosure-to-release
lag); gvisor-product's downstream lag is \textasciitilde0d by virtue of
the rolling-channel consumption pattern.

This is also where the class-level isolation strength (microVM \rkbest{} vs
container \rkworst{} in 1.4) breaks as a predictor. arrakis is a strong-isolation
microVM (Cloud Hypervisor) with a frozen-pin policy; daytona is the
lowest-isolation container engine but is \texttt{current} on the
headline measure via Docker's bundled runc. The strong-isolation product
is operator-visible-worse on patch propagation. Class-level isolation
does not predict downstream lag; the two must be combined.

What 1.5 does not measure: patch speed under uncoordinated disclosure
(the dataset has none in window --- when the disclosure population
diversifies, the column becomes meaningful); cloud-hosted instance state
(e2b cloud is \texttt{opaque} per construction, not estimated);
forward-looking bug recurrence (that's 1.4's job; 1.5 reads cadence on
the bugs 1.4 catalogues); the kernel CVE backport stream (kernel.org CNA
filings are post-publication, so negative disclosure-to-release lags are
not meaningful for the kernel).

\subsubsection{2.6 Axis 1.6 --- Upstream fuzzing
posture}\label{axis-1.6-upstream-fuzzing-posture}

The forward-looking residual-bug signal that 1.1--1.5 cannot express.
The intuition is classical: saturated continuous fuzzing is the
strongest empirical evidence that ``no easy bugs left.'' The methodology
v2 promotes the binary ``documented continuous fuzzer y/n + dashboard
link'' to the primary operator-facing signal and demotes the per-CVE
attribution percentage to a best-effort secondary read, after the
2026-05-21 rerun made the percentile column structurally meaningless
under both coordinated and silent-fix-first disclosure.

\paragraph{2.6.1 Per-engine inventory and
ranking}\label{per-engine-inventory-and-ranking}

{\def\LTcaptype{none} % do not increment counter
\begin{longtable}[]{@{}
  >{\raggedright\arraybackslash}p{(\linewidth - 8\tabcolsep) * \real{0.2000}}
  >{\raggedright\arraybackslash}p{(\linewidth - 8\tabcolsep) * \real{0.2000}}
  >{\raggedright\arraybackslash}p{(\linewidth - 8\tabcolsep) * \real{0.2000}}
  >{\raggedright\arraybackslash}p{(\linewidth - 8\tabcolsep) * \real{0.2000}}
  >{\raggedright\arraybackslash}p{(\linewidth - 8\tabcolsep) * \real{0.2000}}@{}}
\toprule\noalign{}
\begin{minipage}[b]{\linewidth}\raggedright
Rank
\end{minipage} & \begin{minipage}[b]{\linewidth}\raggedright
Engine
\end{minipage} & \begin{minipage}[b]{\linewidth}\raggedright
Fuzzing
\end{minipage} & \begin{minipage}[b]{\linewidth}\raggedright
Public dashboard
\end{minipage} & \begin{minipage}[b]{\linewidth}\raggedright
Per-CVE attribution (24mo)
\end{minipage} \\
\midrule\noalign{}
\endhead
\bottomrule\noalign{}
\endlastfoot
\rkbest{} & \textbf{gVisor} & continuous syzkaller + in-tree \texttt{secfuzz}
library (seccomp-bpf, 2024-2025) + release-process syzkaller smoke test
(\texttt{9a4feec24}) & \textbf{yes} ---
\href{https://syzkaller.appspot.com/gvisor}{\seqsplit{syzkaller.appspot.com/gvisor}}
& 0/3 from commit-trailer evidence (baseline
\bttt{\textasciitilde{}70\%} unsupported on public artefacts; caveat:
silent-fix-first hides attribution by construction) \\
\rkbest{} & \textbf{Linux kernel} & syzbot continuous + KASAN / KMSAN / KFENCE /
KCSAN sanitisers across subtrees & \textbf{yes} ---
\href{https://syzkaller.appspot.com/upstream}{\seqsplit{syzkaller.appspot.com/upstream}}
& not inventoried per-CVE (infeasible at \textasciitilde3,500 in-window
CVEs); widely cited as \bttt{\textasciitilde{}80\%} via syzbot status
reports \\
\rkmid{} & \textbf{Cloud Hypervisor} & cargo-fuzz workspace at \texttt{fuzz/},
18 targets (\texttt{balloon}, \texttt{block}, \texttt{cmos},
\texttt{console}, \texttt{http\_api}, \texttt{iommu},
\texttt{linux\_loader}, \bttt{linux\_loader\_cmdline}, \texttt{mem},
\texttt{net}, \texttt{pmem}, \texttt{qcow}, \texttt{rng},
\texttt{serial}, \texttt{vhdx}, \texttt{vsock}, \texttt{watchdog},
\texttt{x86emul}); CI \texttt{fuzz-build} job runs
\bttt{cargo\ fuzz\ build} + \bttt{cargo\ fuzz\ check} on every PR
--- compile-gate only, no execution & no & 0/2 (CVE-2026-27211 fix by
Rob Bradford @ Meta --- internal review; CVE-2026-45782 fix by Dylan
Reid @ Meta, reporter DemiMarie / Qubes OS security audit --- manual not
fuzzer) \\
\rkmid{} & \textbf{runc} & OSS-Fuzz, 2 go-fuzz targets (\texttt{FuzzUser},
\texttt{FuzzUnmarshalJSON}) in
\bttt{tests/fuzzing/oss\_fuzz\_build.sh}; no CI run & indirect via
OSS-Fuzz issue tracker & 0/4 attributable from commit logs (shallow
clone --- see caveats) \\
\rkworst{} & \textbf{Firecracker} & \textbf{none} in repo;
\bttt{-\/-features\ fuzzing} is a deterministic-build hook for
\emph{external} fuzzers (TCP ISN hardcoded; balloon stats inline),
documented in \texttt{docs/fuzzing.md} (32 lines, banner:
\bttt{not\ for\ production\ use}); no \texttt{fuzz/}, no
\texttt{fuzz\_targets/}, no OSS-Fuzz integration & no & 0/2 attributable
from commit logs (shallow clone --- see caveats); methodology baseline
\bttt{\textasciitilde{}60\%} \textbf{not supported} by repo evidence
--- see §2.6.4 \\
\rkworst{} & \textbf{libkrun} & \textbf{none documented}; repo-wide search for
\texttt{fuzz} returns 5 hits, all GUI-input variables
(\bttt{krun\_gtk\_display}, \texttt{gui\_vm}) & no & n/a (0 published
CVEs in window --- see 1.4 negative finding) \\
\end{longtable}
}

The ``Public dashboard'' and ``Fuzzing'' columns together are the
operator-relevant signal. Of the engines an operator can run today, only
gVisor (syzkaller against runsc) and the Linux kernel (syzbot) expose a
continuous, externally-verifiable fuzzing operation. For the other four
engines, residual-bug confidence has to come from other sources
(architectural review, CVE history shape, vendor statements about
internal QA).

\textbf{Academic-literature corroboration for the gVisor row.} Li et
al.'s G-Fuzz (TDSC 2024 / arXiv:2409.13139, Ant Group + Zhejiang
University) is the only academic directed-fuzzer published for gVisor in
the literature and reports a max \textbf{131× speedup over Syzkaller} on
general targets. The paper's existence --- a published, peer-reviewed,
industrially-deployed directed fuzzer against the specific engine under
test --- is independent evidence the gVisor fuzzing posture is actively
invested at multiple layers; it supplements the upstream syzkaller
dashboard and in-tree \texttt{secfuzz} library on the gVisor \rkbest{} row.

\textbf{Linux-kernel fuzzer state-of-the-art context.} Four 2024--2025
papers anchor the rate at which the Linux-kernel fuzzing tier (the
dependency layer for every engine in this set) is advancing. Liu et
al.~(Psyzkaller, arXiv:2510.08918) report branch coverage +4.6--7.0\%
and crashes +110.4--187.2\% over Syzkaller with SDR-biased seed
generation and disclose 8 previously unknown bugs via CNVD. Sun et
al.~(SyzParam, arXiv:2501.10002) report +32.57\% average edge coverage
over Syzkaller across 8 drivers with sysfs-parameter-write mutation and
30 unique bugs / 9 CVE IDs. Yang et al.'s KernelGPT (ASPLOS 2025 /
arXiv:2401.00563) uses GPT-4 three-stage prompting to generate 532 new
syscall specifications (+13.6\% over Syzkaller's 3,903 baseline)
yielding 24 previously unknown bugs / 11 CVE IDs. The 107-paper SoK from
Xu et al.~(TOSEM / arXiv:2501.16165) reports Syzkaller has reported
\emph{``over 5,000 bugs''} across the surveyed body of work. The
landscape framing: the kernel fuzzing tier is advancing at the
seed-generation and harness-augmentation layers; the implication for
sandbox engines is that bugs absent from today's published CVE record
under coordinated disclosure can become measurable under any of the
augmented Syzkaller variants whenever the next mainstream tooling
generation lands.

\paragraph{2.6.2 Three-tier investment
distribution}\label{three-tier-investment-distribution}

The six engines split cleanly into three pairs by continuous-fuzzing
investment:

{\def\LTcaptype{none} % do not increment counter
\begin{longtable}[]{@{}
  >{\raggedright\arraybackslash}p{(\linewidth - 4\tabcolsep) * \real{0.3333}}
  >{\raggedright\arraybackslash}p{(\linewidth - 4\tabcolsep) * \real{0.3333}}
  >{\raggedright\arraybackslash}p{(\linewidth - 4\tabcolsep) * \real{0.3333}}@{}}
\toprule\noalign{}
\begin{minipage}[b]{\linewidth}\raggedright
Tier
\end{minipage} & \begin{minipage}[b]{\linewidth}\raggedright
Engines
\end{minipage} & \begin{minipage}[b]{\linewidth}\raggedright
Characterisation
\end{minipage} \\
\midrule\noalign{}
\endhead
\bottomrule\noalign{}
\endlastfoot
\textbf{Continuous public fuzzer + dashboard} & gVisor, Linux kernel &
Externally-observable fuzz operations; bugs filed via public dashboards;
in-tree harness depth varies (gVisor adds \texttt{secfuzz} in-tree for
seccomp-bpf; Linux adds sanitisers across subtrees) \\
\textbf{In-tree harness, no continuous upstream run} & Cloud Hypervisor,
runc & Harness exists and is testable; CI gates compilability (Cloud
Hypervisor) or relies on external infrastructure for execution
(runc/OSS-Fuzz); corpus does not accumulate upstream \\
\textbf{No upstream fuzzer} & Firecracker, libkrun & No in-tree fuzzer;
Firecracker has a determinism \emph{hook}
(\bttt{-\/-features\ fuzzing}) for external fuzzer drivers but no
fuzzer itself; libkrun has nothing \\
\end{longtable}
}

A secondary in-tree-vs-external-only distinction matters operationally.
\textbf{External-only fuzzing} (runc + OSS-Fuzz): the harness can go
stale silently --- if the OSS-Fuzz build script breaks, the fuzzer stops
running and the project may not notice; findings are filed out-of-band
to the project's issue tracker. \textbf{In-tree-but-not-executed
fuzzing} (Cloud Hypervisor): the harness lives with the code, CI gates
compilability on every PR, reviewers see fuzz-target changes alongside
production-code changes --- this prevents bit-rot but does not advance
corpus coverage. The natural progression --- in-tree harness + CI
build-gate + scheduled execution + public dashboard + corpus
accumulation --- is the path that gets a project from Cloud Hypervisor's
posture to gVisor's. Cloud Hypervisor has the first two; gVisor has all
five.

\paragraph{2.6.3 Per-CVE attribution --- the full 0/5
walk}\label{per-cve-attribution-the-full-05-walk}

The two engines with full local git history yield a full per-CVE walk
for the 24-month window:

{\def\LTcaptype{none} % do not increment counter
\begin{longtable}[]{@{}
  >{\raggedright\arraybackslash}p{(\linewidth - 12\tabcolsep) * \real{0.1429}}
  >{\raggedright\arraybackslash}p{(\linewidth - 12\tabcolsep) * \real{0.1429}}
  >{\raggedright\arraybackslash}p{(\linewidth - 12\tabcolsep) * \real{0.1429}}
  >{\raggedright\arraybackslash}p{(\linewidth - 12\tabcolsep) * \real{0.1429}}
  >{\raggedright\arraybackslash}p{(\linewidth - 12\tabcolsep) * \real{0.1429}}
  >{\raggedright\arraybackslash}p{(\linewidth - 12\tabcolsep) * \real{0.1429}}
  >{\raggedright\arraybackslash}p{(\linewidth - 12\tabcolsep) * \real{0.1429}}@{}}
\toprule\noalign{}
\begin{minipage}[b]{\linewidth}\raggedright
Engine
\end{minipage} & \begin{minipage}[b]{\linewidth}\raggedright
CVE
\end{minipage} & \begin{minipage}[b]{\linewidth}\raggedright
Fix commit(s)
\end{minipage} & \begin{minipage}[b]{\linewidth}\raggedright
Author
\end{minipage} & \begin{minipage}[b]{\linewidth}\raggedright
Reporter (from GHSA/NVD)
\end{minipage} & \begin{minipage}[b]{\linewidth}\raggedright
Fuzzer trailer?
\end{minipage} & \begin{minipage}[b]{\linewidth}\raggedright
Discovery route
\end{minipage} \\
\midrule\noalign{}
\endhead
\bottomrule\noalign{}
\endlastfoot
Cloud Hypervisor & CVE-2026-27211 (HostLeak, CVSS v4 9.1) &
\texttt{f93340d33} (``vmm: Add option to control backing files'') & Rob
Bradford \bttt{\textless{}rbradford@meta.com\textgreater{}} (Meta) &
(internal design review) & no & landlock-interaction design flaw,
identified internally \\
Cloud Hypervisor & CVE-2026-45782 (Escape (potential), CVSS v4 8.9) &
\texttt{3bf94535d} + \texttt{68feea7cb} + \texttt{3ba8e92c6}
(virtio-block async-I/O UAF chain) & Dylan Reid
\bttt{\textless{}dgreid@fb.com\textgreater{}} (Meta) & DemiMarie
(Qubes OS security audit) & no & external manual security audit \\
gVisor & CVE-2025-2713 (file-capabilities mishandling) &
\texttt{586c38d70081} (``Apply image's file capabilities'') &
\texttt{chjing@google.com} & (internal) & no & internal review \\
gVisor & CVE-2024-10603 (weak RNG in TCP source port) &
\texttt{cbdb2c61b1f7} (``Randomize TCP source port'') + follow-ups &
\texttt{zeling@google.com} & (academic researchers, see CVE-2024-10026
chain) & no & academic disclosure as part of \texttt{math/rand}
cluster \\
gVisor & CVE-2024-10026 (weak RNG class, netstack) &
\texttt{83f75082e5b0} (``netstack: cryptographically secure RNG'') +
chain & \bttt{krakauer@google.com} & \textbf{Inon Kaplan, Ron Even,
Amit Klein} (Hebrew University \& Bar Ilan) & no & \textbf{academic
researchers, named in fix commit} \\
\end{longtable}
}

The closest analogs to fuzzer-attribution in the walked sample are: one
external manual audit (CVE-2026-45782, Qubes OS), one
academic-researcher disclosure (CVE-2024-10026, Hebrew University \& Bar
Ilan), three internal-review attributions. \textbf{None are
fuzzer-found} by any direct evidence in the public commit log or the
linked GHSA / NVD entries. This is the 0/5 rate the methodology now
treats as the empirical anchor for demoting the percentage column from
primary to secondary read.

Two structural caveats explain why the 0/5 is not by itself an
indictment. \textbf{(a) Coordinated disclosure hides the discovery
channel in the commit-trailer view} --- the GHSA prose names the
reporter, but the fix commit typically does not (none of the Cloud
Hypervisor commits walked carried a \texttt{Reported-by:} trailer at
all). \textbf{(b) Silent-fix-first disclosure produces a structural
blind spot} --- gVisor's process (fix → ship → CVE) means fuzzer-found
bugs caught by syzkaller before a CVE is filed are invisible to a
per-CVE walk by construction. The bugs that \emph{do} receive a CVE are
a biased sample --- by definition, they are the ones the fuzzer either
didn't find or didn't find first. The 0/3 gVisor figure is compatible
with multiple underlying realities and the commit-trailer view alone
cannot disambiguate them (the cross-axis read in §3.2 carries this point
further).

\paragraph{2.6.4 Baseline-shifting findings this
rerun}\label{baseline-shifting-findings-this-rerun-1}

Two live findings from the 2026-05-21 rerun re-set the methodology's
prior baselines.

\textbf{Firecracker baseline rewritten --- the largest factual gap
across all six axes to date.} Methodology §1.6's expected baseline read
\texttt{cargo-fuzz\ +\ OSS-Fuzz\ /\ high\ /\ yes\ /\ \textasciitilde{}60\%}.
Empirical survey contradicts every part of the string: no \texttt{fuzz/}
directory, no \texttt{fuzz\_targets/}, no \texttt{cargo-fuzz}
configuration; no OSS-Fuzz integration (Firecracker is not listed in
\texttt{google/oss-fuzz} build configurations); no public dashboard; the
\texttt{\textasciitilde{}60\%} per-CVE attribution figure is unsupported
a priori because no fuzzer is present. The only fuzz-related artefact in
the repo is \texttt{-\/-features\ fuzzing}, documented in
\texttt{docs/fuzzing.md} (32 lines) as a determinism \emph{hook} for
external fuzzers --- the same docs file carries a banner warning that
the feature is \texttt{not\ for\ production\ use}, and the two
integration tests
(\texttt{tests/integration\_tests/build/test\_fuzzing.py}, 21 lines;
\texttt{tests/integration\_tests/functional/test\_fuzzing.py}) verify
the feature compiles in debug mode and refuses to compile in release
mode --- neither runs a fuzzer. AWS's internal QA process likely
includes fuzzers against Firecracker, but that is not in scope per the
methodology's upstream-public-only definition. The baseline in
\texttt{sandbox-isolation-methodology-v2.md} §1.6 has been rewritten to
reflect actual upstream state.

\textbf{Cloud Hypervisor --- CVE-2026-45782 lived in the \texttt{block}
target already in the harness list.} The fault path (virtio-block
async-I/O UAF in the default io\_uring backend) was reachable from
existing harness coverage --- \texttt{block} is one of the 18 targets in
\texttt{fuzz/Cargo.toml}. But compile-gating alone (CI runs
\texttt{cargo\ fuzz\ build} + \texttt{cargo\ fuzz\ check}, not
\texttt{cargo\ fuzz\ run}) does not exercise the corpus and did not
surface the UAF before DemiMarie's manual audit found it. This is the
strongest empirical case in the 1.6 walk for the §2.6.2 takeaway:
in-tree harness + CI build-gate raises the floor for harness
\emph{bit-rot} but does not advance \emph{corpus coverage}. The natural
next step --- a scheduled \texttt{cargo\ fuzz\ run} job against the 18
targets with corpus accumulation --- is filed in the fixture's
open-items list as a recommended upstream issue.

\paragraph{2.6.5 Class hierarchy does not predict fuzzing
posture}\label{class-hierarchy-does-not-predict-fuzzing-posture}

The cross-cutting structural finding: the engine-class hierarchy from
1.4 (microVM \rkbest{}, userspace kernel \rkmid{}, container \rkworst{}) \textbf{does not}
predict fuzzing investment.

The userspace-kernel engine has the strongest upstream fuzzing posture
(gVisor's multi-layered investment --- continuous syzkaller + in-tree
\texttt{secfuzz} + release-process smoke test --- is the most thorough
in the set). Under 1.4's class ordering, microVMs sit above gVisor for
class-level isolation strength; under 1.6, gVisor sits above all three
microVMs on fuzzing posture. The microVMs split internally: Cloud
Hypervisor invests in an 18-target in-tree harness (Tier 2); Firecracker
has no upstream fuzzer (Tier 3); libkrun has nothing (Tier 3). Three
engines, same isolation class, three different fuzzing tiers.

The strongest combination --- class-level isolation strength ×
continuous fuzzing posture --- would be a microVM with a continuous
public fuzzer. \textbf{No engine in the set occupies that combination.}
The closest is Cloud Hypervisor (microVM, in-tree harness, but no
continuous execution). The actual operator-facing strongest combination
on this axis is gVisor (userspace kernel, continuous fuzzer,
multi-layered) --- but its class-level isolation strength sits below the
microVMs. An operator weighting ``shallowest residual bugs'' against
``strongest isolation class'' faces a real tradeoff: there is no Tier-1
engine that wins both criteria.

\paragraph{2.6.6 Structural caveats}\label{structural-caveats-2}

\textbf{Per-CVE fuzzer attribution is best-effort, not authoritative.}
The walk reads public commit-trailer evidence in the local repo. Two of
five engine repos (\texttt{runc}, \texttt{firecracker},
\texttt{libkrun}) are shallow clones with only the HEAD commit visible,
so per-CVE walks are not possible for those engines. For the two with
full history (\texttt{cloud-hypervisor}, \texttt{gvisor}), fix commits
had no fuzzer trailers --- but absence of a trailer does not prove
absence of a fuzzer in the discovery chain. Authoritative attribution
requires reading each linked GHSA/NVD advisory, the linked patch PR, and
the linked vendor security writeup; that depth is out of scope.

\textbf{Internal vendor fuzzer effort is not visible from upstream
public repos.} AWS likely runs internal fuzzers against Firecracker;
Google certainly runs internal effort against gVisor beyond the public
syzkaller dashboard; Meta likely runs internal effort against Cloud
Hypervisor. None of this is reachable from the upstream public repos.
The findings are a lower bound on each engine's fuzzing effort, not an
estimate.

\textbf{Coverage \% is not comparable across fuzzers.} syzkaller's
coverage metric (basic-block coverage at the syscall surface) is a
different number from cargo-fuzz's \texttt{corpus\ coverage}
(libfuzzer-style edge coverage); neither is comparable to OSS-Fuzz's
reported coverage percentages for a Go codebase. Where dashboards expose
a number, treat it as a within-project trend, not a cross-project
comparison.

\textbf{The absence of CVEs + absence of fuzzing combination is the
riskiest posture for libkrun.} libkrun has 0 published CVEs (1.4
negative finding) and no upstream fuzzing investment. Neither channel is
providing data --- residual-bug depth is structurally unmeasured for
microsandbox operators (no upstream CVE history, no upstream fuzzer, no
published academic study to cross-check). The risk profile is genuinely
``unmeasured'' rather than ``low'' or ``high''; the methodology should
not infer either direction from the absence of both signals. Where the
analysis compares engine postures on residual-bug confidence, libkrun
rows are reported as \texttt{n/a\ —\ no\ signal} rather than aggregated
with the others.

\textbf{External/independent academic validation is unavailable for the
modern Rust VMMs.} The published hypervisor-fuzzing literature is robust
(Nyx 2021, HYPER-CUBE 2020, V-Shuttle 2021, HYPERPILL 2024) but targets
QEMU/KVM; none of the surveyed venues (USENIX Security, NDSS, CCS,
arXiv) contain a study specifically of Cloud Hypervisor, Firecracker,
runc, or gVisor's fuzzing posture. Academic literature lags upstream
availability by years; an engine adopting cargo-fuzz today will not show
up in published USENIX/NDSS work until at least 2027.

What 1.6 does not measure: internal vendor fuzzer effort (AWS, Google,
and Meta likely all run internal corpora; none of it is reachable from
upstream public repos --- the findings are a lower bound, not an
estimate); forward-looking bug-class prediction (1.6 is a snapshot of
posture, not a prediction of where the next bug lands); coverage \%
comparability across fuzzer types; operator-side fuzzing of deployed
sandboxes (out of scope for the upstream-only read here ---
operator-side fuzzing is a Tier-2/Tier-3 concern).

\subsection{3. Cross-axis reads}\label{cross-axis-reads}

Five interactions across the six axes that no single axis paper can
express.

\subsubsection{3.1 Small surface (1.1) is necessary but not sufficient
for fewer escape CVEs
(1.4)}\label{small-surface-1.1-is-necessary-but-not-sufficient-for-fewer-escape-cves-1.4}

The correlation between attack surface size and Escape-class CVE count
holds at the \emph{extremes} of the engine set but breaks in the middle.

At the bottom of the ladder, runc's unfiltered host-kernel surface (1.1:
mode-0, all 14 primitives reachable except 3 blocked by
\texttt{devices.deny}) co-occurs with the highest Escape-class CVE count
(1.4: 4 in window --- the 2025-11-05 trio CVE-2025-52565 /
CVE-2025-52881 / CVE-2025-31133 recurs the procfs mount-race pattern of
historical CVE-2019-19921 / CVE-2023-27561 / CVE-2023-28642, plus
CVE-2024-45310 as `Escape (limited)'). At the top, gVisor's 5/14
primitive-reachability score (1.1: tightest of the four mediated
engines) co-occurs with 0 Escape-class CVEs in window (1.4: 2 HostLeak +
1 InternalEsc, no host-boundary breach).

The middle is where the correlation breaks. Firecracker has the
\emph{tightest} seccomp ceiling in the set (55 syscalls allowlisted,
mode-2 across all threads --- 1.1 \rkbest{}) but \emph{both} of its in-window
CVEs are Escape-class (1.4: CVE-2026-5747 OOB write in virtio-pci,
CVE-2026-1386 jailer symlink host-write). The bugs were in the kernel
paths Firecracker's own VMM exercises, not in the wider kernel surface
that the seccomp ceiling forecloses. The seccomp filter does its job;
the bug was inside the part of the kernel the filter still allows.

Read this as: \textbf{a small attack surface bounds \emph{which} kernel
paths the in-sandbox actor can exercise, not \emph{how many bugs} live
in those paths.} An operator deciding from 1.1 alone would predict
Firecracker as the safest microVM and is contradicted by 1.4. The same
logic explains why 1.4 cannot, on its own, be read as a forward-looking
quality signal --- runc's CVE count includes the procfs mount-race
recurrences that 1.1 cannot exclude architecturally, but the kernel
paths Firecracker exposes (virtio-pci ring handling, jailer mount setup)
are not enumerated by any 1.1 sub-probe.

The cross-axis claim: \textbf{1.1 × 1.4 together are needed to read
either}, and neither suffices alone.

\subsubsection{3.2 Disclosure model (1.5) hides fuzzer attribution (1.6)
by
construction}\label{disclosure-model-1.5-hides-fuzzer-attribution-1.6-by-construction}

gVisor's downstream disclosure pattern is silent-fix-first: the fix
commit lands and ships in a tagged release months before the CVE is
assigned (1.5: --165d to --458d across post-2023 CVEs). The 1.6 per-CVE
attribution walk reads the fix commit's trailers looking for
\texttt{Reported-by:\ syzbot} or analogous fuzzer credit. \textbf{By
construction, that walk is biased.}

The CVEs the 1.6 walk reads are the bugs that received a CVE --- by
definition, the ones the fuzzer either did not find or did not find
\emph{first}. The fuzzer-found bugs that syzkaller caught before a CVE
was filed (silently fixed by Google, never CVE-assigned) are not in the
1.6 sample at all. The 0/3 gVisor attribution result is compatible with
both ``fuzzer caught many bugs but credit didn't reach the in-window CVE
set'' and ``fuzzer didn't catch these specifically'' --- and the
commit-trailer view cannot disambiguate them.

The 1.6 paper documents this caveat in its §1.3 (``Why a binary signal,
not a percentile'') and §5.6. The synthesis surfaces it again because it
explains a finding that would otherwise read as a contradiction: gVisor
is \rkbest{} on 1.6's binary signal (continuous syzkaller dashboard, in-tree
\texttt{secfuzz}, release-process smoke test) but 0/3 on the per-CVE
attribution. Both numbers are consistent under the silent-fix-first
model.

The cross-axis claim: \textbf{the disclosure model classification in 1.5
is a prerequisite for reading 1.6's per-CVE attribution.} Stripped of
the disclosure-model context, the 0/3 gVisor result could be misread as
evidence that gVisor's fuzzer doesn't work.

\subsubsection{3.3 Stackability (1.3) and patch cadence (1.5) are
independent operator
levers}\label{stackability-1.3-and-patch-cadence-1.5-are-independent-operator-levers}

Two of the six axes measure things an operator can change
post-deployment without forking the product: 1.3 (which layers the
operator can stack on top) and 1.5 (the pin policy the operator
inherits, with the option to override on the self-hosted path). The two
are independent:

\begin{figure}
\centering
\includegraphics[width=1\linewidth,height=\textheight,keepaspectratio,alt={Stackability (1.3) vs.~patch propagation (1.5): independent operator levers}]{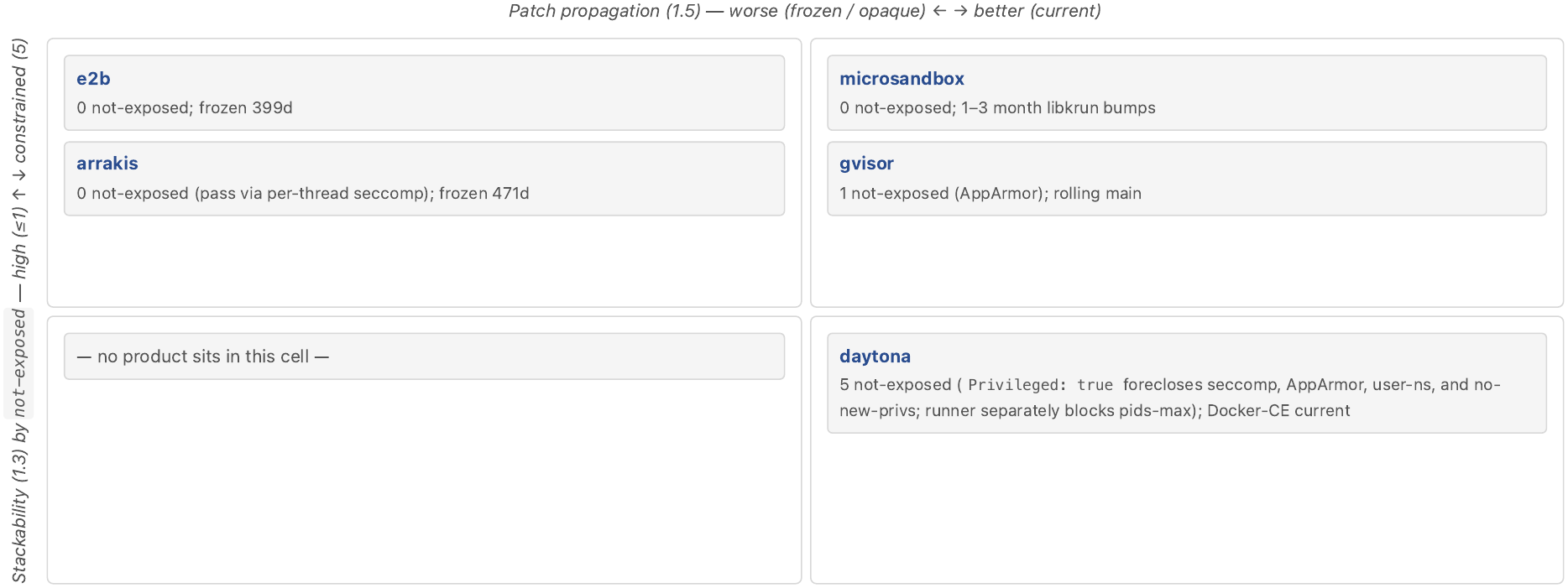}
\caption{Stackability (1.3) vs.~patch propagation (1.5): independent
operator levers}
\end{figure}

daytona ranks at the top of 1.5 (\texttt{current} on the default Docker
Compose deploy, runc 1.3.5 via Docker-CE 29.x includes all in-window
fixes) and at the bottom of 1.3 (5 \texttt{not-exposed} cells,
concentrated downstream of the \texttt{Privileged:\ true} choice in the
runner that disables AppArmor and re-adds capabilities). The patches
arrive; the hardening cannot be stacked on top. Conversely, arrakis
ranks at the bottom of 1.5 (\texttt{frozen} for 471+ days at Cloud
Hypervisor \texttt{v44.0}) but holds a \texttt{pass} on the 1.3 seccomp
cell (the per-thread BPF filter Cloud Hypervisor applies to 32/33 worker
threads). The hardening is in place; the patches are not arriving.

The operator's lever set is not unified. Choosing daytona means
accepting that the in-product \texttt{Privileged:\ true} blocks five
Linux hardening layers an operator might otherwise want, even if the
patches arrive on time. Choosing arrakis means accepting that the engine
ships with sound per-thread seccomp at defaults, but the operator
inherits a pin policy that has missed two critical/high CVE fixes inside
the in-window stretch.

The cross-axis claim: \textbf{1.3 and 1.5 measure separate
post-deployment operator levers, and neither substitutes for the other.}
A product that scores well on one and badly on the other is not
``average'' --- it is good on one operational concern and bad on
another, and the operator must read both.

A note on the per-vendor variance reading. Heng et
al.~(arXiv:2411.07480) report fix-first as the dominant disclosure
pattern across 312 Java-ecosystem CVEs at 85.89\%, framing it as a
vendor / project behaviour. The five products in this set sit across two
distinct disclosure ecosystems --- Rust-VMM coordinated for Firecracker
/ Cloud Hypervisor / runc; Google silent-fix-first for gVisor --- and
the Java-ecosystem proportion does not transport. The cross-vendor
variance in our set is between \emph{ecosystems} (Rust-VMM vs Google
CNA), not within a single vendor's track record over time. The
implication for operators: pin-policy is the dominant lever on the
operator side, but disclosure-model selection is the dominant lever on
the \emph{engine} side, and an engine in the Google silent-fix-first
model carries a structurally different downstream lag profile than an
engine in the coordinated model.

\subsubsection{3.4 Per-thread seccomp is a methodology improvement that
affects both 1.1 and
1.3}\label{per-thread-seccomp-is-a-methodology-improvement-that-affects-both-1.1-and-1.3}

The 1.3 paper's per-thread seccomp matrix lifted arrakis's seccomp cell
from \texttt{stack-redeploy} to \texttt{pass}. The same finding appears
in 1.1's primitive-reachability probe: the per-thread matrix shows that
32 of arrakis's 33 worker threads carry a mode-2 BPF filter; only the
leader thread is mode-0. Under a leader-only read, the arrakis seccomp
posture would be ``mode 0 / no filter applied''; under the per-thread
read, it is ``mode 2 on 32 threads.''

This matters because the leader thread is not the path the
kernel-surface attacks travel through. The Cloud Hypervisor architecture
has worker threads doing the virtqueue and device-emulation work; the
leader handles control. Under the per-thread matrix, the
operationally-relevant threads are filtered.

The two axis papers (1.1 and 1.3) document the same observation from
different angles. The 1.1 paper presents it as ``the
primitive-reachability count would be 14/14 under a leader-only read and
is 12/14 under the per-thread read''; the 1.3 paper presents it as ``the
seccomp cell would be \texttt{stack-redeploy} under a leader-only read
and is \texttt{pass} under the per-thread read.'' Both are correct; the
synthesis names the observation once.

The cross-axis claim: \textbf{the per-thread matrix is a methodology
improvement, not a per-axis improvement.} Future axes that read
per-thread state (2.1's default config audit, in particular) should
adopt the same reading.

\subsubsection{3.5 The libkrun cell is structurally unmeasured (1.4 ×
1.6)}\label{the-libkrun-cell-is-structurally-unmeasured-1.4-1.6}

Five products, five engines, six axes. Libkrun is the only cell where
two axes return absence-of-signal and the methodology cannot infer
either direction from the combination:

\begin{figure}
\centering
\includegraphics[width=1\linewidth,height=\textheight,keepaspectratio,alt={libkrun is structurally unmeasured (1.4 x 1.6, with 1.5 as derived consequence)}]{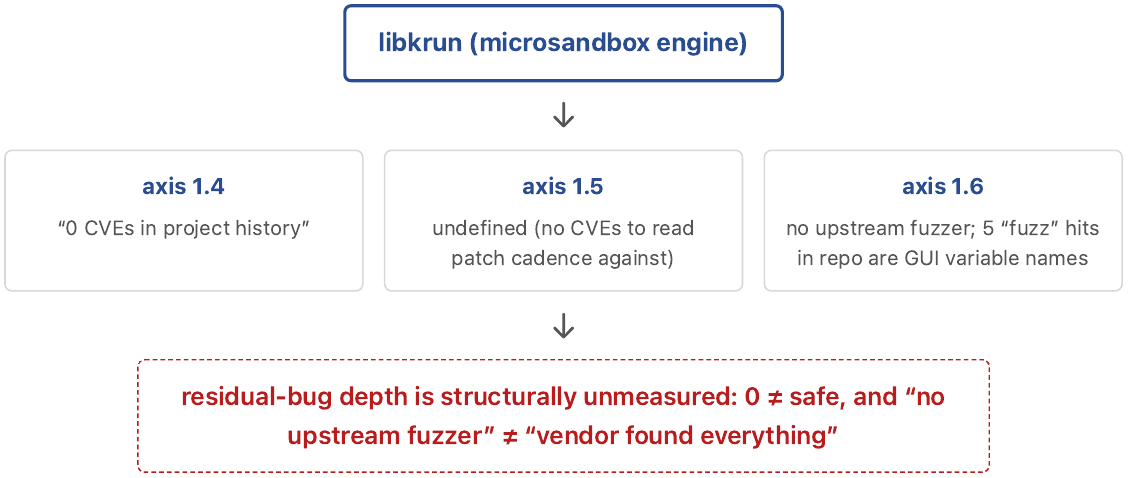}
\caption{libkrun is structurally unmeasured (1.4 x 1.6, with 1.5 as
derived consequence)}
\end{figure}

The 1.1 and 1.2 axes return their own numbers (microsandbox/libkrun:
11/14 primitives reachable, 0 leaks) --- but those are architectural
posture, not bug-rate signal.

The 1.6 paper documents the consequence in §5.6: ``for operators running
microsandbox (which sits on libkrun), residual-bug depth is structurally
unmeasured. There is no upstream CVE history to read, no upstream fuzzer
to consult, and no published academic study to cross-check.'' The
methodology does not call this \texttt{inconclusive} (the verdict ladder
doesn't have a slot for two-axis absence-of-signal); the cell is
reported as the intersection of two negative findings.

This is the \emph{riskiest posture in the set} on the residual-bug axis
--- not because libkrun is known to be buggy, but because no channel is
providing data. An operator reading libkrun's 0 CVEs as ``safest
engine'' is inverting the signal: zero is the absence of finding, not
the presence of soundness. An operator reading the no-fuzzer finding as
``vendor didn't invest'' is also missing the point: the vendor may or
may not have invested internally, but neither investment nor absence is
publicly observable.

The cross-axis claim: \textbf{libkrun and microsandbox should be read
with a wider confidence interval than the other four engine-product
pairs.} Until either a CVE channel produces signal, or upstream fuzzing
lands, or an external audit publishes, the residual-bug confidence on
this product is bounded by neither axis individually but by the joint
absence of both.

\textbf{Why ``measure real engines'' is the load-bearing methodological
choice.} Two earlier precedents anchor the contrast. ToolEmu (Ruan et
al., ICLR 2024 / arXiv:2309.15817) demonstrated that AI-agent risk
evaluation could run against an LM-emulated sandbox where a GPT-4-based
emulator fabricates tool outputs --- cheap, no engine required, but the
sandbox layer is fictional. SANDBOXESCAPEBENCH (Marchand et al.,
arXiv:2603.02277) measures agent capability to exploit a sandbox given
that a vulnerability exists, exercising real OCI containers but seeding
the vulnerability rather than measuring residual-bug rates. Our axes 1.4
/ 1.5 / 1.6 sit in a third position --- measuring \emph{engine-side}
residual-bug signal directly without seeding, against real upstreams ---
and §3.5's libkrun finding is the case where that approach returns
\emph{the absence of signal} as a result and the methodology must read
it as such rather than as soundness. The fuzzing-tier advances
catalogued in §2.6 (Psyzkaller, SyzParam, KernelGPT --- arXiv:2510.08918
/ 2501.10002 / 2401.00563) are the channel through which today's
``unmeasured'' cells become tomorrow's \emph{measured} cells; the
libkrun row stays \texttt{n/a\ —\ no\ signal} until one of those
channels lands on a libkrun harness.

\subsection{4. Threat-model qualification
matrix}\label{threat-model-qualification-matrix}

The methodology defines a single threat model (\texttt{T0.H2.N2} per
AISI taxonomy --- single-tenant operator running actively hostile code
on their own infrastructure). For the synthesis, that single threat is
split into four operator-facing sub-questions, each anchored to the two
or three axes that primarily inform it:

\begin{enumerate}
\def\labelenumi{\arabic{enumi}.}
\tightlist
\item
  \textbf{A. Escape resistance} --- given untrusted code in the sandbox,
  how hard is host kernel escape? (Primarily 1.1 + 1.4 + 1.6)
\item
  \textbf{B. Reconnaissance resistance} --- given untrusted code in the
  sandbox, what can it learn about the host passively? (Primarily 1.2)
\item
  \textbf{C. Hardening compatibility} --- can the operator layer
  additional Linux hardening on top? (Primarily 1.3)
\item
  \textbf{D. Patch propagation} --- does the operator receive fixes
  promptly once upstream patches? (Primarily 1.5)
\end{enumerate}

Three sources inform the form of the sub-questions and the urgency
framing. The dangerous-capability cohort cited in the Background section
--- Phuong et al.~(arXiv:2403.13793, threshold-crossing windows
2025--2029), Folkerts et al.~(arXiv:2603.11214, log-linear capability
scaling with inference compute through Opus 4.6, no observed plateau),
Wei et al.~(arXiv:2505.18384, \textgreater40\% relative improvement on
InterCode CTF in 8 GPU-hours / \textless\$36) --- establishes that A
(escape resistance) is the load-bearing column on a present-tense
timeline rather than a future-decade one. Wei et al.'s
stateful-vs-non-stateful distinction further refines A: stateful
environments (login lockouts, IP blocks) impose non-reversibility that
bounds adversarial iteration; the four products in this set are stateful
with respect to the operator's host (one successful escape persists) and
non-stateful with respect to a fresh sandbox spawn. An
alternative-architecture line --- Yan's transactional / fault-tolerant
sandboxing proposal (arXiv:2512.12806) --- sits outside the four columns
by construction: transactional rollback addresses \emph{post-action
revert} rather than \emph{prevention of host kernel escape}, and is
cited as a fifth-architecture candidate the matrix here does not score
against.

Each cell below is \texttt{qualifies} (Q),
\texttt{qualifies\ with\ caveats} (Q\emph{), or
\texttt{does\ not\ qualify} (\ding{55}) for the named sub-question against a
}reasonable* operator default. The thresholds are spelled out in the
table footnotes; \textbf{the matrix is not a composite score}, and no
overall row score is computed.

{\def\LTcaptype{none} % do not increment counter
\begin{longtable}[]{@{}
  >{\raggedright\arraybackslash}p{(\linewidth - 8\tabcolsep) * \real{0.2000}}
  >{\raggedright\arraybackslash}p{(\linewidth - 8\tabcolsep) * \real{0.2000}}
  >{\raggedright\arraybackslash}p{(\linewidth - 8\tabcolsep) * \real{0.2000}}
  >{\raggedright\arraybackslash}p{(\linewidth - 8\tabcolsep) * \real{0.2000}}
  >{\raggedright\arraybackslash}p{(\linewidth - 8\tabcolsep) * \real{0.2000}}@{}}
\toprule\noalign{}
\begin{minipage}[b]{\linewidth}\raggedright
Product
\end{minipage} & \begin{minipage}[b]{\linewidth}\raggedright
A. Escape resistance
\end{minipage} & \begin{minipage}[b]{\linewidth}\raggedright
B. Recon resistance
\end{minipage} & \begin{minipage}[b]{\linewidth}\raggedright
C. Hardening compatibility
\end{minipage} & \begin{minipage}[b]{\linewidth}\raggedright
D. Patch propagation
\end{minipage} \\
\midrule\noalign{}
\endhead
\bottomrule\noalign{}
\endlastfoot
\textbf{arrakis} & Q* --- per-thread seccomp filter on 32/33 worker
threads, but \rkworst{} on primitive reachability (12/14) anchored by live
\texttt{/dev/kvm} ioctl surface, plus 1 Escape-class + 1 HostLeak CVE in
window & Q* --- 1 leak (CPU brand via \texttt{cpu\ =\ host} default;
configurable) & Q --- 0 not-exposed cells; per-thread seccomp already
applied & \ding{55} --- \texttt{frozen} 471+ days, CVE-2026-45782 unpatched
≥7d \\
\textbf{e2b (self-hosted)} & Q* --- tightest seccomp ceiling, but no
upstream fuzzer and 2 fresh Escape-class CVEs in 2026 & Q --- 0 leaks &
Q --- 0 not-exposed cells; Firecracker seccomp + \texttt{no\_new\_privs}
applied at defaults & \ding{55} --- \texttt{frozen} 399 days, CVE-2026-5747
unpatched ≥44d \\
\textbf{e2b (cloud-hosted)} & Q* --- same engine as self-hosted; cloud
pin not observable & Q & Q & \ding{55} --- \texttt{opaque}, can't verify \\
\textbf{microsandbox} & Q* --- 11/14 primitives reachable, \emph{but} no
fuzzer + 0 CVEs = unmeasured residual depth (§3.5) & Q --- 0 leaks & Q
--- 0 not-exposed cells; libkrun ships no hardening of its own but every
layer is operator-reachable & Q --- 1--3 month bump cadence; libkrun CVE
exposure undefined \\
\textbf{gvisor} & Q --- 5/14 primitives reachable, continuous syzkaller,
0 Escape-class CVEs in window & Q* --- 2 leaks (RAM total, BIOS product)
& Q* --- 1 not-exposed (AppArmor --- runsc strips the OCI spec field);
of the remaining six, four already applied as Sentry's primary isolation
(seccomp, cap-drop, \texttt{no\_new\_privs}, \texttt{pids.max}), one
reachable via stack-redeploy (user-ns), one skipped on this AppArmor
host (SELinux) & Q --- rolling main; silent-fix-first means operator
receives the fix before the CVE exists \\
\textbf{daytona (default)} & \ding{55} --- full host kernel surface, 4
Escape-class in-window CVEs (3 procfs-mount-race recurrences + 1 `Escape
(limited)') & \ding{55} --- 10 leaks (full shared-kernel signature plus disk
identifiers) & \ding{55} --- 5 not-exposed cells: 4 (seccomp, AppArmor, user-ns,
\texttt{no\_new\_privs}) downstream of \texttt{Privileged:\ true}; the
fifth (\texttt{pids-max}) independently blocked by the runner not
populating \bttt{HostConfig.PidsLimit} & Q --- Docker-CE 29.x pulls
runc 1.3.5; all in-window CVEs included \\
\end{longtable}
}

Threshold definitions (per the methodology and the per-axis paper
rules):

\begin{itemize}
\tightlist
\item
  \textbf{A. qualifies} = no in-window Escape-class CVEs \emph{and}
  either continuous fuzzing or tight seccomp ceiling.
\item
  \textbf{A. qualifies with caveats} = either no Escape-class CVEs but
  other gaps, or Escape-class CVEs with mitigating evidence (small
  surface, fast patches).
\item
  \textbf{A. does not qualify} = ≥3 in-window Escape-class CVEs
  \emph{and} unfiltered host-kernel surface.
\item
  \textbf{B. qualifies} = 0 leaks on the 28-probe matrix. \textbf{B.
  qualifies with caveats} = 1--2 leaks, all configurable or
  implementation-gap. \textbf{B. does not qualify} = ≥10 leaks.
\item
  \textbf{C. qualifies} = 0 \texttt{not-exposed} cells (every applicable
  layer reachable via operator-side stacking). \textbf{C. qualifies with
  caveats} = 1 \texttt{not-exposed} cell. \textbf{C. does not qualify} =
  ≥3 \texttt{not-exposed} cells. C is scored on the operator's lever
  (what they can stack) rather than the engine's defaults --- the count
  of \texttt{applied} layers is not cross-class comparable per §2.3.
\item
  \textbf{D. qualifies} = \texttt{current} verdict per 1.5. \textbf{D.
  does not qualify} = \texttt{frozen},
  \texttt{won\textquotesingle{}t-fix}, or \texttt{opaque}.
\end{itemize}

Read the matrix in two ways:

\begin{enumerate}
\def\labelenumi{\arabic{enumi}.}
\tightlist
\item
  \textbf{Operator with a specific concern.} An operator who weights
  escape resistance most heavily (A column) finds gvisor as the only
  unambiguous \texttt{Q}; everyone else carries caveats and daytona is
  excluded. An operator who weights patch propagation (D column) finds
  three products \texttt{Q} (microsandbox, gvisor, daytona-default), and
  the two \texttt{frozen}-pin products excluded. The matrix is the
  qualification step \emph{before} any ranking, not a substitute for
  ranking.
\item
  \textbf{Operator without a single concern.} Reading across each row,
  no product is \texttt{Q} on all four columns. \textbf{gvisor} reaches
  Q on two columns (A escape resistance via tight surface + continuous
  fuzzing + 0 in-window Escape-class CVEs; D patch propagation via the
  silent-fix-first model) with Q* on the other two (B from 2 information
  leaks; C from the AppArmor \texttt{not-exposed} cell).
  \textbf{microsandbox} formally reads Q on three columns (B / C / D)
  and Q* on A, but the A caveat --- residual-bug depth structurally
  unmeasured per §3.5 --- is the load-bearing claim; reading the row as
  ``best overall'' inverts the §3.5 caveat into a recommendation.
  \textbf{arrakis} and \textbf{e2b} (self-hosted / cloud-hosted) sit at
  Q* / Q* / Q / \ding{55} and Q* / Q / Q / \ding{55} respectively --- strong on A
  through C but blocked by D under their pin policies.
  \textbf{daytona-default} is the inverse extreme, with \ding{55} / \ding{55} / \ding{55} / Q
  --- strong on the one column it qualifies on (patches arrive via
  Docker-CE), weak everywhere else.
\end{enumerate}

The matrix does not collapse into a single score because the columns are
not commensurable. Patch propagation and escape resistance are different
operational concerns with different remediation paths, and an operator
who weights them equally is reasoning about a different threat model
than an operator who treats one as the load-bearing axis. The
methodology's ``no composite score'' rule lives here for a reason.

\subsection{5. Per-product portraits}\label{per-product-portraits}

Five short paragraphs reading the six axes against each product in turn.
Each paragraph names the operationally-relevant claim and cites the axis
evidence that supports it.

\subsubsection{5.1 arrakis (Cloud
Hypervisor)}\label{arrakis-cloud-hypervisor}

arrakis is the product where the engine investment (Cloud Hypervisor's
18-target in-tree fuzz harness, per-thread seccomp on worker threads,
small attack surface) is partially undone by product-level defaults. The
per-thread BPF filter is real and is the reason the 1.3 seccomp cell is
\texttt{pass}; the in-tree harness is the largest of any microVM in the
set. But the live \texttt{/dev/kvm} ioctl surface (1.1 §4.7 reachability
+ §5.1a ABI confirmation) means guest userland could call
\texttt{KVM\_CREATE\_VM} and would run an L2 guest with no privilege
escalation --- a meaningfully larger kernel-LPE surface than the other
microVMs expose. The pin policy compounds the picture: arrakis has held
Cloud Hypervisor \texttt{v44.0} for 471+ days across 12 upstream
releases, and that pin missed CVE-2026-45782 (virtio-block UAF, 8.9 v4,
escape-class --- patched in \texttt{v52.0} / \texttt{v51.2} on
2026-05-14, ≥7d unpatched in arrakis) and CVE-2026-27211 (QCOW HostLeak,
≥90d unpatched). The engine has been doing its job; the product has not
been propagating the fixes. Operator implication: arrakis is the
candidate that improves the most under a pin-policy change, and the
candidate where the 2.1 default-config audit (product-level) is most
likely to reveal whether nested-virt is intentional or a documentation
gap.

\subsubsection{5.2 e2b (Firecracker)}\label{e2b-firecracker}

e2b inherits the tightest seccomp ceiling in the set (55 syscalls
allowlisted under Firecracker's mode-2 filter), and the 1.1
primitive-reachability score (7/14) is the second-best of the five
products. The 1.3 stackability cell shows 2 layers applied at defaults
with 0 not-exposed cells --- the cleanest operator-side surface in the
microVM class. Two findings cut against this. (a) 2026 produced
Firecracker's first two Escape-class CVEs (CVE-2026-5747 OOB write in
virtio-pci, 8.7 v4; CVE-2026-1386 jailer symlink host-write, 6.0 v4);
the methodology's prior baseline of ``no published hypervisor-escape''
was broken twice in four months. (b) 1.6 rewrote the Firecracker fuzzing
baseline to ``no upstream fuzzer'' --- the prior baseline of
``\texttt{cargo-fuzz\ +\ OSS-Fuzz\ /\ \textasciitilde{}60\%}'' was
unsupported by any repo artefact; \texttt{-\/-features\ fuzzing} is a
deterministic-build hook, not a fuzzer. The self-hosted product has held
Firecracker \texttt{v1.14.1\_458ca91} for 399 days, leaving
CVE-2026-5747 unpatched at the orchestrator-default flag for ≥44d. The
cloud-hosted product is \texttt{opaque} on the pin question. Operator
implication: e2b's engine-level posture is strong, but the absence of an
upstream fuzzer means residual-bug confidence relies on AWS's internal
effort (not publicly observable), and the pin policy reproduces the
arrakis pattern at smaller magnitude.

\subsubsection{5.3 microsandbox (libkrun)}\label{microsandbox-libkrun}

microsandbox is the product where the most signals are absent. libkrun
has 0 published CVEs (1.4 negative finding --- see §3.5 for the
interpretation), no upstream fuzzer (1.6 --- five ``fuzz'' hits in the
repo are GUI variable names), and no published academic study (1.6 §3
web-search returned nothing). On the axes where signal \emph{is}
present, microsandbox's posture is mixed: 1.1 returns 11/14 primitives
reachable (mode-0 across all 16 threads, libkrun ships no engine-side
seccomp), 1.2 returns 0 leaks (the cleanest microVM result, tied with
e2b), and 1.3 returns 0 layers applied at defaults (microsandbox is the
only product where every Linux hardening layer requires operator-side
stacking). 1.5 returns \texttt{current} (1--3 month bump cadence), but
the CVE exposure is undefined. Operator implication: microsandbox
carries the riskiest posture in the residual-bug sense (§3.5) and the
cleanest data-leakage posture; the operator who picks microsandbox is
implicitly trusting that the absence of CVE and fuzzing signal reflects
vendor diligence rather than absence of search, and is stacking
\emph{all} the hardening layers themselves.

\subsubsection{5.4 gvisor (runsc)}\label{gvisor-runsc}

gvisor is the product where engine investment, product defaults, and
operator-side composability all align in the same direction --- and is
the closest the set comes to ``qualifies on every column'' on the
threat-model matrix. The 1.1 primitive-reachability score (5/14) is the
tightest in the set; 1.3 ships 4 layers applied at defaults (seccomp
mode-2, cap-drop, \texttt{no\_new\_privs}, \texttt{pids.max}) with 1
not-exposed (the AppArmor cell --- runsc strips the OCI spec AppArmor
field, which is a real cost on hosts that depend on AppArmor profiles);
1.4 returns 0 Escape-class CVEs in the rollup window across 3 in-window
CVEs (2 HostLeak + 1 InternalEsc); 1.6 is the only entry in the engine
set with a continuous public dashboard
(\texttt{syzkaller.appspot.com/gvisor}), in-tree \texttt{secfuzz}
library, and release-process syzkaller smoke test. The 1.2 cell carries
2 leaks (host RAM total via \texttt{/proc/meminfo}, BIOS product name
via \texttt{/sys/class/dmi/*}) which are implementation gaps in the
Sentry's kernel emulation rather than architectural concessions. 1.5
ranks gvisor as \texttt{current} but with the correct interpretation:
the silent-fix-first disclosure model means the operator receives the
fix months before the CVE exists (--165d to --458d disclosure-to-release
lags), which is operationally a strict improvement over coordinated
disclosure even though it reads as a ``negative number'' on the upstream
cadence axis. The cross-axis caveat (§3.2): silent-fix-first hides
fuzzer attribution in commit trailers; gvisor's 0/3 per-CVE fuzzer
attribution is \emph{compatible with} the syzkaller dashboard catching
everything else, not evidence that the dashboard fails. Operator
implication: gvisor pays the surface-tightness cost in workload
compatibility (the Sentry's kernel API surface is real but bounded, and
applications using io\_uring or userfaultfd will hit \texttt{ENOSYS});
the security posture is the strongest in the set on five of six axes.

\subsubsection{5.5 daytona (runc)}\label{daytona-runc}

daytona is the only product in the set where the host kernel is exposed
verbatim to the in-sandbox actor. 1.1 returns 11/14 primitives reachable
with mode-0 across all 15 threads and no engine-side seccomp filter; 1.2
returns 10 leaks (full shared-kernel signature, four disk-hardware
identifiers); 1.3 returns 1 layer applied (runc's empty-cap default on
the container init PID) and 5 not-exposed cells (four --- seccomp,
AppArmor, user-ns, \texttt{no\_new\_privs} --- downstream of the runner
hardcoding \texttt{Privileged:\ true}, which disables Docker's default
seccomp profile, disables AppArmor, makes user-ns incompatible, and pins
\texttt{no-new-privileges=false}; the fifth --- \texttt{pids-max} ---
independently blocked by the runner not populating
\texttt{HostConfig.PidsLimit}); 1.4 returns 4 Escape-class in-window
CVEs --- the 2025-11-05 trio (CVE-2025-52565, CVE-2025-52881,
CVE-2025-31133) recurs the procfs mount-race pattern, plus
CVE-2024-45310 as `Escape (limited)'. The single axis where daytona is
\texttt{current} is 1.5: the default install path is
\texttt{docker\ compose\ -f\ docker/docker-compose.yaml\ up\ -d}, which
pulls Docker-CE 29.x with runc 1.3.5 bundled via \texttt{containerd.io}
--- every in-window runc CVE fix is included. The
\texttt{won\textquotesingle{}t-fix} finding (Ubuntu's ``Ignored ---
backport too intrusive'' posture on the 2025-11 runc trio) survives only
for operators who deviate from daytona's docs and
\texttt{apt\ install\ runc} from the Ubuntu archive instead. 1.6 returns
OSS-Fuzz with 2 targets covering a narrow JSON/user-parsing surface ---
fuzzing investment exists but is narrow. Operator implication: daytona
is the product where engine class (container, not microVM) is the
dominant security signal, and product defaults amplify it; the
\texttt{Privileged:\ true} choice forecloses five of seven hardening
layers an operator might otherwise want. The single redeeming axis is
patch propagation (Docker-CE bundles runc and bumps it on its own
release cadence); the rest of the threat-model matrix is the inverse
extreme of gvisor.

Two external precedents corroborate the framing. The historical seccomp
baseline that the \texttt{Privileged:\ true} choice forecloses comes
from Wan et al.~(arXiv:1712.05493, ``Mining Sandboxes for Linux
Containers''): \emph{``By default, Docker disallows 44 system calls out
of 300+''} --- quoted as historical 2017 baseline; cross-checked
2026-05-30 against moby/moby
\texttt{vendor/github.com/moby/profiles/seccomp/default.json} at tag
\texttt{docker-v29.5.2} (the Docker-CE 29.5.2 release this row pins via
\texttt{containerd.io}), which carries 361 syscalls unconditionally
allowed and 426 unique syscall names across all ALLOW blocks once
capability / kernel-version / architecture gating is unioned in ---
above D.2's ``300+'' floor, with the allow-list shape
(\texttt{defaultAction:\ SCMP\_ACT\_ERRNO}) unchanged. Separately,
Bühler et al.'s AgentBound (FSE 2026 / arXiv:2510.21236) builds MCP-side
access control on Docker + iptables + env whitelists and explicitly
names the engine layer (Docker, Linux kernel features) as their trusted
computing base in their §5 Discussion --- the runc cluster that this row
measures is the TCB their policy enforcement engine inherits.

\subsection{6. Open questions and
carry-forwards}\label{open-questions-and-carry-forwards}

Five items carry forward from the engine-level reading: two are concrete
findings flagged for follow-up work, two are entire bodies of work whose
absence currently bounds the conclusions a defensible overall ranking
could draw, and one is the set of adjacent architectures and protocol
layers that this paper's scope excludes.

\subsubsection{6.1 arrakis nested-KVM ioctl surface (product-level lead
from
1.1)}\label{arrakis-nested-kvm-ioctl-surface-product-level-lead-from-1.1}

The 1.1 primitive-reachability probe only opens \texttt{/dev/kvm} and
does not exercise any ioctl. A follow-up one-off probe called three
read-only, no-allocation KVM ioctls from inside a fresh arrakis guest
and confirmed: \texttt{KVM\_GET\_API\_VERSION} returns 12 (current
upstream KVM API), \texttt{KVM\_GET\_VCPU\_MMAP\_SIZE} returns 12288 (3
× 4 KiB pages, modern KVM ABI),
\texttt{KVM\_CHECK\_EXTENSION(KVM\_CAP\_NR\_VCPUS)} returns 6 (nested
KVM can host an L2 VM with up to 6 vCPUs). A guest userland process
could call \texttt{KVM\_CREATE\_VM} and would run an L2 guest with no
privilege escalation required on the host (KVM\_CREATE\_VM was
deliberately not exercised). The \texttt{0o660} device node is owned by
the kvm group; the guest's default user is on the host-side kvm group
(uid/gid 0 inside the guest). This expands the kernel-LPE surface of the
arrakis product compared to the rest of the shortlist (CVE-2023-3640
class and the broader KVM kernel-LPE family); the product-level
default-config audit (axis 2.1) should determine whether the nested-virt
exposure is intentional product behaviour or a documentation gap that
the Cloud Hypervisor default config does not surface.

\subsubsection{6.2 Cloud Hypervisor fuzz-execution proposal (operational
remediation from
1.6)}\label{cloud-hypervisor-fuzz-execution-proposal-operational-remediation-from-1.6}

1.6 names the Cloud Hypervisor compile-gate-only fuzz pipeline as the
largest gap between the engine's in-tree investment (18 cargo-fuzz
targets across virtio, x86 emulation, image parsers, and the HTTP API)
and the operational maturity of gVisor's continuous syzkaller. Three
remediation options ordered by effort: (A) scheduled GitHub Actions job
running \texttt{cargo\ fuzz\ run} nightly against each target with
cache-backed corpus persistence (low effort, demonstrates value, hits
GitHub-runner wall-clock limits at \textasciitilde20 min per target per
night); (B) ClusterFuzzLite integration for durable corpus accumulation
(medium effort, requires storage); (C) OSS-Fuzz application for
long-term continuous infrastructure (high effort, right answer).
CVE-2026-45782 (virtio-block async-I/O UAF, 8.9 v4) lived in the
\texttt{block} target already in the harness list --- compile-gating did
not surface it; corpus-execution might have. This is the highest-impact
engine-side action this body of work surfaced.

\subsubsection{6.3 Live probing of real product
deployments}\label{live-probing-of-real-product-deployments}

The methodology splits the six axes into a desk-research subset (1.4 CVE
history, 1.5 patch cadence, 1.6 fuzzing posture --- read repos,
advisories, dashboards) and a probing subset (1.1 host attack surface,
1.2 information leakage, parts of 1.3 stackability --- spawn a sandbox,
run probe scripts, capture output). The 1.3 axis sits awkwardly between
the two: the \texttt{applied} cell is observed from \texttt{/proc}
evidence on the live deployment, but the \texttt{stack-effort} cell is
\emph{inferred from product source} per the methodology's cell shape. A
probing pass would upgrade the inferred cells to measured cells by
applying each layer to the live deployment and recording whether the
engine still functions.

The 1.6 paper's structural precedent applies here: what the docs say
about the project ≠ what the deployment actually does. Compile-gating ≠
continuous execution at the engine level; inferred stack-effort ≠
measured stack-effort at the product level. A probing pass closes the
gap.

\subsubsection{6.4 Per-product configuration
audits}\label{per-product-configuration-audits}

Per the methodology, seven product-level configuration axes (2.1 default
config audit, 2.2 inside-the-sandbox enumeration, 2.3 image / template
supply chain, 2.4 network egress controls, 2.5 secrets handling, 2.6 SDK
/ API surface, 2.7 compositional attack scenarios). The engine-level
axes measure structural properties; the product-level axes measure what
the deployed product actually does. The arrakis nested-KVM lead (§6.1)
is an example of a product-level observation surfacing inside an
engine-level primitive-reachability sub-probe; every product-level axis
is expected to produce that kind of carry-forward.

\subsubsection{6.5 Adjacent architectures and protocol layers --- out of
scope
here}\label{adjacent-architectures-and-protocol-layers-out-of-scope-here}

Three lines of work share the ``AI-agent sandbox'' framing but sit
outside the five-engine × six-axis scope of this paper; flagged here so
an operator surveying the literature for adjacent designs can locate the
right starting point.

\textbf{Alternative engine architectures.} Moore \& Zenla (Edera,
arXiv:2501.04580) propose a type-1 Xen-paravirtualised-per-pod runtime
as an adjacent execution-environment architecture. The paper is a
\textbf{vendor-authored whitepaper} by the Edera team and is cited here
for landscape completeness with the vendor-framing caveat noted
explicitly --- the vendor-reported performance numbers comparing Edera
against Docker / gVisor / Kata / Firecracker are not quoted as
authoritative in this work. An engine-level measurement against Edera
under the methodology in this paper would be a natural extension once
the runtime ships a stable upstream public artefact.

\textbf{MCP-protocol-layer access control.} Bühler et al.'s AgentBound
(FSE 2026 / arXiv:2510.21236) implements \texttt{AgentBox} policy
enforcement at the MCP (Model Context Protocol) layer using Docker +
iptables + environment whitelists. AgentBound's scope is
\emph{protocol-layer access control over tool invocations}, sitting one
layer above the engine substrate this paper measures; their §5
Discussion explicitly names the engine layer as their trusted computing
base (cited again in §5.5). MCP-protocol containment is a complementary
defense layer, not a substitute for engine-level isolation.

\textbf{Alternative containment paradigms.} Yan's transactional /
fault-tolerant sandboxing proposal (arXiv:2512.12806) frames containment
as \emph{post-action transactional rollback} rather than
\emph{prevention of host kernel escape}. The paper tests only Gemini
CLI; no empirical comparison against Docker / gVisor / Firecracker is
published. Cited as an alternative-architecture candidate for the
operator who can tolerate the rollback semantics; the engine-level
escape prevention this paper measures and the transactional revert Yan
proposes address different operator concerns and would compose rather
than substitute.

\subsection{7. Caveats and limitations of the
synthesis}\label{caveats-and-limitations-of-the-synthesis}

\begin{enumerate}
\def\labelenumi{\arabic{enumi}.}
\tightlist
\item
  \textbf{All findings are inherited.} Every number in this paper traces
  back to a published number in one of the six source papers. Where the
  synthesis aggregates, the aggregation is explicit. Errors in the
  underlying numbers propagate; the synthesis adds no error-correction
  layer.
\item
  \textbf{The threat-model matrix splits a single threat into four
  sub-questions.} The methodology defines exactly one threat
  (\texttt{T0.H2.N2} per AISI taxonomy). The four sub-questions in §4
  are an analytical convenience for reading the matrix; they are not a
  re-definition of the threat model. An operator with a different threat
  model (multi-tenant SaaS, microarchitectural side channels,
  supply-chain) reads a different methodology.
\item
  \textbf{The \rkbest{}/\rkmid{}/\rkworst{} scheme is per-axis.} It is not portable across axes
  and is not summable. A product with three \rkbest{} marks is not ``above'' a
  product with three \rkmid{} marks unless the operator's threat model weights
  the three axes equally --- which the methodology forbids inferring.
\item
  \textbf{The synthesis aggregates evidence from the six axis papers;
  the cross-axis reads in §3 are observational.} The underlying axes mix
  probe-based measurements on live deployments (1.1 strace and
  primitive-reachability, 1.2 leak probes, 1.3 \texttt{/proc} reads,
  plus the §6.1 follow-up KVM ioctl probe) with desk research (1.4 CVE
  history, 1.5 patch cadence, 1.6 fuzzing posture). The cross-axis reads
  in §3 are observational (``here are two findings that align'' / ``here
  are two findings in tension''); they are not experimentally verified.
\item
  \textbf{Per-product portraits are time-bounded.} The synthesis reads
  the engine set against a fixed dataset. Pin policies change; upstream
  advisories land; product orchestrators bump or pin. The portraits will
  require re-reading as the underlying data shifts.
\item
  \textbf{No live exploits, no replay.} The methodology explicitly
  dropped CVE replay (see methodology §``What changed from v1'') and the
  synthesis inherits that scope. No claim here is verified by exploit
  chain; all claims are inferred from architectural and historical
  evidence.
\item
  \textbf{Multi-product comparisons assume the methodology's single
  threat model.} A multi-tenant operator reading these portraits will
  need different evidence; nothing in this paper bounds
  tenant-A-to-tenant-B exposure. The methodology's threat-model scope is
  explicit on this point.
\end{enumerate}

\subsection{8. Responsible disclosure}\label{responsible-disclosure}

Three findings in this paper carry a disclosure dimension. Each is
classified below per its disclosure pathway.

\textbf{arrakis nested-KVM ioctl reachability (§6.1).} A guest userland
process inside an arrakis sandbox can open \texttt{/dev/kvm} and
successfully invoke read-only KVM ioctls
(\texttt{KVM\_GET\_API\_VERSION}, \texttt{KVM\_GET\_VCPU\_MMAP\_SIZE},
\texttt{KVM\_CHECK\_EXTENSION(KVM\_CAP\_NR\_VCPUS)}); the guest's
default user is on the host-side kvm group, so \texttt{KVM\_CREATE\_VM}
would also succeed. The probe sequence is read-only and no exploit chain
is demonstrated. The arrakis maintainers will be notified with the §6.1
contents prior to public release of this paper, and this section will be
updated post-acknowledgement with the contact date, vendor response, and
any embargo terms. No embargo is currently in effect: the observation is
a structural reachability claim derived from read-only ioctl calls, not
a working exploit. Upstream Cloud Hypervisor maintainers were not
contacted separately --- the observation describes a product-level
configuration of the engine (kvm-group membership of the guest's default
user), not an engine-level vulnerability in Cloud Hypervisor.

\textbf{e2b frozen Firecracker pin under public Escape-class CVEs (§2.5,
§5.2).} The self-hosted e2b orchestrator pins Firecracker at
\texttt{v1.14.1\_458ca91} (≥399 days; last bumped 2025-04-17), leaving
CVE-2026-5747 (Firecracker virtio-pci OOB-write, 8.7 v4, Escape-class)
unpatched at the orchestrator-default flag for ≥44 days. \textbf{No
embargo required.} Both the pin manifest
(\href{https://github.com/e2b-dev/fc-versions}{\texttt{e2b-dev/fc-versions}})
and the CVE (Firecracker GHSA, NVD) are public artefacts; the paper
reports their conjunction, not new information about either.

\textbf{daytona runner hardcoded \texttt{Privileged:\ true} (§5.5).} The
daytona runner binary creates the container with
\texttt{HostConfig.Privileged:\ true}, which the runc engine class then
propagates into capability and security-policy defaults. \textbf{No
embargo required.} The configuration choice is documented in the daytona
runner's public source
(\href{https://github.com/daytonaio/daytona}{\texttt{daytonaio/daytona}}).

The methodology explicitly excludes ``live exploitation of unpublished
bugs'' from scope (§Threat model; methodology §1.0 ``out of scope''). No
exploit chain is demonstrated in this paper; the strongest claim --- the
arrakis observation above --- is a structural reachability statement
derived from three read-only ioctl calls, classified in §6.1 as a
``product-level lead from 1.1'' pending vendor disclosure and the 2.1
default-config audit (per the methodology's seven product-level
configuration axes).

\subsection{9. Conclusion}\label{conclusion}

The cross-axis reading shows that engine class is the dominant signal
where the axis measures an architectural property (1.1, 1.2, 1.4, 1.6)
but does not finish the job --- within each engine class, product-level
choices (arrakis's nested-KVM exposure, daytona's
\texttt{Privileged:\ true}, microsandbox's 0 applied hardening layers,
e2b's frozen pin) split the products in operationally consequential
ways. The operator-facing variance on patch propagation (1.5) is
dominated by product-side pin policy, not by engine-side patch speed.
The strongest combination on the residual-bug axis (microVM × continuous
public fuzzer) is unoccupied in this set, which forces an explicit
tradeoff between ``shallowest residuals'' (gvisor) and ``strongest
isolation class'' (the microVMs) that no single axis can resolve.

No overall ranking is proposed. The threat-model matrix in §4 is a
qualification step, not a substitute for ranking --- and two further
bodies of work pend: (a) a \textbf{live-probing pass over the existing
six engine-level axes} to upgrade 1.3's inferred \texttt{stack-effort}
cells to measured cells (§6.3), and (b) \textbf{Tier 2 --- seven
product-level configuration axes} (2.1 default config audit, 2.2
inside-sandbox enumeration, 2.3 image / template supply chain, 2.4
network egress controls, 2.5 secrets handling, 2.6 SDK / API surface,
2.7 compositional attack scenarios) measuring what the deployed product
actually does rather than what the engine architecturally permits
(§6.4). Each of the per-product portraits in §5 carries the same
operator-actionable form: here is what the product does well, here is
what it does badly, here is the cross-axis read that explains the
combination. The deferred-ranking note is not caution but a description
of the state of the evidence: until live probing complements desk
research and until Tier 2 lands, a defensible overall ranking would
invert under the missing data.

Two leads carry forward concretely. The arrakis nested-KVM ioctl surface
(§6.1) is the highest-impact product-level carry-forward from 1.1 and
\textbf{becomes the first input to Tier 2's axis 2.1 (default-config
audit)} --- it determines whether the nested-virt exposure is
intentional product behaviour or a documentation gap. The Cloud
Hypervisor fuzz-execution proposal (§6.2) is the highest-impact
engine-side remediation this work surfaced; it targets upstream
\texttt{cloud-hypervisor/cloud-hypervisor} and \textbf{sits outside Tier
2's product-level scope}.

\subsection{10. Reproducibility}\label{reproducibility}

\textbf{Companion repository.} Harness, fixtures, axis-paper sources,
and rerun logs ship at
\url{https://github.com/orbitalab/RnD-ai-sandboxes-sec-study-part-1}
(code under Apache-2.0). The measurements in this synthesis were taken
against \texttt{main} HEAD
\texttt{c7c8484b2da243f773327554cd2f693d5ea6678c} (2026-05-24); reruns
should pin the same SHA to reproduce these numbers.

\textbf{Host configuration.} Single bare-metal Linux node: kernel
\texttt{6.8.0-100-generic}, Ubuntu 24.04 LTS, x86\_64. No virtualisation
underneath the host. SDK versions, container images, and engine pin
manifests are pinned per the methodology §``Pinning and
reproducibility''.

\textbf{Harness invocation.}

\begin{Shaded}
\begin{Highlighting}[]
\ExtensionTok{pnpm}\NormalTok{ install}
\FunctionTok{cp}\NormalTok{ .env.example .env  }\CommentTok{\# fill credentials for the products you intend to measure}
\ExtensionTok{pnpm}\NormalTok{ sec }\OperatorTok{\textless{}}\NormalTok{product}\OperatorTok{\textgreater{}} \OperatorTok{\textless{}}\NormalTok{axis}\OperatorTok{\textgreater{}}
\end{Highlighting}
\end{Shaded}

Products: \texttt{arrakis}, \texttt{e2b}, \texttt{microsandbox},
\texttt{gvisor}, \texttt{daytona}. The CLI lists available pairs via
\texttt{pnpm\ sec\ -\/-list}; results land in
\texttt{./results/\textless{}product\textgreater{}-\textless{}axis\textgreater{}-\textless{}check\textgreater{}-\textless{}timestamp\textgreater{}.json},
large artefacts (strace dumps, LinPEAS reports, pcaps) in
\texttt{./evidence/}. Two runs within an hour on the same host must
produce bitwise-identical verdicts per the methodology's
replay-determinism rule; flaky checks are marked \texttt{inconclusive}.

\textbf{Per-axis entry points.} Source paper, fixture report, and
reproduction target line up one-to-one. Axes 1.1--1.3 are probe /
measurement axes reproducible via the harness; 1.4--1.6 are
desk-research axes whose inputs are external public artefacts
consolidated in §References below.

{\def\LTcaptype{none} % do not increment counter
\begin{longtable}[]{@{}
  >{\raggedright\arraybackslash}p{(\linewidth - 6\tabcolsep) * \real{0.2500}}
  >{\raggedright\arraybackslash}p{(\linewidth - 6\tabcolsep) * \real{0.2500}}
  >{\raggedright\arraybackslash}p{(\linewidth - 6\tabcolsep) * \real{0.2500}}
  >{\raggedright\arraybackslash}p{(\linewidth - 6\tabcolsep) * \real{0.2500}}@{}}
\toprule\noalign{}
\begin{minipage}[b]{\linewidth}\raggedright
§
\end{minipage} & \begin{minipage}[b]{\linewidth}\raggedright
Source paper
\end{minipage} & \begin{minipage}[b]{\linewidth}\raggedright
Fixture report
\end{minipage} & \begin{minipage}[b]{\linewidth}\raggedright
Reproduce via
\end{minipage} \\
\midrule\noalign{}
\endhead
\bottomrule\noalign{}
\endlastfoot
1.1 &
\href{https://github.com/orbitalab/RnD-ai-sandboxes-sec-study-part-1/blob/main/.docs/papers/1.1-host-attack-surface.md}{\bttt{1.1-host-attack-surface.md}}
& \bttt{./fixtures/host-surface-step-1.1-report.md} &
\bttt{pnpm\ sec\ \textless{}product\textgreater{}\ tier1-host-surface} \\
1.2 &
\href{https://github.com/orbitalab/RnD-ai-sandboxes-sec-study-part-1/blob/main/.docs/papers/1.2-information-leakage.md}{\bttt{1.2-information-leakage.md}}
& \bttt{./fixtures/info-leak-step-1.2-report.md} &
\bttt{pnpm\ sec\ \textless{}product\textgreater{}\ tier1-info-leak} \\
1.3 &
\href{https://github.com/orbitalab/RnD-ai-sandboxes-sec-study-part-1/blob/main/.docs/papers/1.3-defense-in-depth-stackability.md}{\bttt{1.3-defense-in-depth-stackability.md}}
& \bttt{./fixtures/stackability-step-1.3-report.md} &
\bttt{pnpm\ sec\ \textless{}product\textgreater{}\ tier1-stackability} \\
1.4 &
\href{https://github.com/orbitalab/RnD-ai-sandboxes-sec-study-part-1/blob/main/.docs/papers/1.4-public-cve-history.md}{\bttt{1.4-public-cve-history.md}}
& \bttt{./fixtures/cve-history-step-1.4-report.md} & desk research ---
CVE Program, NVD, GHSA, upstream advisories (see §References) \\
1.5 &
\href{https://github.com/orbitalab/RnD-ai-sandboxes-sec-study-part-1/blob/main/.docs/papers/1.5-patch-cadence.md}{\bttt{1.5-patch-cadence.md}}
& \bttt{./fixtures/patch-cadence-step-1.5-report.md} & desk research
--- engine release tags, product pin manifests, USN, gVisor
commit→release crosswalk (see §References) \\
1.6 &
\href{https://github.com/orbitalab/RnD-ai-sandboxes-sec-study-part-1/blob/main/.docs/papers/1.6-fuzzing-posture.md}{\bttt{1.6-fuzzing-posture.md}}
& \bttt{./fixtures/fuzzing-posture-step-1.6-report.md} & desk research
--- syzbot dashboard, OSS-Fuzz, cargo-fuzz upstreams (see
§References) \\
\end{longtable}
}

Per-check options, output JSON schema, verdict semantics, and the
negative-control fixtures under \texttt{./fixtures/} are documented in
\texttt{./README.md} of the companion repo.

\subsection{References}\label{references}

\subsubsection{Companion source papers}\label{companion-source-papers}

The six engine-level measurement papers ship as companion artefacts in
the public repo; §2.4--§2.6 inline their load-bearing material so this
synthesis reads standalone.

\begin{itemize}
\tightlist
\item
  \href{https://github.com/orbitalab/RnD-ai-sandboxes-sec-study-part-1/blob/main/.docs/papers/1.1-host-attack-surface.md}{1.1
  Host attack surface}
\item
  \href{https://github.com/orbitalab/RnD-ai-sandboxes-sec-study-part-1/blob/main/.docs/papers/1.2-information-leakage.md}{1.2
  Information leakage}
\item
  \href{https://github.com/orbitalab/RnD-ai-sandboxes-sec-study-part-1/blob/main/.docs/papers/1.3-defense-in-depth-stackability.md}{1.3
  Defense-in-depth stackability}
\item
  \href{https://github.com/orbitalab/RnD-ai-sandboxes-sec-study-part-1/blob/main/.docs/papers/1.4-public-cve-history.md}{1.4
  Public CVE history}
\item
  \href{https://github.com/orbitalab/RnD-ai-sandboxes-sec-study-part-1/blob/main/.docs/papers/1.5-patch-cadence.md}{1.5
  Patch cadence}
\item
  \href{https://github.com/orbitalab/RnD-ai-sandboxes-sec-study-part-1/blob/main/.docs/papers/1.6-fuzzing-posture.md}{1.6
  Fuzzing posture}
\item
  \href{https://github.com/orbitalab/RnD-ai-sandboxes-sec-study-part-1/blob/main/.docs/sandbox-isolation-methodology-v2.md}{Sandbox
  isolation methodology v2.3} --- §1.1--§1.6 govern this paper;
  revisions v2.1--v2.3 summarised in §``Methodology in brief''.
\end{itemize}

\subsubsection{Threat model}\label{threat-model}

\begin{itemize}
\tightlist
\item
  UK AI Security Institute (AISI). \emph{Technical guidance on
  sandboxing for AI evaluations}, August 2025. Threat-model taxonomy
  \texttt{T0.H2.N2} (single-tenant operator, actively hostile guest, no
  network egress). \url{https://www.aisi.gov.uk/}
\end{itemize}

\subsubsection{Related work --- AI-agent sandbox
precedents}\label{related-work-ai-agent-sandbox-precedents}

\begin{itemize}
\tightlist
\item
  Rabin, R., Hostetler, J., McGregor, S., Weir, B., Judd, N.
  \emph{SandboxEval: Towards Securing Test Environment for Untrusted
  Code.} arXiv:2504.00018, March 2025. 51-property test suite for
  LLM-generated-code execution environments; the closest
  ``test-suite-for-sandboxes'' precedent. SandboxEval probes
  per-environment properties; this paper measures engine-level
  architectural properties. \url{https://arxiv.org/abs/2504.00018}
\item
  Marchand, R., O Cathain, A., Wynne, J., Giavridis, P. M., Deverett,
  S., Wilkinson, J., Gwartz, J., Coppock, H. \emph{Quantifying Frontier
  LLM Capabilities for Container Sandbox Escape} (SANDBOXESCAPEBENCH).
  arXiv:2603.02277, March 2026. Inspect-AI CTF benchmark measuring
  LLM-agent capability to exploit seeded vulnerabilities and escape
  nested OCI containers. Orthogonal to this paper: SANDBOXESCAPEBENCH
  measures \emph{agent capability to exploit a sandbox given a
  vulnerability exists}; this paper measures \emph{engine properties
  that bound how many vulnerabilities exist and how reachable they are}.
  Their benchmark currently covers Docker/OCI only --- engines beyond
  that class (gVisor, Firecracker, Cloud Hypervisor, libkrun) are not
  yet exercised by an equivalent agent-capability benchmark.
  \url{https://arxiv.org/abs/2603.02277}
\item
  Yan, B. \emph{Fault-Tolerant Sandboxing for AI Coding Agents: A
  Transactional Approach to Safe Autonomous Execution.}
  arXiv:2512.12806, December 2025. Proposes transactional filesystem
  snapshots as an alternative-architecture to containers/microVMs; cited
  here as alternative-architecture candidate, not as empirical
  comparator (no empirical comparison against Docker/gVisor/Firecracker;
  tests only Gemini CLI). \url{https://arxiv.org/abs/2512.12806}
\item
  Bühler, C., Biagiola, M., Di Grazia, L., Salvaneschi, G.
  \emph{AgentBound: Securing Execution Boundaries of AI Agents.} Proc.
  ACM Softw. Eng. 3, FSE, Article FSE096 (FSE 2026). arXiv:2510.21236.
  MCP-side access control with \texttt{AgentBox} policy enforcement
  implemented on Docker + iptables + env whitelists; their §5 Discussion
  acknowledges the engine layer (Docker, Linux kernel features) as their
  trusted computing base, hand-off to the runc cluster measured here.
  \url{https://arxiv.org/abs/2510.21236}
\item
  Ruan, Y., Dong, H., Wang, A., Pitis, S., Zhou, Y., Ba, J., Dubois, Y.,
  Maddison, C. J., Hashimoto, T. \emph{Identifying the Risks of LM
  Agents with an LM-Emulated Sandbox} (ToolEmu). 12th ICLR (Vienna, May
  2024). arXiv:2309.15817v2. Earlier, widely-cited precedent for
  LM-emulated sandboxes for agent risk evaluation. Disambiguation:
  ToolEmu's ``sandbox'' is a GPT-4-based emulator that fabricates tool
  outputs from tool specifications --- explicitly \textbf{not}
  engine-level isolation; cite as lineage paper for the LM-emulated
  test-environment line. \url{https://arxiv.org/abs/2309.15817}
\end{itemize}

\subsubsection{Related work --- surveys}\label{related-work-surveys}

\begin{itemize}
\tightlist
\item
  Chhabra, A., Datta, S., Nahin, S. K., Mohapatra, P. \emph{Agentic AI
  Security: Threats, Defenses, Evaluation, and Open Challenges.} IEEE
  Access, Vol. 2026, 29 pp., DOI 10.1109/ACCESS.2026.3675554.
  arXiv:2510.23883v3, April 2026. Umbrella survey with four-family
  defense taxonomy carving out \emph{Sandboxing \& Capability
  Confinement} as one family (§IV.C) and an explicit §VI.C(4) open
  challenge calling for sandbox/emulated-environment fidelity
  measurement --- direct hand-off for this paper.
  \url{https://arxiv.org/abs/2510.23883}
\item
  Dehghantanha, A., Homayoun, S. \emph{SoK: The Attack Surface of
  Agentic AI --- Tools, and Autonomy.} arXiv:2603.22928, March 2026.
  7-goal × 10-surface × 5-path taxonomy with sandboxing as a single
  bullet in Appendix A.3 and one cell in Table 2 --- shallower than B.1
  on sandboxing, sibling-survey-positioning.
  \url{https://arxiv.org/abs/2603.22928}
\item
  Sroor, M., Das, T., Mohanani, R., Mikkonen, T. \emph{A Systematic
  Mapping Study on Risks and Vulnerabilities in Software Containers.}
  arXiv:2512.11940, December 2025. Systematic mapping of 129 primary
  studies (2000--2024) across a six-phase life-cycle taxonomy with 66
  distinct risks; verbatim observation \emph{``Image, host, and runtime
  phases receiving comparatively less attention''} and 9-of-129
  empirical-method rate motivate the engine-empirical contribution of
  this paper. The mapping enumerates no CVE IDs --- cited as landscape,
  not CVE-comparator. \url{https://arxiv.org/abs/2512.11940}
\end{itemize}

\subsubsection{Related work --- engine-level security
analyses}\label{related-work-engine-level-security-analyses}

The following four papers measure engine-level security on substrates
overlapping with our five-engine set. C.1 and C.2 are out of scope per
AISI T0.H2.N2 (microarchitectural side channels --- single-tenant
operator, malicious-output threat model excludes cross-tenant
micro-architectural leakage); cited for completeness and
engine-population overlap.

\begin{itemize}
\tightlist
\item
  Weissman, Z., Tiemann, T., Eisenbarth, T., Sunar, B.
  \emph{Microarchitectural Security of AWS Firecracker VMM for
  Serverless Cloud Platforms.} arXiv:2311.15999v1, November 2023.
  Empirical Spectre/MDS PoC study on Firecracker v1.0.0 and v1.4.0;
  load-bearing finding is the Medusa Cache-Indexing-Block-Write variant
  leaking across Firecracker microVMs but not on bare metal (Table 1).
  \textbf{Out of scope per AISI T0.H2.N2} (microarchitectural side
  channel). \url{https://arxiv.org/abs/2311.15999}
\item
  Dipta, D. R., Tiemann, T., Gulmezoglu, B., Marin, E., Eisenbarth, T.
  \emph{Dynamic Frequency-Based Fingerprinting Attacks against Modern
  Sandbox Environments.} arXiv:2404.10715v3, April 2024 (last revised
  May 2024). CNN-based two-phase fingerprinting via \texttt{cpufreq}
  against Docker, gVisor, Firecracker, Gramine, AMD-SEV --- same engine
  population as ours, different channel. \textbf{Out of scope per AISI
  T0.H2.N2} (microarchitectural side channel).
  \url{https://arxiv.org/abs/2404.10715}
\item
  Li, Y., Chen, Y., Ji, S., Zhang, X., Yan, G., Liu, A. X., Wu, C., Pan,
  Z., Lin, P. \emph{G-Fuzz: A Directed Fuzzing Framework for gVisor.}
  IEEE Transactions on Dependable and Secure Computing 21(1):168--185,
  January--February 2024. arXiv:2409.13139v1. The only academic
  directed-fuzzer for gVisor in the literature, industrially deployed at
  Ant Group; load-bearing for §2.6 gVisor row as evidence the fuzzing
  posture is actively invested (max \textbf{131× speedup over Syzkaller}
  on general targets in the published evaluation). Same syscall-level
  channel --- no OoS caveat. \url{https://arxiv.org/abs/2409.13139}
\item
  Moore, M., Zenla, A. \emph{Goldilocks Isolation: High Performance VMs
  with Edera.} arXiv:2501.04580, January 2025. \textbf{Vendor-authored
  whitepaper} proposing the Edera type-1 Xen-paravirtualized-per-pod
  runtime --- adjacent engine architecture outside the five-product set.
  Cited in §6 for landscape completeness with explicit vendor-framing
  caveat; vendor-reported performance numbers (Edera vs Docker / gVisor
  / Kata / Firecracker) not quoted as authoritative.
  \url{https://arxiv.org/abs/2501.04580}
\item
  Marin, E., Perino, D., Di Pietro, R. \emph{Serverless Computing: A
  Security Perspective.} arXiv:2107.03832v2, January 2022 (v1 July
  2021). Earliest widely-cited execution-environment survey spanning
  containers / gVisor / Firecracker; Table 2 is the four-class engine
  taxonomy (VMs · containers · g-Visor · microVMs) we anchor against in
  §1. Predates the AISI T0.H2.N2 framing; Cloud Hypervisor / libkrun
  absent from their taxonomy (consistent with 2021/2022 publication
  date). \url{https://arxiv.org/abs/2107.03832}
\end{itemize}

\subsubsection{Related work --- kernel attack surface, seccomp,
syscalls}\label{related-work-kernel-attack-surface-seccomp-syscalls}

\begin{itemize}
\tightlist
\item
  Zhan, D., Yu, Z., Yu, X., Zhang, H., Ye, L. \emph{Shrinking the Kernel
  Attack Surface Through Static and Dynamic Syscall Limitation.} IEEE
  Transactions on Services Computing, 2023. arXiv:2510.03720v1, October
  2025 preprint. Hybrid static+dynamic syscall-limitation methodology
  over 100 Docker-Hub utilities --- distinct from our axis 1.1
  (engine-level \emph{primitives reachable} vs application-level
  \emph{syscalls actually invoked}) but a methodology-precedent
  candidate for a future runtime-instrumented refinement.
  \url{https://arxiv.org/abs/2510.03720}
\item
  Wan, Z., Lo, D., Xia, X., Cai, L., Li, S. \emph{Mining Sandboxes for
  Linux Containers.} arXiv:1712.05493v1, December 2017. Foundational
  two-phase sandbox-mining for Docker containers; the verbatim baseline
  quote \emph{``By default, Docker disallows 44 system calls out of
  300+''} anchors the §2.1 Docker-default-seccomp framing and the §5.5
  daytona \texttt{Privileged:\ true} observation. Cited as historical
  2017 baseline; cross-checked 2026-05-30 against
  \texttt{vendor/github.com/moby/profiles/seccomp/default.json} at
  moby/moby tag \texttt{docker-v29.5.2} (the Docker-CE 29.5.2 release
  daytona pins via \texttt{containerd.io}): 361 syscalls unconditionally
  allowed and 426 unique syscall names across all ALLOW blocks once
  capability / kernel-version / architecture gating is unioned in ---
  \texttt{defaultAction:\ SCMP\_ACT\_ERRNO} (allow-list shape unchanged
  from D.2); the ``300+'' floor still holds.
  \url{https://arxiv.org/abs/1712.05493}
\item
  Ghimire, S., Bhurtel, N., Sahani, R., Jha, S. \emph{eBPF-PATROL:
  Protective Agent for Threat Recognition and Overreach Limitation using
  eBPF in Containerized and Virtualized Environments.}
  arXiv:2511.18155v1, November 2025 preprint (institutional affiliations
  not listed in the document). Single-engine (Docker) eBPF
  runtime-monitoring agent; load-bearing for §2.4.4 shared-kernel
  framing via verbatim §III: \emph{``As containerized workloads share
  the same kernel, any vulnerability that can be triggered via system
  calls \ldots{} can potentially compromise the host or other
  co-resident containers.''} \url{https://arxiv.org/abs/2511.18155}
\end{itemize}

\subsubsection{Related work --- fuzzing \& CVE
discovery}\label{related-work-fuzzing-cve-discovery}

\begin{itemize}
\tightlist
\item
  Liu, B., Zhang, Y., Cheng, L., Zhang, Y., Fan, J., Fu, Y.
  \emph{Psyzkaller: Learning from Historical and On-the-Fly Execution
  Data for Smarter Seed Generation in OS kernel Fuzzing.}
  arXiv:2510.08918v1, October 2025. Syzkaller \emph{extension} applying
  SDR-biased seed generation; \textbf{branch coverage +4.6--7.0\% over
  Syzkaller}, \textbf{crashes +110.4--187.2\%}, 8 previously unknown
  Linux-kernel bugs disclosed via CNVD (no MITRE/NVD IDs as of
  publication). Linux-kernel target only; cited as evidence the fuzzing
  tier is actively advancing at the seed-generation layer.
  \url{https://arxiv.org/abs/2510.08918}
\item
  Sun, Y., Kang, Y., Wu, C., Lu, K., Wang, J., Li, X., Hu, Y., Ren, J.,
  Lai, Y., Xie, M., Wang, Z. \emph{SyzParam: Incorporating Runtime
  Parameters into Kernel Driver Fuzzing.} arXiv:2501.10002v1, January
  2025. Syzkaller \emph{extension} incorporating sysfs parameter writes;
  \textbf{+32.57\% average edge coverage} over Syzkaller across 8
  drivers, \textbf{+34.5\% unique crashes} with concurrent-modification
  mutation, \textbf{30 unique bugs / 9 CVE IDs assigned}
  (CVE-2024-26622, CVE-2024-26933/26934, CVE-2024-27055, CVE-2024-36896,
  CVE-2024-37356, CVE-2024-38619, CVE-2024-39472, CVE-2024-39487 ---
  Linux-kernel row only by attribution policy).
  \url{https://arxiv.org/abs/2501.10002}
\item
  Yang, C., Zhao, Z., Zhang, L. \emph{KernelGPT: Enhanced Kernel Fuzzing
  via Large Language Models.} Proc. 30th ASPLOS, Rotterdam, March--April
  2025. arXiv:2401.00563v3. GPT-4 (T=0.1) three-stage prompting with
  MAX\_ITER=5 validate-then-repair loops; \textbf{+13.6\% (532 new
  syscalls + 294 types over Syzkaller's 3,903 baseline)}, \textbf{+2.3\%
  basic-block coverage over Syzkaller / +4.0\% over SyzDescribe} (Linux
  6.7 commit \texttt{d2f51b}), \textbf{24 previously unknown bugs / 21
  confirmed / 12 fixed / 11 CVE IDs} (CVE-2023-52429,
  CVE-2024-23848/23849/23850/23851/25739/25740/25741/26655/50277/50291).
  UIUC single-institution, public GitHub artifact,
  upstream-mainline-merged specs. \url{https://arxiv.org/abs/2401.00563}
\item
  Xu, J., Sun, H., Jiang, S., Wang, Q., Zhang, M., Li, X., Shen, K.,
  Zhang, C., Ji, S., Cheng, P., Chen, J. \emph{A Survey of Operating
  System Kernel Fuzzing.} ACM Transactions on Software Engineering and
  Methodology (TOSEM), accepted. arXiv:2501.16165v3, December 2025.
  107-paper SoK (2017--2025), 67\% security venues, 59\% Linux-target
  concentration, Syzkaller \emph{``has reported over 5,000 bugs''} ---
  landscape anchor for the §2.6 fuzzing-tier subsection. Taxonomy is
  functional (Environment Preparation × Input Model × Fuzzing Loop),
  orthogonal to our deployment-status tier split.
  \url{https://arxiv.org/abs/2501.16165}
\end{itemize}

\subsubsection{Related work --- patch
cadence}\label{related-work-patch-cadence}

\begin{itemize}
\tightlist
\item
  Heng, Y. W., Ma, Z., Zhang, H., Li, Z., Chen, T.-H. (P.).
  \emph{Discovery of Timeline and Crowd Reaction of Software
  Vulnerability Disclosures.} arXiv:2411.07480v3, November 2024. Manual
  two-author coding of 312 Java-ecosystem CVEs post-2017 into a
  six-event lifecycle (Reserved · Vendor-Fix · Vendor-Disclose ·
  CVE-Published · Community-Fix · Community-Disclose), establishing the
  fix-first disclosure pattern as dominant (85.89\%). Their corpus is
  disjoint from ours (Java libraries vs container engines); the authors'
  own External Validity disclaimer reads \emph{``the findings are
  confined to the Java software ecosystem, suggesting that the observed
  trends and reactions may not be applicable to vulnerabilities in other
  platforms.''} Cited as methodology-precedent for manual labeling of
  vulnerability-disclosure events, not as numerical comparator.
  \url{https://arxiv.org/abs/2411.07480}
\end{itemize}

\subsubsection{Related work --- dangerous-capability
framing}\label{related-work-dangerous-capability-framing}

\begin{itemize}
\tightlist
\item
  Phuong, M., Aitchison, M., Catt, E., Cogan, S., Kaskasoli, A.,
  Krakovna, V., Lindner, D., Rahtz, M., et al.~(27 authors, Google
  DeepMind). \emph{Evaluating Frontier Models for Dangerous
  Capabilities.} arXiv:2403.13793v2, April 2024. Four-area
  dangerous-capability programme (persuasion + cyber-security +
  self-proliferation + self-reasoning) on Gemini 1.0 Ultra/Pro/Nano
  without safety filters; abstract finding \emph{``We do not find
  evidence of strong dangerous capabilities in the models we evaluated,
  but we flag early warning signs.''} Expert-forecast threshold-crossing
  windows 2025--2029 (§8) anchor the front-matter
  ``containment-now-before-capability-arrives'' framing.
  \url{https://arxiv.org/abs/2403.13793}
\item
  Wei, B., Stroebl, B., Xu, J., Zhang, J., Li, Z., Henderson, P.
  \emph{Dynamic Risk Assessments for Offensive Cybersecurity Agents.}
  arXiv:2505.18384v5, October 2025 (v1 May 2025). Risk-assessment
  methodology for offensive AI agents --- five degrees-of-freedom
  framework + stateful-vs-non-stateful environment distinction +
  headline finding \textbf{\textgreater40\% relative improvement on
  InterCode CTF in \textless\$36 / 8 H100 GPU-hours}, \emph{``without
  any external assistance''}. \url{https://arxiv.org/abs/2505.18384}
\item
  Folkerts, L., Payne, W., Inman, S., Giavridis, P., Skinner, J.,
  Deverett, S., Aung, J., Zorer, E., Schmatz, M., Ghanem, M., Wilkinson,
  J., Steer, A., Hong, V., Wang, J. \emph{Measuring AI Agents' Progress
  on Multi-Step Cyber Attack Scenarios.} arXiv:2603.11214v3, March 2026
  (all 14 authors at the UK AI Security Institute, confirmed via v3 PDF
  title page; Deverett listed as work-done-while-at-AISI). Two
  cyber-attack ranges (\emph{``The Last Ones''} 32 steps,
  \emph{``Cooling Tower''} 7 steps) run under Inspect AI;
  \textbf{performance scales log-linearly with inference-time compute,
  with no observed plateau}; average steps trajectory 1.7 → 9.8 (5.8×)
  over $\approx$18 months at fixed 10M-token budgets; Opus 4.6 best run
  completes \textbf{22 of 32} corporate-network steps at 100M tokens.
  \url{https://arxiv.org/abs/2603.11214}
\end{itemize}

\subsubsection{Upstream engine projects}\label{upstream-engine-projects}

\begin{itemize}
\tightlist
\item
  \href{https://github.com/opencontainers/runc}{opencontainers/runc} ---
  OCI runtime.
\item
  \href{https://github.com/firecracker-microvm/firecracker}{firecracker-microvm/firecracker}
  --- microVM.
\item
  \href{https://github.com/cloud-hypervisor/cloud-hypervisor}{cloud-hypervisor/cloud-hypervisor}
  --- microVM.
\item
  \href{https://github.com/google/gvisor}{google/gvisor} --- userspace
  kernel (runsc).
\item
  \href{https://github.com/containers/libkrun}{containers/libkrun} ---
  microVM library.
\item
  Linux kernel --- shared host dependency for axes 1.4 and 1.6.
  \url{https://www.kernel.org/} · CVE announcement channel:
  \url{https://lore.kernel.org/linux-cve-announce/}
\end{itemize}

\subsubsection{Continuous-fuzzing
infrastructure}\label{continuous-fuzzing-infrastructure}

\begin{itemize}
\tightlist
\item
  \href{https://github.com/google/syzkaller}{google/syzkaller} ---
  kernel fuzzer.
\item
  \href{https://syzkaller.appspot.com/upstream}{syzbot --- Linux
  upstream dashboard}.
\item
  \href{https://syzkaller.appspot.com/gvisor}{syzkaller --- gVisor
  instance}.
\item
  \href{https://github.com/google/oss-fuzz}{google/oss-fuzz} ---
  continuous-fuzzing platform; hosts runc's two go-fuzz targets.
\item
  \href{https://github.com/rust-fuzz/cargo-fuzz}{rust-fuzz/cargo-fuzz}
  --- Cloud Hypervisor's 18-target in-tree harness compiles against it.
\item
  gVisor \texttt{secfuzz} --- in-tree seccomp-BPF fuzzing library;
  source under \texttt{pkg/secfuzz} in the gVisor tree.
\end{itemize}

\subsubsection{Product source artefacts cited in §5 and
§8}\label{product-source-artefacts-cited-in-5-and-8}

\begin{itemize}
\tightlist
\item
  \href{https://github.com/e2b-dev/fc-versions}{e2b-dev/fc-versions} ---
  e2b Firecracker version manifest.
\item
  \href{https://github.com/daytonaio/daytona}{daytonaio/daytona} ---
  daytona runner source (the \texttt{HostConfig.Privileged:\ true}
  setting cited in §5.5 and §8 lives here).
\end{itemize}

\subsubsection{Vulnerability advisories cited in
§2.4--§2.5}\label{vulnerability-advisories-cited-in-2.42.5}

CVE records are canonically resolvable via the
\href{https://www.cve.org/}{CVE Program} and
\href{https://nvd.nist.gov/}{NVD}; per-repository GHSA entries are
linked from each engine's GitHub repository under
\texttt{security/advisories}.

\textbf{Firecracker (§2.4, §2.5).}

\begin{itemize}
\tightlist
\item
  CVE-2026-5747 --- virtio-pci OOB write (Escape, potential; CVSS v4
  8.7).
\item
  CVE-2026-1386 --- jailer symlink host-write (Escape; CVSS v4 6.0).
\end{itemize}

\textbf{Cloud Hypervisor (§2.4, §2.5).}

\begin{itemize}
\tightlist
\item
  CVE-2026-45782 --- virtio-block async-I/O UAF (Escape, potential; CVSS
  v4 8.9).
\item
  CVE-2026-27211 --- QCOW backing-file HostLeak (CVSS v4 9.1; first
  concrete firing of the CVE.org tie-breaker rule, §2.4.5).
\end{itemize}

\textbf{runc (§2.4, §2.5).}

\begin{itemize}
\tightlist
\item
  CVE-2025-52565, CVE-2025-52881, CVE-2025-31133 --- 2025-11-05 procfs /
  mount-race trio (Escape).
\item
  CVE-2024-45310 --- filesystem-trick container escape (Escape,
  limited).
\item
  CVE-2024-21626 --- OCI mount-target traversal (Escape; ``Leaky
  Vessels'').
\item
  Ubuntu Security Notice USN-7851-1 --- snap-runc backport of the
  2025-11 trio, 2025-11-04. Ubuntu CVE tracker:
  \url{https://ubuntu.com/security/cves}.
\end{itemize}

\textbf{gVisor (§2.4, §2.5).}

\begin{itemize}
\tightlist
\item
  CVE-2025-2713 --- file-capabilities mishandling (silent-fix-first).
\item
  CVE-2024-10026 --- netstack weak-RNG class; academic disclosure by
  Inon Kaplan, Ron Even, and Amit Klein (Hebrew University \& Bar-Ilan).
\item
  CVE-2024-10603 --- weak RNG in TCP source port.
\item
  CVE-2023-7258 --- mount-state propagation.
\end{itemize}

\subsubsection{Kernel-LPE families referenced in §2.1 and
§2.4.4}\label{kernel-lpe-families-referenced-in-2.1-and-2.4.4}

The kernel CVEs listed in §2.1's primitive-reachability rationale column
(BPF JIT/verifier, io\_uring UAF, user-namespace LPE, ptrace credential
injection, perf subsystem, OCI mount traversal, FUSE path-confusion,
cgroup-v1 \texttt{release\_agent}, KVM nested-virt) are cited as
illustrative bug-class anchors per primitive, not as in-window findings.
Three syzkaller-typical kernel UAFs (CVE-2024-50264, CVE-2025-21756,
CVE-2026-31431) appear in §2.4.4 to illustrate the per-class
transitivity of the shared-kernel dependency. Each CVE is resolvable by
ID via NVD or CVE.org.

\end{document}